\documentclass[%
 preprint, 
 superscriptaddress,
nofootinbib,
 amsmath,amssymb,
 aps, physrev,
longbibliography 
]{revtex4-2}

\usepackage{graphicx}
\usepackage{bm}

\makeatletter
\def\switch@array{}
\makeatother
\usepackage{multirow}
\usepackage{booktabs}
\usepackage{tabularx}

\usepackage[pdftex,
  bookmarks=false,
  bookmarksopen=false,
  pdfstartpage=1,
  pdfstartview={FitH},
  colorlinks=true,
  linkcolor=blue,
  urlcolor=blue,
  citecolor=blue,
  linktoc=page  
  ]{hyperref}   

\begin{document}

\title{\textbf{Production of Magic States via $\boldsymbol{Z}$ Bosons and Dark Photons} 
}%

\author{Carlos Alvarado}
\email{calvarado@vassar.edu}
\affiliation{Department of Physics and Astronomy, Vassar College, Poughkeepsie, NY 12604, USA}
\affiliation{Dual CP Institute of High Energy Physics, C.P. 28045, Colima, M\'exico}

\author{Alfredo Aranda}
\email{fefo@ucol.mx}
\affiliation{Dual CP Institute of High Energy Physics, C.P. 28045, Colima, M\'exico}
\affiliation{Facultad de Ciencias, Universidad de Colima, C.P. 28045, Colima, M\'exico}

\author{C\'esar Bonilla}
\email{cesar.bonilla@ucn.cl}
\affiliation{Departamento de F\'isica, Universidad Cat\'olica del Norte, Avenida Angamos 0610, 1240000, Antofagasta, Chile}

\author{Yahir Lua}
\email{ylua@ucol.mx}
\affiliation{Facultad de Ciencias, Universidad de Colima, C.P. 28045, Colima, M\'exico}

\author{Ethan~Rodr\'iguez-Mart\'inez}
\email{ethan.rodriguez@alumnos.ucn.cl}
\affiliation{Departamento de F\'isica, Universidad Cat\'olica del Norte, Avenida Angamos 0610, 1240000, Antofagasta, Chile}

\begin{abstract}
The production of magic states is studied in two settings. The first is the electroweak (EW) sector of the Standard Model (SM). The second is an extension featuring a new broken $U(1)$ gauge symmetry and a Dirac fermion charged under it. This setup resembles a dark $U(1)$ scenario, with the additional fermion playing the role of a dark matter candidate that annihilates into SM particles through its coupling to the new gauge boson. In the EW sector, the low-energy regime reproduces earlier magic production results obtained for Quantum Electrodynamics, whereas the high-energy and $Z$-resonance regimes generate new magic distribution functions and non-trivially reorganize the stabilizer state classes, with Bhabha scattering exhibiting the strongest sensitivity to electroweak effects. Also, a subset of fixed stabilizer states is identified, for which the magic distributions remain unchanged across the different energy regimes. In the dark sector, the main effect of the new massive mediator is the appearance of new magic distributions functions for M\o ller-like, Bhabha-like, and inverse pair-annihilation processes in the low-energy limit. These reach the maximal magic value at the SM-to-dark fermion mass ratios $m_f/m_\chi \to  0$ and $m_f/m_\chi\to 1.83929$.
\end{abstract}

\maketitle

\newpage 

\tableofcontents

\section{Introduction}
\label{sec:intro}

Entanglement of states is absent from classical physics, as it is a truly quantum-mechanical phenomenon. Harnessing quantum phenomena in computing technologies has been envisioned since the 1980s \cite{QuantumComputerBenioff,QuantumComputerFeynman}, with the goal of utilizing them to solve problems that their classical counterparts cannot ---the so-called \textit{quantum advantage}---. Since then, the accumulated progress in information theory and developments in areas such as superconductivity (see the review in Ref. \cite{Superconduction}), quantum photonics \cite{PhotonicComputer}, and trapped ion manipulation \cite{IonTrapComputer}, has brought us to the so-called Noisy Intermediate-Scale Quantum (NISQ) computing era \cite{NISQ}, a time where quantum processors have been built and served beyond merely a proof-of-concept. Yet, the order of magnitude in the number of qubits employed in NISQ processors and their error-correction capabilities are still distant from those seeked by the quantum advantage \cite{quantum_error}.

Over the years, entanglement\footnote{Specifically, suitable measures of it, for instance the Von Neumann entropy.} has been one of the key resources exploited when looking for advantages of circuits over their classical counterparts, for example in algorithms such as Grover’s database search and Shor’s factorization \cite{Grover,Shor}. The rationale is that preparing certain maximally entangled states\footnote{Also known as Bell states.} in key locations through a circuit significantly reduces the number of queries needed to reach the algorithm's objective. Beyond quantum computing, developments around entanglement generation include estimating the feasibility of the usage of quantum bits to simulate a gauge lattice and measure non-perturbative effects \cite{Lattice_Gauge,Collider_phys}, the dynamics of quantum many-body systems \cite{ManyBody}, and even the information transport in a black holes \cite{BlackHoleInfo}.

One crucial finding towards the physical realization of a quantum computer is that a finite set of quantum gates turns out to be enough to simulate an arbitrary quantum operation on $n$-qubits, that is, the set constitutes a \textit{universal set of gates} \cite{DiVincenzoUniversal,Gottesman2}. On the other hand, according to the Gottesman-Knill theorem \cite{Gottesman3}, any quantum circuit composed solely of the CNOT, Hadamard ($H$), and phase ($S$) gates can be simulated with a classical computer at low computational cost (that is, in polynomial time \cite{Gottesman1,Gottesman3}), as long as it acts on states (dubbed \textit{stabilizers}) that are invariant under the set of three gates above (named the Clifford set, which is non-universal). An unavoidable implication is that entanglement alone is insufficient to achieve a genuine quantum advantage, as it is the action of non-Clifford gates who captures effects impossible to simulate with classical gates.

To quantify the effects of operations outside the Clifford set, an additional quantum resource was introduced in Ref. \cite{magic_states} which is known as \textit{magic} (or \textit{non-stabilizerness}). Magic is straightforward to calculate and quantifies the deviation of a quantum state from a classically simulable set, i.e. a state with zero magic is a stabilizer state, carrying no computational advantage. A state could saturate the maximum magic value, in fact for an $n$-qubit system the upper bound on magic depends on the dimension of the corresponding Hilbert space and the order of the \textit{R\'enyi} entropy chosen in its definition. In the same way that entanglement entropy diagnoses non-classical correlations between subsystems, magic diagnoses the non-classical character of the state itself. Together with entanglement, magic is a necessary ingredient for universal fault-tolerant quantum computation, and its production, distillation, and characterization have been studied across a wide range of platforms \cite{distillation,Tao}. Both resources can be extracted from the density matrix of a two-qubit final state, and for particle scattering systems whose spins are chosen as the qubit degrees of freedom, one could interpret particle colliders as natural sandboxes to study quantum information across a wide range of energies \cite{QuantumInfCollider}. For 2-level, spin-1/2 states the corresponding qubits can be defined either through spin projections $|\pm\rangle$ \cite{spins} or helicities $|R/L\rangle$ \cite{Tao2}, mapped onto the 2-qubit computational basis $\{|00\rangle,|01\rangle,|10\rangle,|11\rangle\}$. We deem pertinent to remind the reader that entanglement and magic, while related, are different quantities. In particular, it is possible for a 2-qubit state to simultaneously display null magic and maximal entanglement. The interplay between both resources, with a state's entanglement quantified in terms of its \textit{concurrence} \cite{concurrence}, has been recently explored in Ref. \cite{RegionsEntVERSUSMag} where it was conveniently represented in a two-dimensional histogram in the entanglement-magic plane.

While implementing non-stabilizer quantum circuitry in a way that is both fault-tolerant and large-scale remains an enourmous technical challenge, it is of interest to determine how non-stabilizerness itself arises at the level of fundamental interactions in high energy physics. A list of previous endeavors in this direction, by no means comprehensive, includes the determination of magic production in purely gauge spin-1 and spin-2 theories \cite{MagicGluonGraviton}, in top-antitop quark production at colliders \cite{ttbarMagic}, in low-energy baryon collisions \cite{baryonMagic}, and in electron/muon scattering in Quantum Electrodynamics (QED) \cite{magic_in_qed}. This last paper presented a detailed look into the capability of $2\to2$ fermion scattering for producing magic under photon exchange. One could ask how the incorporation of the $Z$ boson exchange alters this capability, and if so, what is its energy dependence. Moreover, one could also ask how this resource behaves when the center-of-mass (CM) energy hits the $Z$ resonance.

In the present work, we study the production of magic in 2-qubit, spin-1/2 systems via $2\to2$ tree-level scattering processes mediated by electrically neutral gauge bosons. In preparation for that task, the notation is established in Sec.~\ref{sec:notation}, as well as its connection with earlier works. Next, in Sec.~\ref{sec:magicEW}, magic generation is looked at under the exchange of neutral electroweak bosons between Standard Model fermions, comparing it to the case of photon-only exchange. Before addressing the dark symmetry setup, a short discussion on sets of stabilizer states with particular behavior under the different energy regimes is presented in Sec. \ref{sec:FixedStabStates}. Later, in Secs. \ref{sec:DarkU1} and \ref{sec:DarkMagic}, we turn our attention at the exchange of a massive mediator of a dark $U(1)$ symmetry between SM fermions and hypothetical SM-singlet, $U(1)$-charged fermions. There, the new physics effect on the magic resource of the final states is parametrized. Once that the SM electroweak and beyond-the-SM dark $U(1)$ scenarios have been investigated numerically, comparisons are drawn, and the emergence of maximal magic functions is commented on in Sec. \ref{sec:MaxMagic}. Finally, the conclusions are stated.

\section{Notation and conventions}
\label{sec:notation}

Following Ref. \cite{magic_in_qed}, we adopt the stabilizer formalism and determine the set of initial states that undergo scattering. There are exactly 60 stabilizer states $|\psi_{\text{stab}}^{(k)}\rangle$ of two qubits, including both separable and maximally entangled configurations. In the computational basis defined by the two fermion spins, $\{ |1\rangle,|2\rangle,|3\rangle,|4\rangle \}\equiv\{ \mid\uparrow\uparrow\rangle,\mid\uparrow\downarrow\rangle,\mid\downarrow\uparrow\rangle,\mid\downarrow\downarrow\rangle \}$, these states admit a decomposition $|\psi_{\text{stab}}^{(k)}\rangle=\sum_{i=1}^{4}c_i^{(k)}|i\rangle$ whose coefficients are found at the end of Ref. \cite{magic_in_qed}. The various $|\psi_{\text{stab}}^{(k)}\rangle$ are invariant under the action of $\mathcal{P}_2$, the set of two-qubit Pauli strings with phase $+1$, i.e.\ $\mathcal{P}_2 = \{P_1 \otimes P_2\}$, $P_i \in \{I,X,Y,Z\}$, where $I$, $X$, $Y$, and $Z$ are the $2\times2$ identity and three usual Pauli matrices. The (final) scattered states $|\psi\rangle$ are obtained by the action of the transition operator $S$,
\begin{equation}
    S = 1 + iT \qquad \langle f|T|i\rangle = (2\pi)^4 \delta^{(4)}\left(\sum p_i 
    - \sum p_f \right) \mathcal{M}_{i\to f},
    \label{eq:scattering}
\end{equation}
as $|\psi\rangle=S|\psi_{\text{stab}}^{(k)}\rangle$, and their magic is obtained by computing the second stabilizer R\'enyi entropy (SRE) under $\mathcal{P}_2$ \cite{TwoRenyiEntropy} ,
\begin{equation}
    M_2(|\psi \rangle) = -\log \Xi_2(|\psi \rangle), \qquad 
    \Xi_2(|\psi \rangle) \equiv \sum_{P \in \mathcal{P}_2} 
    \frac{\langle \psi|P|\psi\rangle^4}{4}.
    \label{eq:Second-Reny-entropy}
\end{equation}
For two-qubit states, it turns out that the upper bound on the corresponding SRE \cite{max_magic} is
\begin{equation}
    M_2 \leq \log\left(\frac{16}{7}\right) \approx 0.827,
    \label{eq:maximal_magic}
\end{equation}
a numerical value that will be used as a benchmark throughout the analysis. In the following sections, we will obtain the magic distribution functions of the stabilizer states under the various $2\to 2$ scatterings as functions of kinematical parameters.

\section{Magic production in EW theory}
\label{sec:magicEW}

Ref. \cite{magic_in_qed} studied the production of quantum magic for $2\rightarrow 2$ fermion scattering mediated by tree-level photon exchange, as a function of the scattering angle $\theta$ and the mass ratio $\lambda\equiv m_{e}/m_{\mu}$. In the Standard Model, neutral-current scattering receives an additional contribution from $Z$-boson exchange at the same perturbative order. The present section extends the QED analysis by including the $Z$-boson contribution to tree-level neutral current scattering, under the same functional measure of quantum magic. Though only electrons and muons will participate in the scatterings, the inclusion of the tau lepton would be straightforward as it simply introduces slightly smaller electron-to-tau and muon-to-tau mass ratios.

The scattering amplitude is thus written as the coherent sum
\begin{equation}
    \mathcal{M}=\mathcal{M}_\gamma+\mathcal{M}_Z,
\end{equation}
where $\mathcal{M}_\gamma$ alone reproduces the QED results mentioned before, and $\mathcal{M}_Z$ accounts for the electroweak interaction. As opposed to photon exchange, the $Z$-boson contribution involves a massive propagator and chiral couplings to fermions, leading to effects that in principle modify the spin structure of the scattered two-fermion state. This boson exchange is described by the following massive propagator
\begin{equation}
    \frac{-i}{q^2-m_Z^2+im_Z\Gamma_Z}\left( g^{\mu\nu}-\frac{q^\mu q^\nu}{m_Z^2}\right), \label{eq:ew_propagator}
\end{equation}
where $m_Z$ and $\Gamma_Z$ denote the mass and decay width of the $Z$ boson, respectively, and are set to the world-average values $m_Z=91.188\,\text{GeV}$ and $\Gamma_Z=2.4955\,\text{GeV}$ reported by the Particle Data Group \cite{PDG2024}. Likewise, at energies near $m_{Z}$ and higher the weak angle is taken as $s_{W}^{2}\approx 0.231$ in accordance with results from LHCb $pp$ collisions \cite{s2wLHC_2015}, and as $s_{W}^{2}\approx 0.239$ otherwise \cite{s2wLowE}.

\subsection{Low-energy limit}
\label{subsec:low_Elimit}

We first consider the low-energy regime below the $Z$-boson resonance, defined by $\sqrt{s}<m_{Z}$. The full scattering amplitude must incorporate the neutral-current contribution, thus the amplitude becomes the sum of the photon and $Z$-boson exchange amplitudes, as previously stated.

Given that in the low-energy regime $q^2\ll m_Z^2$, the massive propagator of the $Z$ boson simplifies considerably: neglecting terms suppressed by $q^2/m_Z^2$, it reduces to
\begin{equation}
     \frac{-i}{q^2-m_Z^2+im_Z\Gamma_Z}\left( g^{\mu\nu}-\frac{q^\mu q^\nu}{m_Z^2}\right)\approx \frac{-ig^{\mu\nu}}{m_Z^2}.
\end{equation}
The expression above is strongly suppressed compared to the propagator for photon exchange, which scales as $1/s$ instead.

The analysis of quantum magic in the low-energy regime follows the same procedure as that in Ref. \cite{magic_in_qed}: for each scattering process, the spin amplitudes are constructed in the spin basis described earlier, and grouped according to the magic function induced by each initial stabilizer state. The only modification is the inclusion of the $Z$-boson exchange contribution.

\subsubsection*{Pair annihilation $e^{-}e^{+}\rightarrow \mu^{-}\mu^{+}$}
\label{subsubsec:PairAnniZ}

The analysis begins with electron-positron pair annihilation, which proceeds exclusively through the $s$-channel. This amplitude is non-vanishing only when the CM energy satisfies the kinematic threshold condition $\sqrt{s}\geq 2m_\mu$. For this reason, the low-energy limit fixes $s$ at the muon-antimuon threshold, where the scattering angle $\theta$ is not well-defined, and the magnitude of the three-momentum satisfies
\begin{equation*}
    \frac{|\vec{p}|}{m_\mu}\stackrel{\text{threshold}}{=}\sqrt{1-\lambda^2}.
\end{equation*}
The neutral-current contribution $\mathcal{M}_{Z}$ introduces dependence on the weak angle, as well as an additional mass ratio $\lambda_Z\equiv m_{Z}/m_{\ell}$ depending on the charged lepton $\ell$ involved. The matrix of the total amplitude reduces to a matrix that is constant in $s$ in the corresponding computational basis.

As an aside, while an overall $e^2$ factor appears in the full amplitude entries, it is a global factor that cancels out once the scattered state is normalized. Since the magic measure depends only on the normalized state, global constants do not affect the resulting stabilizer R\'enyi entropy. For this reason, these factors will be omitted throughout this analysis. 

\begin{table}[ht]
    \centering
    \begin{tabularx}{\textwidth}{cXcc}
        \toprule
        Magic distribution & \   Stabilizer state & $(M_2)_{\text{max}}$ & $\theta_{\text{max}}$ or $(\lambda_\chi)_{\text{max}}$\\
        \toprule
        $\mathcal{F}_1$ & 1, 2, 3, 4, 5, 6, 9, 10, 37, 38, 39, 40, 42, 43, 44, 45, 48, 49, 50 & 0 & ---\\
        \hline
        $\mathcal{G}_1$ & 7, 8, 11, 12, 46, 47, 59, 60  & $\log(4/3)$ & $-1 + \sqrt{2}$\\
        \hline
        $\mathcal{G}_2$ & 13, 14, 15, 16, 17, 18, 19, 20, 21, 22, 23, 24, 25, 26, 27, 28, 29, 30, 31, 32, 33, 34, 35, 36, 51, 52, 53, 54, 55, 56, 57, 58 & $\log(9/5)$ & 1 \\
        \hline
        Vanishing & 41 & --- & ---\\
        \bottomrule
    \end{tabularx}
    \caption{Magic distributions and corresponding stabilizers obtained from $\mathcal{M}_{\gamma}+\mathcal{M}_{Z}$ under pair annihilation in the low-energy limit.}
    \label{tab:AnniComplete}
\end{table}

The resulting magic distribution functions exactly coincide with those obtained in the photon-exchange case in Ref. \cite{magic_in_qed} as shown in Table \ref{tab:AnniComplete}. In this table, the magic distributions functions of $\theta$ with no $\lambda$ dependence are denoted by $\mathcal{F}$, while those with  both $\theta$ and $\lambda$ dependence are denoted by $\mathcal{G}$. These three functions \footnote{The analytical form can be found in Ref. \cite{magic_in_qed}.}, later found again in a few processes in the dark $U(1)$ theory, have been graphed in Fig. \ref{fig:D_Anni_SML}.

One naively expects in the full amplitude a dependence on the weak angle and the mass ratio $\lambda_Z$, and hence a corresponding deviation from the pure QED behavior, but this turns out not to be true, After all, the $Z$-exchange contribution is known to be subleading with respect to the photon amplitude.

\subsubsection*{Inverse pair annihilation $\mu^-\mu^+ \rightarrow e^-e^+$}
\label{subsubsec:InvPairAnn}

The inverse annihilation process is next considered, which at tree level, involves only the $s$-channel too. Only the low-energy regime is discussed explicitly for this process, since at high energies its amplitude structure and thus its magic behavior coincides with that of direct pair annihilation. Unlike $e^-e^+\rightarrow \mu^-\mu^+$, this scattering does not posses a kinematic threshold, then in order to investigate its low-energy behavior, the nonrelativistic limit for the incoming muons is taken,
\begin{equation*}
    \frac{|\vec{p}|}{m_\mu}\rightarrow 0,
\end{equation*}
which leads to a modified structure of the scattering amplitude.

In contrast to its corresponding reverse process, whose $\sqrt{s}$ sits at the muon-antimuon threshold, the full $\mathcal{M}$ exhibits matrix elements that depend on the scattering angle $\theta$. Still, after constructing the scattered state and evaluating the corresponding R\'enyi entropy, the angular magic distribution functions are found to be identical to those obtained in the photon-exchange case. The results are organized in Table \ref{tab:invAnniComplete}, whose corresponding magic distributions are found in Ref. \cite{magic_in_qed}.

\begin{table}[ht]
    \centering
    \begin{tabularx}{\textwidth}{cX}
        \toprule
        Magic distribution & Stabilizer state \\
        \toprule
        $\mathcal{G}_3$ & 1, 2, 39,40 \\
        \hline
         $\mathcal{G}_4$ & 3, 4, 42, 43, 44 \\
        \hline
         $\mathcal{G}_5$ & 5, 6, 37, 49, 50 \\
        \hline
         $\mathcal{G}_6$ & 7, 8, 59, 60 \\
        \hline
         $\mathcal{F}_1$ & 9, 10, 38, 45, 48 \\
        \hline
         $\mathcal{G}_7$ & 11, 12 \\
        \hline
         $\mathcal{G}_8$ & 13, 14, 15, 16, 17, 18, 19, 20, 21, 22, 23, 24, 25, 26, 27, 28 \\
        \hline
         $\mathcal{G}_9$ & 29, 30, 31, 32, 52, 54, 57, 58 \\
        \hline
         $\tilde{\mathcal{G}}_9$ & 33, 34, 35, 36, 51, 53, 55, 56 \\
        \hline
        --- & 41 \\
        \hline
         $\mathcal{G}_{10}$ & 46 \\
        \hline
         $\tilde{\mathcal{G}}_{10}$ & 47 \\
        \bottomrule
    \end{tabularx}
    \caption{Magic distributions and corresponding stabilizers obtained from $\mathcal{M}_{\gamma}+\mathcal{M}_{Z}$ under inverse pair annihilation in the low-energy limit.}
    \label{tab:invAnniComplete}
\end{table}

\subsubsection*{M\o ller, Bhabha, and elastic scattering}
\label{subsubsec:remaining}

Next we analyze the remaining processes, which are M\o ller scattering $e^-e^-\rightarrow e^-e^-$, Bhabha scattering $e^-e^+\rightarrow e^-e^+$, and elastic scattering $e^-\mu^-\rightarrow e^-\mu^-$.
By organizing the amplitudes of these three processes in terms of the parameter $\mu\equiv|\vec{p}|/m_\mu$, and working in the low-energy regime, one finds that the photon-mediated contribution scale as $O(\mu^{-2})$, while the $Z$-boson exchange contributions are $O(\mu^0)$. 
Since in the low-energy regime $\mu\ll 1$, the dominant contribution to the scattering amplitude arises from the photon exchange term and the $Z$-boson contribution is therefore subleading and unable to modify the leading structure of the scattering matrix. Consequently, to leading order in $\mu$, $\mathcal{M}_{\gamma}+\mathcal{M}_{Z}\approx \mathcal{M}_{\gamma}$ and these three processes do not generate new angular magic distribution functions beyond those obtained from photon-exchange \cite{magic_in_qed}.

\subsection{High-energy limit}
\label{subsec:high_Elimit}

The study of quantum magic in this regime is carried out for the same set of scattering processes considered in the low-energy case.

In the regime of high energy, the momentum transfer satisfies $q^2 \gg m_Z^2$. Thus, both the mass and width of the $Z$ boson become negligible in the propagator's denominator and thus Eq. (\ref{eq:ew_propagator}) simplifies to
\begin{equation}
    \frac{-i}{q^2}\left( g^{\mu\nu}-\frac{q^\mu q^\nu}{m_Z^2}\right).
\end{equation}
This behavior contrasts with the low-energy limit, where the propagator reduces to an energy-independent interaction proportional to $1/m_Z^2$. Instead, in the high-energy regime it reduces to an interaction proportional to $1/q^2$.

Unlike the low-energy case, the term proportional to $q^\mu q^\nu$ cannot be neglected based on $m_{Z}$ suppression, and it must be looked at explicitly. The $q^\mu q^\nu$ piece introduces an additional contribution to the amplitude of the form
\begin{equation*}
    \frac{1}{m_Z^2}j_{\text{NC}}^\mu q_\mu q_\nu j_{\text{NC}}^\nu.
\end{equation*} 
Here, $j_{NC}^\mu$ denotes the fermionic neutral current,
\begin{equation*}
    j_{NC}^\mu=\bar{f}\gamma^\mu\left(g_V^f-g_A^f\gamma^5 \right)f,
\end{equation*}
associated with the $Z$-boson interaction, where $g_V^f$ and $g_A^f$ are the vector and axial-vector coupling constants for the specific fermion respectively.

However, in the high-energy limit the external fermions can be treated as effectively massless in virtue of the Dirac equation,
\begin{equation*}
    (p-m)u(p)=0\quad\longrightarrow\quad \gamma^\mu p_\mu u(p)=0.
\end{equation*}
As a result, $j_{\text{NC}}^\mu q_\mu\approx0$ and the $q^\mu q^\nu$ term can be disregarded  effectively reducing the propagator  to
\begin{equation}
    \frac{-i}{q^2}g^{\mu\nu} \ .
\end{equation}

\subsubsection*{Pair annihilation $e^-e^+\rightarrow \mu^-\mu^+$}

\begin{table}[ht]
    \centering
    \begin{tabularx}{\textwidth}{cXcc}
        \toprule
        Magic distribution & Stabilizer state & $(M_2)_{\text{max}}$ & $\theta_{\text{max}}$ (rad)\\
        \toprule
        $\mathcal{F}_{\text{HE -}1}$ & 1, 2, 13, 14, 15, 16, 17, 18, 19, 20, 21, 22, 23, 24, 25, 26, 27, 28, 29, 30, 31, 32, 33, 34, 35, 36, 39, 40, 51, 52, 53, 54, 55, 56, 57, 58 & 0.581\dots & 0.790\dots\\
        \hline
        $\mathcal{F}_5$ & 3, 4, 42, 43, 44  & $\log(4/3)$ & $\pi/8$, $3\pi/8$, $5\pi/8$, $7\pi/8$ \\
        \hline
       $\mathcal{F}_{\text{HE -}2}$ & 5, 6, 11, 12, 37, 46, 47, 49, 50 & 0.534\dots & 1.94\dots\\
        \hline
        $\mathcal{F}_1$ & 7, 8, 9, 10, 38, 45, 48, 59, 60 & 0 & Arbitrary \\
        \hline
        --- & 41 & --- & ---\\
        \bottomrule
    \end{tabularx}
    \caption{Magic distributions and corresponding stabilizers obtained from $\mathcal{M}_{\gamma}+\mathcal{M}_{Z}$ under pair annihilation in the high-energy limit.}
    \label{tab:HEAnniComplete}
\end{table}
In the high-energy limit, the incoming states are ultrarelativistic and 
\begin{equation*}
    \frac{|\vec{p}|}{m_\mu}\rightarrow\infty,
\end{equation*}
implying a series expansion in the small parameter $1/\mu$ for the full amplitude $\mathcal{M}_{\gamma}+\mathcal{M}_{Z}$. When reporting the amplitude analytic expressions, terms will be retained up to order $O(1/\mu)$, an approximation that will suffice for our analysis. For some stabilizer states, the leading contribution to the amplitude is of order $O(\mu^0)$, while for others the expansion starts at order $O(1/\mu)$, validating our truncation in the power expansion.

\begin{figure}[ht]
    \centering
    \includegraphics[width=\linewidth]{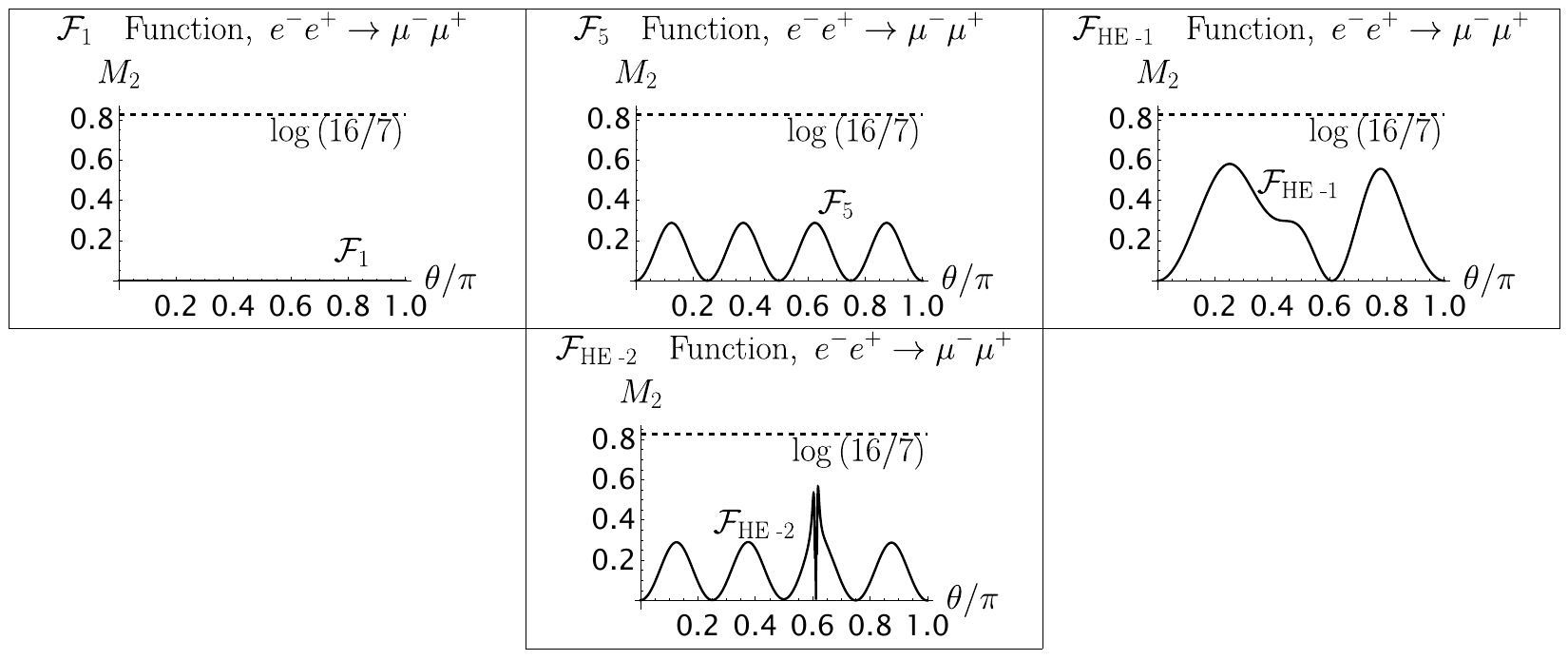}
    \caption{Angular dependence of the magic distributions obtained from $\mathcal{M}_{\gamma}+\mathcal{M}_{Z}$ under pair annihilation in the high-energy limit.}
    \label{fig:HEAnni}
\end{figure}

The resulting angular magic distribution functions (named as $\mathcal{F}_{\text{HE -}n}$) are summarized in Table \ref{tab:HEAnniComplete} and displayed in Fig. \ref{fig:HEAnni}. These new magic functions, which did not arise in the pure photon-exchange case, encode the additional contribution of the neutral-current and reflect the modified spin structure of $\mathcal{M}$ at high energies.

Compared to the pure QED case, the inclusion of the $Z$-boson exchange preserves the trivial stabilizer classes associated with the magic functions $\mathcal{F}_1$ and the vanishing distribution of stabilizer $\#41$, but modifies the structure of the nontrivial magic distributions. In particular, the QED magic function $\mathcal{F}_4$ is replaced by a new distribution, $\mathcal{F}_{\text{HE -}1}$, which keeps only one of the two original maxima of the photon-exchange. In addition, the class of stabilizers leading to $\mathcal{F}_5$ is only partially preserved, because part of it gives rise to a new class defined by the function $\mathcal{F}_{\text{HE -}2}$ exhibiting an additional sharp peak compared to pure QED. One observes that the neutral-current does not simply replicates the QED results, but also generates new magic distributions functions due to $\mathcal{M}_{Z}$ becoming non-negligible.

\subsubsection*{M\o ller scattering $e^-e^-\rightarrow e^-e^-$}

\begin{figure}[ht]
    \centering
    \includegraphics[width=1\linewidth]{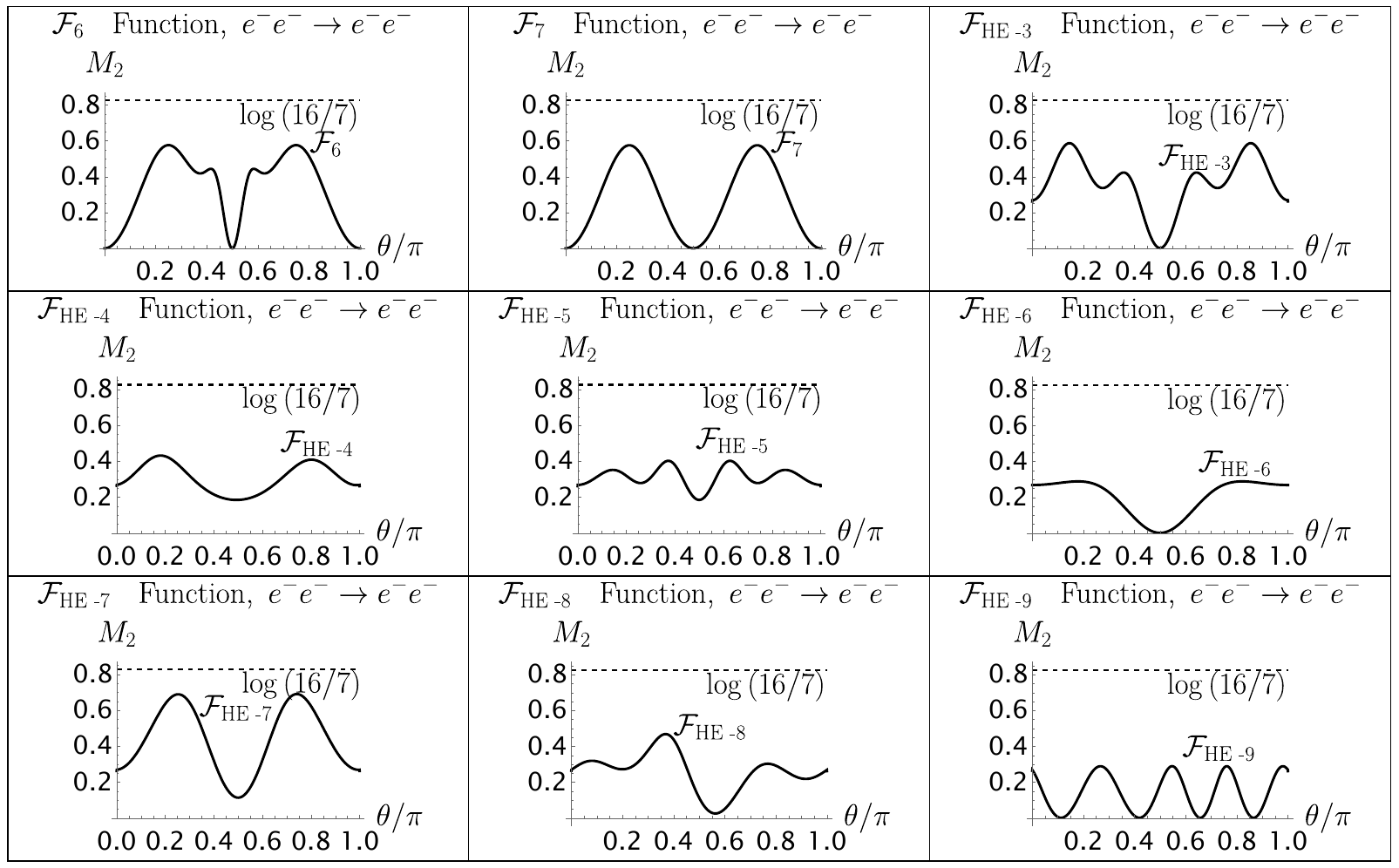}
    \caption{Angular dependence of the magic distributions obtained from $\mathcal{M}_{\gamma}+\mathcal{M}_{Z}$ under M\o ller scattering in the high-energy limit.}
    \label{fig:HEMoller}
\end{figure}

Evaluation of $\mathcal{M}_{\gamma}+\mathcal{M}_{Z}$ leads to the angular magic distribution functions summarized in Table \ref{tab:HEMollerComplete} and shown in Fig. \ref{fig:HEMoller}.
\begin{table}[ht]
    \centering
    \begin{tabularx}{\textwidth}{cXcc}
        \toprule
        Magic distribution & Stabilizer state & $(M_2)_{\text{max}}$ & $\theta_{\text{max}}$ (rad)\\
        \toprule
        $\mathcal{F}_6$ & 1, 2, 39, 40 & 0.576\dots & $(\pi/2)\pm0.783\dots$ \\
        \hline
        $\mathcal{F}_7$ & 3, 4, 43, 44  & $\log(16/9)$ & $\pi/4$, $3\pi/4$ \\
        \hline
        $\mathcal{F}_{\text{HE -}3}$ & 5, 6, 49, 50 & 0.587\dots & 0.457\dots \\
        \hline
         $\mathcal{F}_{\text{HE -}4}$ & 7, 8, 59, 60 & 0.432\dots & $(\pi/2)\pm0.319\dots$ \\
        \hline
        $\mathcal{F}_{\text{HE -}5}$ & 9, 10 & 0.403\dots & $(\pi/2)\pm0.127\dots$ \\
        \hline
        $\mathcal{F}_{\text{HE -}6}$ & 11, 12, 46, 47 & 0.289\dots & $(\pi/2)\pm0.321\dots$ \\
        \hline
        $\mathcal{F}_{\text{HE -}7}$ & 13, 14, 15, 16, 17, 18, 19, 20, 21, 22, 23, 24, 25, 26, 27, 28 & 0.690\dots & $(\pi/2)\pm0.241\dots$ \\
        \hline
        $\mathcal{F}_{\text{HE -}8}$ & 29, 31, 34, 36, 55, 56, 57, 58 & 0.469\dots & 1.16\dots\\
        \hline
        $\widehat{\mathcal{F}}_{\text{HE -}8}$ & 30, 32, 33, 35, 51, 52, 53, 54 & 0.469\dots & 1.98\dots \\
        \hline
        $\mathcal{F}_1$ & 37, 42 & 0 & Arbitrary \\
        \hline
        $\mathcal{F}_5$ & 38, 41 & $\log(4/3)$ & $\pi/8$, $3\pi/8$, $5\pi/8$, $7\pi/8$ \\
        \hline
        $\mathcal{F}_{\text{HE -}9}$ & 45 & 0.289\dots & 0.838\dots, 1.72\dots, 2.39\dots, 3.08\dots \\
        \hline
       $\widehat{\mathcal{F}}_{\text{HE -}9}$ & 48 & 0.289\dots & 0.0617\dots, 0.751782\dots, 1.42\dots, 2.30\dots\\
        \bottomrule
    \end{tabularx}
    \caption{Magic distributions and corresponding stabilizers obtained from $\mathcal{M}_{\gamma}+\mathcal{M}_{Z}$ under M\o ller scattering in the high-energy limit.}
    \label{tab:HEMollerComplete}
\end{table}

Compared to the pure QED case, the inclusion of the $Z$-boson exchange in high-energy M\o ller scattering furnishes a much richer pattern of angular magic distributions. While the stabilizers associated with distributions $\mathcal{F}_1$, $\mathcal{F}_5$, $\mathcal{F}_6$, $\mathcal{F}_7$ remain unchanged, the rest now fall under new distributions with modified angular features and, in several cases, with different maximal values.

This modification of the angular profiles is not uniform, some stabilizer classes are enhanced and some suppressed. For example, the QED distribution $\mathcal{F}_9$, with maximum approximately at $0.586$, is replaced by $\mathcal{F}_{\text{HE -}4}$ whose maximum is reduced to about $0.432$. On the other hand, the class associated with $\mathcal{F}_{10}$ is enhanced, since $\mathcal{F}_{\text{HE -}5}$ reaches a larger maximum than that of QED.

Furthermore, in several sectors the maximal value remains approximately unchanged, but the shape of the distribution is altered. This happens, for example, when comparing $\mathcal{F}_8$ with $\mathcal{F}_{\text{HE -}3}$, and $\mathcal{F}_2$ with $\mathcal{F}_{\text{HE -}6}$. In such cases, the $Z$-boson contribution does not significantly change the ballpark value of the magic generated, but it does reorganize its angular dependence, whether by producing new peaks or by shifting the location of its minima. 

Finally, angular relations between distributions persists in the complete full amplitude. As in the pure QED analysis, this gives rise to pairs of magic distributions with the same overall profile but displaced by in $\theta$, which are named as $\widehat{\mathcal{F}}_{n,i}\equiv\mathcal{F}_n(\theta+\alpha_i)$. These type of functions are stored in Appendix. \ref{appendix:Angular_shifts}.

The $\theta$ relations are discernible between pairs of distributions such as $\mathcal{F}_{\text{HE -}8}$ and $\widehat{\mathcal{F}}_{\text{HE -}8}$, as well as $\mathcal{F}_{\text{HE -}9}$ and $\widehat{\mathcal{F}}_{\text{HE -}9}$. Similar pairs of magic distribution functions already appeared in QED, for example $\mathcal{F}_{12}$, $\widehat{\mathcal{F}}_{12}$ and $\mathcal{F}_{13}$, $\widehat{\mathcal{F}}_{13}$. Thus, while the inclusion of the $Z$-boson exchange modifies the detailed angular functions and maximal values of the magic distributions, the appearance of angular shifts between the same sectors, already present in QED, is preserved.

\subsubsection*{Bhabha scattering $e^-e^+\rightarrow e^-e^+$}

\begin{figure}[ht]
    \centering
    \includegraphics[width=1\linewidth,trim={0 205 0 0}]{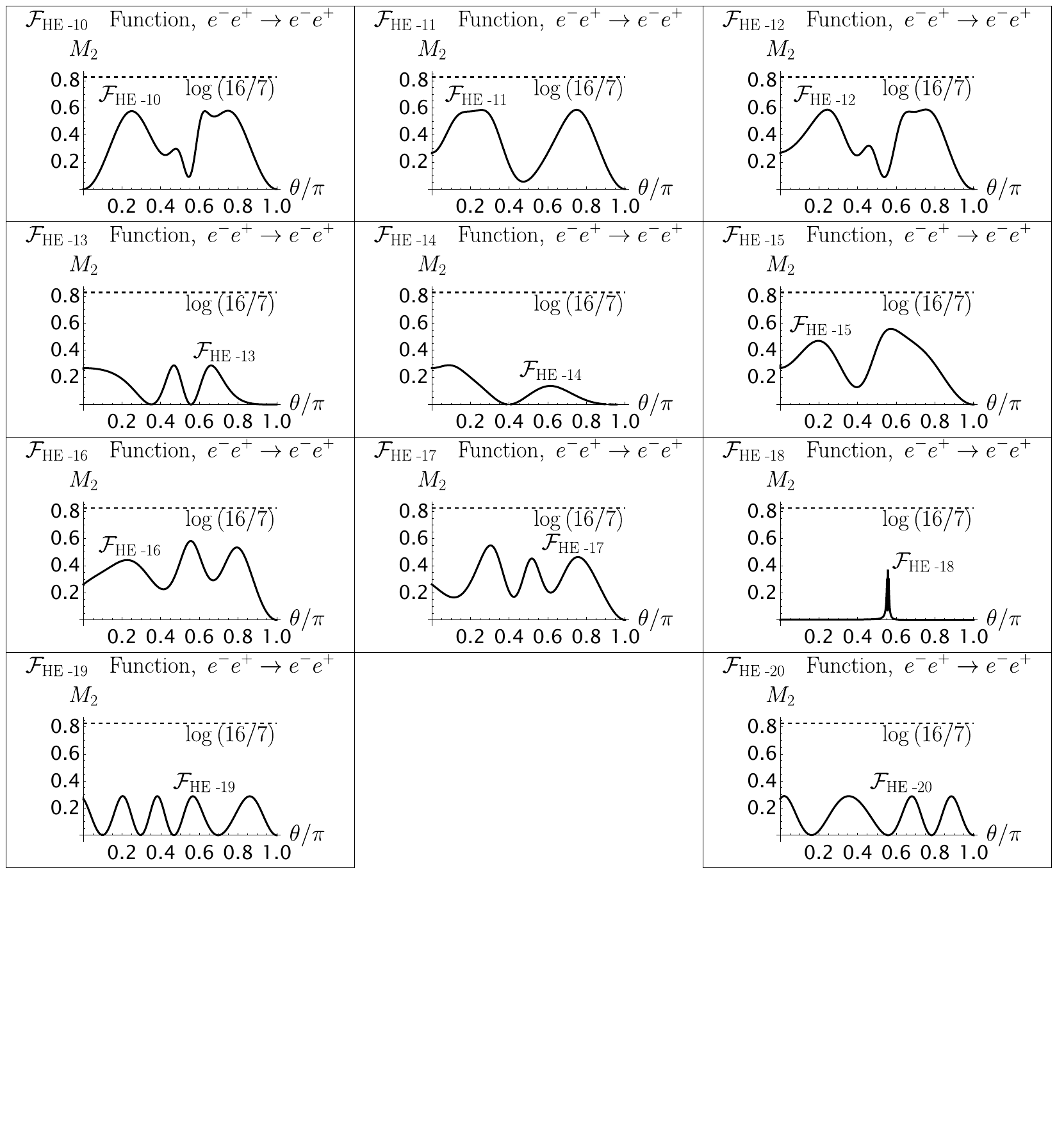}
    \caption{Angular dependence of the magic distributions obtained from $\mathcal{M}_{\gamma}+\mathcal{M}_{Z}$ under Bhabha scattering in the high-energy limit.}
    \label{fig:HEBhabha}
\end{figure}

Bhabha scattering is now considered, whose angular magic functions are collected in Table \ref{tab:HEBhabhaComplete} and portrayed in Fig. \ref{fig:HEBhabha}.
\begin{table}[ht]
    \centering
    \begin{tabularx}{\textwidth}{cXcc}
        \toprule
        Magic distribution & Stabilizer state & $(M_2)_{\text{max}}$ & $\theta_{\text{max}}$ (rad)\\
        \toprule
        $\mathcal{F}_{\text{HE -}10}$ & 1, 2, 39, 40 & 0.578\dots & 0.786\dots, 1.97\dots, 2.35\dots \\
        \hline
         $\mathcal{F}_7$ & 3, 4, 43, 44  & $\log(16/9)$ & $\pi/4$, $3\pi/4$ \\
        \hline
      $\mathcal{F}_{\text{HE -}11}$ & 5, 6, 49, 50 & 0.585\dots & 0.806\dots, 2.35\dots \\
        \hline
        $\mathcal{F}_{\text{HE -}12}$ & 7, 8, 59, 60 & 0.587\dots & 0.763\dots, 2.36\dots \\
        \hline
        $\mathcal{F}_{\text{HE -}13}$ & 9, 10, 45, 48 & 0.288\dots & 1.47\dots, 2.08\dots \\
        \hline
        $\mathcal{F}_{\text{HE -}14}$ & 11, 12 & 0.290\dots & 0.278\dots \\
        \hline
       $\mathcal{F}_{\text{HE -}15}$ & 13, 14, 15, 16, 17, 18, 19, 20, 21, 22, 23, 24, 25, 26, 27, 28 & 0.558\dots & 1.80\dots \\
        \hline
        $\mathcal{F}_{\text{HE -}16}$ & 29, 30, 31, 32, 52, 54, 57, 58 & 0.582\dots & 1.75\dots \\
        \hline
        $\mathcal{F}_{\text{HE -}17}$ & 33, 34, 35, 36, 51, 53, 55, 56 & 0.549\dots & 0.954\dots \\
        \hline
        $\mathcal{F}_5$ & 37, 42 & $\log(4/3)$ & $\pi/8$, $3\pi/8$, $5\pi/8$, $7\pi/8$ \\
        \hline
        $\mathcal{F}_{\text{HE -}18}$ & 38 & 0.389\dots & 1.75\dots \\
        \hline
        $\mathcal{F}_1$ & 41 & 0 & Arbitrary \\
        \hline
        $\mathcal{F}_{\text{HE -}19}$ & 46 & 0.289\dots & 0.638\dots, 1.20\dots, 1.78\dots, 2.70\dots \\
        \hline
       $\mathcal{F}_{\text{HE -}20}$ & 47 & 0.289\dots & 0.0632\dots, 1.11\dots, 2.14\dots, 2.78\dots \\
        \bottomrule
    \end{tabularx}
    \caption{Magic distributions and corresponding stabilizers obtained from $\mathcal{M}_{\gamma}+\mathcal{M}_{Z}$ under Bhabha scattering in the high-energy limit.}
    \label{tab:HEBhabhaComplete}
\end{table}

Compared with the pure QED case, high-energy Bhabha scattering exhibits a much richer structure once the $Z$-boson exchange is included. In the photon-only scenario, the process merely reproduces the same family of magic distributions already found in QED high-energy M\o ller scattering, up to the corresponding reassignment of stabilizer classes. By contrast, once the $Z$-boson exchange is included, the classification not only reorganizes but accomodates a new set of magic distributions. Thus, unlike the pure photon case, the full amplitude generates a richer pattern of magic generation.

Part of the QED structure remains unchanged though. The stabilizers corresponding to $\mathcal{F}_5$ and $\mathcal{F}_7$ remain with the same angular magic values, and so does the trivial distribution $\mathcal{F}_1$. However, in the full amplitude case the group of $\mathcal{F}_1$, which was made of \#38 and \#41, partially splits: while one subset of the stabilizer states continues to produced the flat, vanishing magic function $\mathcal{F}_1$, another now generates $\mathcal{F}_{\text{HE -}18}$. This new distribution retains the flat profile of $\mathcal{F}_1$ but develops a sharp peak for $\theta\approx 1.75$, a feature that is absent in the QED-only case.

The remaining stabilizers are organized into several classes of new angular profiles. Compared with the QED distributions, these new functions exhibit features such as shifted extrema,  broader peaks, and additional oscillatory behavior. For instance, $\mathcal{F}_{\text{HE -}10}$, $\mathcal{F}_{\text{HE -}11}$, and $\mathcal{F}_{\text{HE -}12}$ show different angular features despite having maximal values close to those appearing in the QED case.

One also notices that, while in pure QED the pair $\mathcal{F}_{13}$ and $\widehat{\mathcal{F}}_{13}$ are related by a simple angular shift, in the full $\mathcal{M}_{\gamma}+\mathcal{M}_{Z}$ this pairing relation disappears, with the two functions mildly resembling each other.

\subsubsection*{Elastic scattering $e^-\mu^-\rightarrow e^-\mu^-$}

\begin{figure}[ht]
    \centering
    \includegraphics[width=1\linewidth]{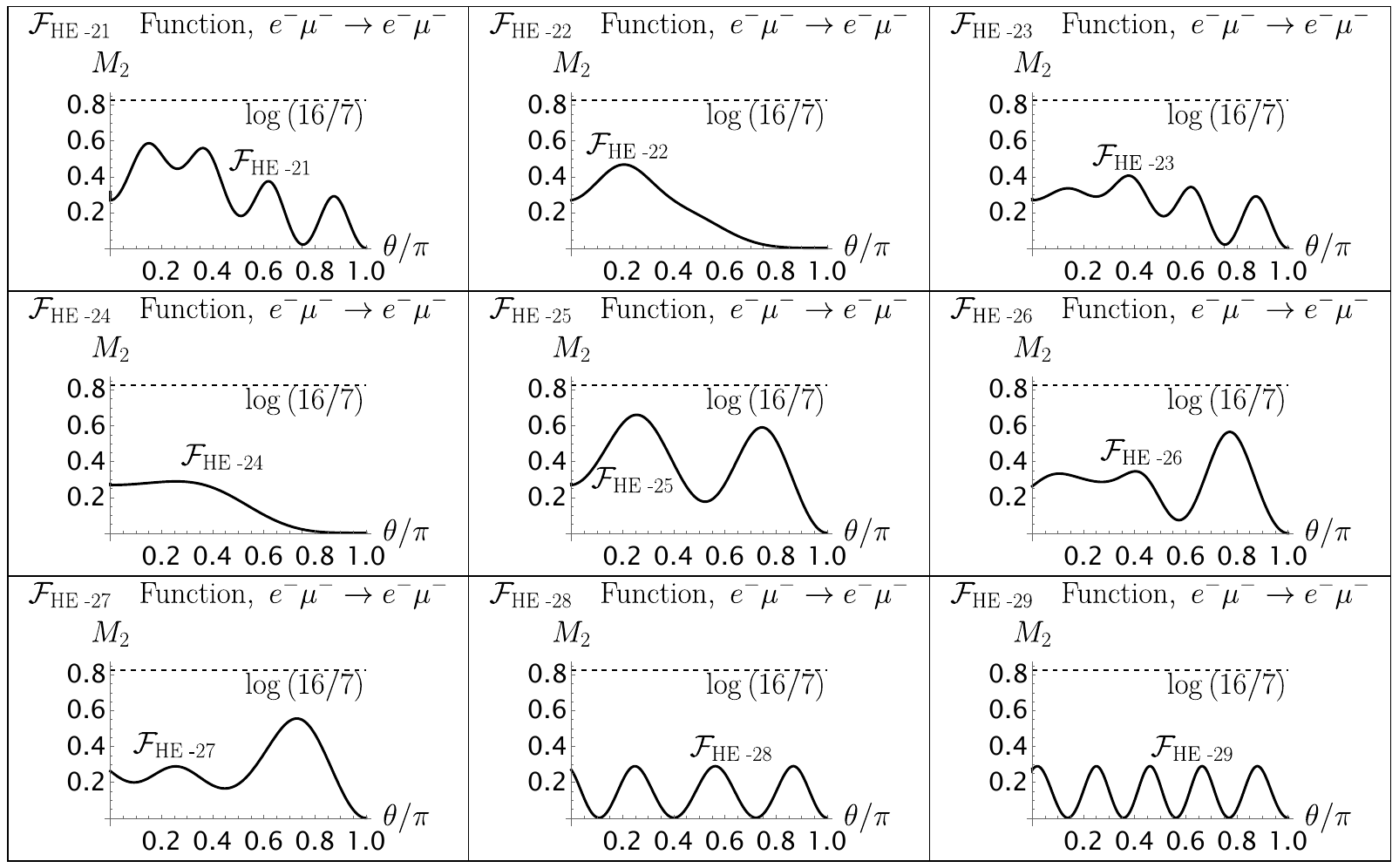}
    \caption{Angular dependence of the magic distributions obtained from $\mathcal{M}_{\gamma}+\mathcal{M}_{Z}$ under elastic scattering $e^-\mu^-\rightarrow e^-\mu^-$ in the high-energy limit.}
    \label{fig:HEemu}
\end{figure}

Finally, the electron-muon scattering process $e^-\mu^-\rightarrow e^-\mu^-$ is examined. Its magic distributions are sorted in Table \ref{tab:HEemuComplete} and plotted in Fig. \ref{fig:HEemu}.

\begin{table}[ht]
    \centering
    \begin{tabularx}{\textwidth}{cXcc}
        \toprule
        Magic distribution & Stabilizer state & $(M_2)_{\text{max}}$ & $\theta_{\text{max}}$ (rad)\\
        \toprule
        $\mathcal{F}_{7}$ & 1, 2, 3, 4, 39, 40, 43, 44 & $\log(16/9)$ & $\pi/4$, $3\pi/4$ \\
        \hline
        $\mathcal{F}_{\text{HE -}21}$ & 5, 6, 49, 50  & 0.587\dots & 0.474\dots \\
        \hline
        $\mathcal{F}_{\text{HE -}22}$ & 7, 8, 59, 60 & 0.468\dots & 0.648\dots \\
        \hline
        $\mathcal{F}_{\text{HE -}23}$ & 9, 10 & 0.406\dots & 1.18\dots \\
        \hline
        $\mathcal{F}_{\text{HE -}24}$ & 11, 12, 46, 47 & 0.289\dots & 0.807\dots \\
        \hline
        $\mathcal{F}_{\text{HE -}25}$ & 13, 14, 15, 16, 17, 18, 19, 20, 21, 22, 23, 24, 25, 26, 27, 28 & 0.660\dots & 0.804\dots \\
        \hline
        $\mathcal{F}_{\text{HE -}26}$ & 29, 31, 34, 36, 55, 56, 57, 58 & 0.565\dots & 2.43\dots \\
        \hline
        $\mathcal{F}_{\text{HE -}27}$ & 30, 32, 33, 35, 51, 52, 53, 54 & 0.555\dots & 2.29\dots \\
        \hline
        $\mathcal{F}_1$ & 37, 42 & 0 & Arbitrary \\
        \hline
        $\mathcal{F}_5$ & 38, 41 & $\log(4/3)$ & $\pi/8$, $3\pi/8$, $5\pi/8$, $7\pi/8$ \\
        \hline
        $\mathcal{F}_{\text{HE -}28}$ & 45 & 0.289\dots & 0.782\dots, 1.77\dots, 2.73\dots \\
        \hline
       $\mathcal{F}_{\text{HE -}29}$ & 48 & 0.289\dots & 0.0621\dots, 0.788\dots, 1.45\dots, 2.09\dots, 2.77\dots \\
        \bottomrule
    \end{tabularx}
    \caption{Magic distributions and corresponding stabilizers obtained from $\mathcal{M}_{\gamma}+\mathcal{M}_{Z}$ under elastic scattering $e^-\mu^-\rightarrow e^-\mu^-$ in the high-energy limit.}
    \label{tab:HEemuComplete}
\end{table}

Compared with the pure QED case, the complete amplitude for high-energy $e^-\mu^-\rightarrow e^-\mu^-$ scattering preserves the stabilizer classes falling under $\mathcal{F}_1$, $\mathcal{F}_5$, and $\mathcal{F}_7$, while replacing the remaining QED distributions $\mathcal{F}_{14}$-$\mathcal{F}_{22}$ by the new $\mathcal{F}_{\text{HE -}21}$ - $\mathcal{F}_{\text{HE -}29}$ ones. In contrast to Bhabha scattering, the number of stabilizer classes is unchanged, with the main effect of the neutral-current contribution limited to deformations of existing angular dependences.

The changes in the maximal value are not uniform though. Some stabilizer classes are only slightly shifted, for example the QED functions $\mathcal{F}_{16}$, $\mathcal{F}_{17}$, $\mathcal{F}_{19}$, $\mathcal{F}_{20}$, $\mathcal{F}_{21}$ now replaced by $\mathcal{F}_{\text{HE -}23}$, $\mathcal{F}_{\text{HE -}24}$, $\mathcal{F}_{\text{HE -}26}$, $\mathcal{F}_{\text{HE -}27}$, $\mathcal{F}_{\text{HE -}28}$, and $\mathcal{F}_{\text{HE -}29}$, respectively. These have maximal values very close to those in QED. For these classes the $Z$-boson exchange changes primarily the shape of the distribution rather than its overall ballpark magic value. Other stabilizer classes exhibit pronounced changes, for example the clearest magic suppression occurs when comparing $\mathcal{F}_{15}$ with $\mathcal{F}_{\text{HE -}22}$, whose maximum is reduced from approximately $0.580$ to $0.468$. By contrast, the stabilizer class corresponding to $\mathcal{F}_{18}$ (maximum of $0.628$) is enhanced, becoming $\mathcal{F}_{\text{HE -}25}$ (maximum of $0.660$). Loosely speaking, the neutral-current appears to redistribute the magic across the different stabilizer classes.

Even when the maxima remain similar, the angular profiles are often greatly modified. For example, $\mathcal{F}_{\text{HE -}21}$, $\mathcal{F}_{\text{HE -}23}$, and $\mathcal{F}_{\text{HE -}26}$ develop additional \textit{shoulders} or oscillations relative to their QED counterparts, while $\mathcal{F}_{\text{HE -}22}$ and $\mathcal{F}_{\text{HE -}24}$ become much broader and more asymmetric.

Lastly, this scattering appears to be intermediate with respect to the previous cases. Unlike pair annihilation, the neutral-current contribution does produce genuinely new angular profiles, but unlike Bhabha scattering it does not enlarge the number of magic classes but it rather deforms the already nontrivial QED functions into their new electroweak counterparts.

\subsection{$Z$-resonance limit}
\label{subsec:ZresonantRegime}

In this regime, the CM energy is pinned down to the $Z$-boson resonance, such that $q^2\approx m_Z^2$. Unlike the high-energy and low-energy limits, the electroweak propagator Eq. (\ref{eq:ew_propagator}) cannot be approximated by a simple scaling such as $1/q^2$ or $1/m_Z^2$. Instead, it is dominated by the resonant denominator, then enhancing the $Z$-boson contribution in the scattering amplitude. All in all, in the $Z$-resonance regime the propagator effectively reduces to 
\begin{equation}
    \frac{-i}{q^2-m_Z^2+im_Z\Gamma_Z}g^{\mu\nu},
\end{equation}
where the term proportional to $q^\mu q^\nu$ is neglected by the same argument invoked in the high-energy regime (i.e. comparatively massless fermions). Since the CM energy is set to the $Z$-boson mass, the entries of the amplitude matrix for the different scattering processes are expanded instead in powers of the small parameter $1/\lambda_Z$ defined in Sec. \ref{subsec:low_Elimit}.

Given that the corresponding magic distribution functions of this section acquire lengthy and complicated forms, only terms up to $O(1/\lambda_Z)$ will be kept. Broadly, many of the resulting magic distributions exhibit very similar angular behavior within the same stabilizer class, differing only by small deformations such as shifts in maxima, minima, or overall shape. 
This motivates $\boldsymbol{\mathcal{F}}_{n,i}$ as our choice of notation, where a bold function indicates that the multiple magic distributions share very similar features even if their $i$ index is different. These are collected in Appendix. \ref{appendix:Variations_magic}.

\subsubsection*{Pair annihilation $e^-e^+\rightarrow \mu^-\mu^+$}

The sixteen entries of the amplitude matrix for the different electron and muon polarizations, up to order $O(1/\lambda_Z)$, are listed in Appendix \ref{appendix:EW_amplitudes} as a function of the scattering angle $\theta$. Do notice that the amplitudes $\mathcal{M}_{\uparrow\downarrow \to\ldots}$ and $\mathcal{M}_{\downarrow\uparrow\ \to\ldots}$ develop a nonzero value only at order $O(1/\lambda_Z)$.

\begin{figure}[ht]
    \centering
    \includegraphics[width=1\linewidth]{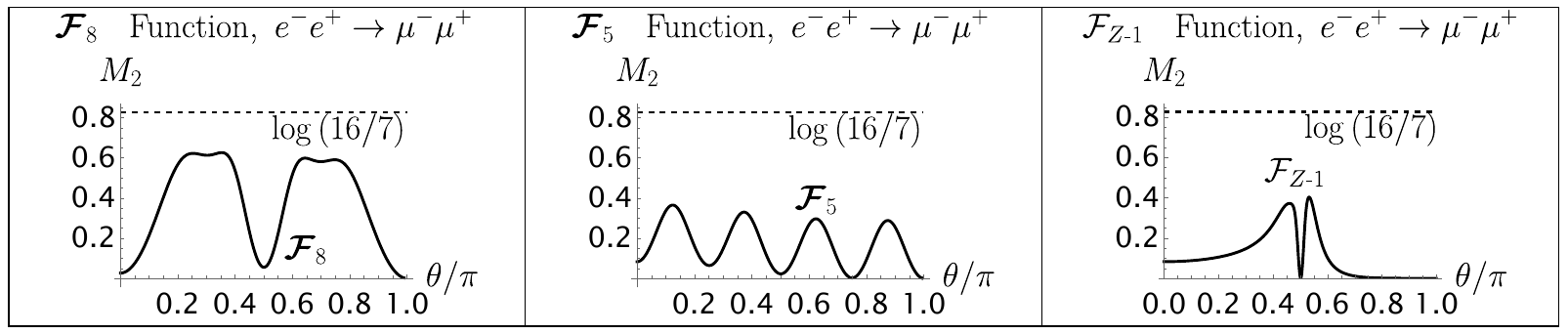}
    \caption{Angular dependence of the magic distributions obtained from $\mathcal{M}_{\gamma}+\mathcal{M}_{Z}$ under pair annihilation in the $Z$-resonant limit.}
    \label{fig:AnniZR}
\end{figure}

The resulting $M_{2}$ functions for resonant $e^{-}e^{+}\to \mu^{-}\mu^{+}$ and their corresponding stabilizers are displayed in Table \ref{tab:eeTOmumu_reson}, and due to their barely-illuminating analytic form, we limit ourselves to graph these functions in the panels of Fig. \ref{fig:AnniZR}. Notice that the new magic distributions functions that arises in the $Z$-resonance regime are named $\mathcal{F}_{Z \text{-}n}$. In extracting these functions numerically, one utilizes $s_{W}^{2}\approx 0.231$ as argued below Eq. (\ref{eq:ew_propagator}). As shown later, certain stabilizer classes turn out to be highly sensitive to the numerical value of weak mixing angle, whereas their sensitivity to the $Z$ width value is only feeble.

\begin{table}[ht]
    \centering
    \begin{tabularx}{\textwidth}{cXcc}
        \toprule
        Magic distribution & Stabilizer state & $(M_2)_{\text{max}}$ & $\theta_{\text{max}}$ (rad)\\
        \toprule
        $\boldsymbol{\mathcal{F}}_{8}$ & 1, 2, 13, 14, 15, 16, 17, 18, 19, 20, 21, 22, 23, 24, 25, 26, 27, 28, 29, 30, 31, 32, 33, 34, 35, 36, 39, 40, 51, 52, 53, 54, 55, 56, 57, 58  & 0.626\dots & 1.11\dots \\
        \hline
         $\mathcal{F}_{7}$ & 3, 4, 42, 43, 44  & $\log(16/9)$ & $\pi/4$, $3\pi/4$  \\
        \hline
        $\boldsymbol{\mathcal{F}}_{5}$ & 5, 6, 11, 12, 37, 46, 47, 49, 50  & 0.365\dots & 0.391\dots \\
        \hline
        $\mathcal{F}_{Z\text{-}1}$ & 7, 8, 9, 10, 38, 45, 48, 59, 60  & 0.403\dots & 1.67\dots \\
        \hline
        Vanishing & 41 & --- & ---\\
        \bottomrule
    \end{tabularx}
    \caption{Magic distributions and corresponding stabilizers obtained from $\mathcal{M}_{\gamma}+\mathcal{M}_{Z}$ under pair annihilation in the $Z$-resonant limit.}
    \label{tab:eeTOmumu_reson}
\end{table}

\begin{figure}[h!]
    \centering
    \includegraphics[scale=0.65]{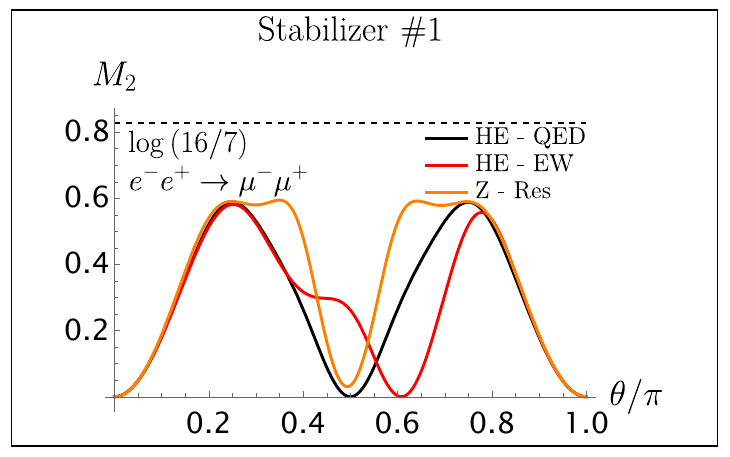}
    \caption{Comparison of $M_{2}(\theta)$ between the $Z$-resonance regime (orange) and the high-energy regime, either in QED (black) or in the full EW case (red), for stabilizer \#1.}
    \label{fig:compareResEWstab1}
\end{figure}

For comparison purposes, Fig. \ref{fig:compareResEWstab1} shows side-by-side the EW resonant-regime ($\boldsymbol{\mathcal{F}}_{8}$), QED high-energy ($\mathcal{F}_{4}$), and EW high-energy ($\mathcal{F}_{\text{HE -}1}$) magic curves for a sample state (stabilizer \#1) as a function of $\theta$. While the highest value of the EW resonant-regime curve (orange) is comparable to that of its QED (black) and EW (red) high-energy counterparts, it generates a higher magic value than them over a wider range of scattering angles, in particular around $\theta=\pi/2$.

\subsubsection*{M\o ller scattering $e^-e^-\rightarrow e^-e^-$}

\begin{figure}[ht]
    \centering
    \includegraphics[width=1\linewidth]{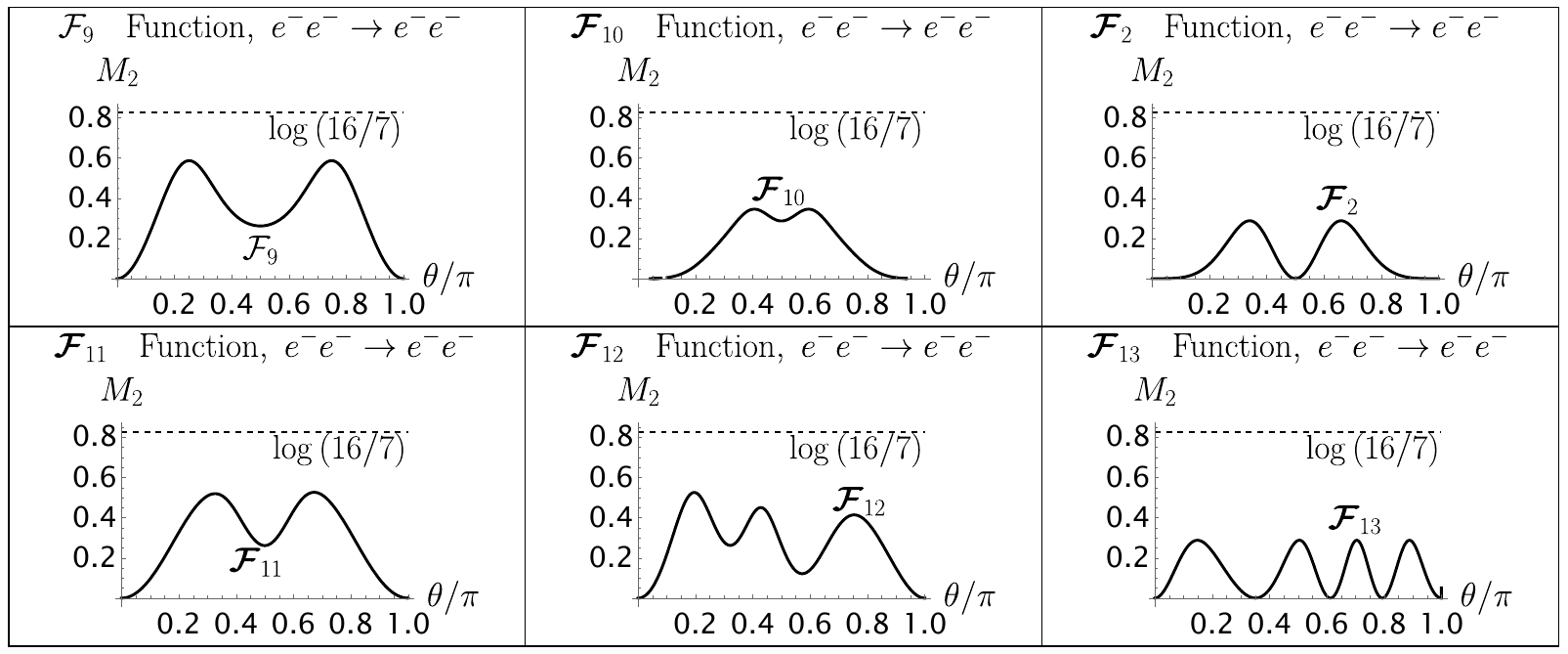}
    \caption{Angular dependence of the magic distributions obtained from $\mathcal{M}_{\gamma}+\mathcal{M}_{Z}$ under M\o ller scattering in the $Z$-resonant limit.}
    \label{fig:MollerZR}
\end{figure}

The magic distribution functions for M\o ller scattering in the $Z$-resonance regime are summarized in Table. \ref{tab:MollerZRComplete}, with their angular behavior illustrated in Fig. \ref{fig:MollerZR}.

\begin{table}[ht]
    \centering
    \begin{tabularx}{\textwidth}{cXcc}
        \toprule
        Magic distribution & Stabilizer state & $(M_2)_{\text{max}}$ & $\theta_{\text{max}}$ (rad)\\
        \toprule
        $\mathcal{F}_6$ & 1, 2, 39, 40 & 0.576\dots & $(\pi/2)\pm0.783\dots$ \\
        \hline
        $\mathcal{F}_7$ & 3, 4, 43, 44  & $\log(16/9)$ & $\pi/4$, $3\pi/4$ \\
        \hline
       $\mathcal{F}_8$ & 5, 6, 49, 50 & $\log(9/5)$ & $\pi/4$, $3\pi/4$, $(\pi/2)\pm \text{arccot}\sqrt{2}$ \\
        \hline
        $\mathcal{F}_9$ & 7, 8, 59, 60 & 0.586\dots & $(\pi/2)\pm0.781\dots$ \\
        \hline
        $\boldsymbol{\mathcal{F}}_{10}$ & 9, 10 & 0.345\dots & $(\pi/2)\pm0.297\dots$ \\
        \hline
        $\boldsymbol{\mathcal{F}}_{2}$ & 11, 12, 46, 47 & 0.289\dots & $(\pi/2)\pm0.501\dots$ \\
        \hline
        $\boldsymbol{\mathcal{F}}_{11}$ & 13, 14, 15, 16, 17, 18, 19, 20, 21, 22, 23, 24, 25, 26, 27, 28 & 0.526\dots & $(\pi/2)\pm0.541\dots$ \\
        \hline
        $\boldsymbol{\mathcal{F}}_{12}$ & 29, 31, 34, 36, 55, 56, 57, 58 & 0.526\dots & 0.615\dots\\
        \hline
        $\boldsymbol{\widehat{\mathcal{F}}}_{12}$ & 30, 32, 33, 35, 51, 52, 53, 54 & 0.526\dots & $\pi-0.615$\dots \\
        \hline
        $\mathcal{F}_1$ & 37, 42 & 0 & Arbitrary \\
        \hline
        $\mathcal{F}_5$ & 38, 41 & $\log(4/3)$ & $\pi/8$, $3\pi/8$, $5\pi/8$, $7\pi/8$ \\
        \hline
        $\boldsymbol{\mathcal{F}}_{13}$ & 45 & $\log(4/3)$ & 0.468\dots, 1.58\dots, 2.21\dots, 2.79\dots \\
        \hline
       $\boldsymbol{\widehat{\mathcal{F}}}_{13}$ & 48 & $\log(4/3)$ & 0.352\dots, 0.932\dots, 1.56\dots, 2.67\dots \\
        \bottomrule
    \end{tabularx}
    \caption{Magic distributions and corresponding stabilizers obtained from $\mathcal{M}_{\gamma}+\mathcal{M}_{Z}$ under M\o ller scattering in the $Z$-resonant limit.}
    \label{tab:MollerZRComplete}
\end{table}

The angular magic distributions obtained in M\o ller scattering in $Z$-resonant regime are now compared with those of the high-energy QED case. In contrast to previous analyses, where the inclusion of the $Z$-boson generated entirely new functions labeled $\mathcal{F}_{\text{HE -}n}$, both the high-energy and resonant regimes here yield essentially the same set of magic distributions. For this reason, the functions of the resonant case are denoted as $\boldsymbol{\mathcal{F}}_n$, emphasizing that they closely resemble their high-energy QED counterparts while still exhibiting small but nontrivial differences.

A first observation is that several magic distributions remain unchanged across the high-energy and resonant regimes. In particular, the stabilizer classes associated with $\mathcal{F}_1$, $\mathcal{F}_5$, $\mathcal{F}_7$ preserve both their angular profiles and maximal values. This indicates that these stabilizer classes are insensitive not only to the inclusion of the $Z$-boson, but also to the transition between the high-energy and resonance regimes. These persistent, or \textit{fixed} features of the scattering process across energies will be discussed in Sec. \ref{sec:FixedStabStates}.

One also observes that the remaining distributions are slightly deformed rather than completely reorganized. To show this effect in detail, each pair $\left(\mathcal{F}_n, \boldsymbol{\mathcal{F}}_n \right)$ is graphed in Fig. \ref{fig:compareMollerZR}, where it is clear that they preserve the same structure (number and localization of peaks) while differing in the exact shape and angle value of their maxima. While many $\boldsymbol{\mathcal{F}}_n$ remain close to their QED counterparts, some exhibit slight enhancements, therefore, unlike the high-energy case where entirely new magic distributions appeared, the $Z$-resonance regime preserves the QED state classification while slightly modifying its features.

\subsubsection*{Bhabha scattering $e^-e^+\rightarrow e^-e^+$}

\begin{figure}[ht]
    \centering
    \includegraphics[width=1\linewidth]{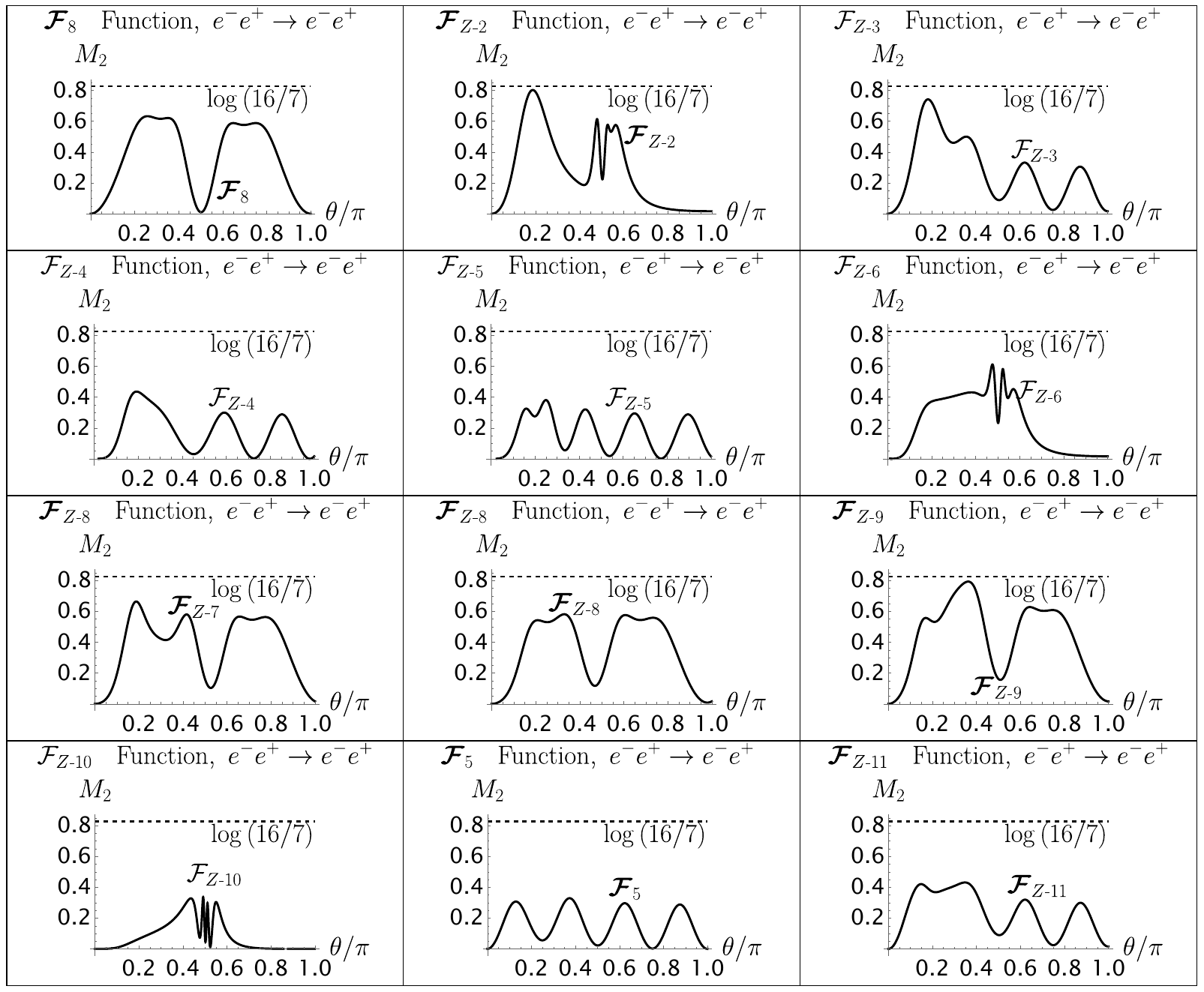}
    \caption{Angular dependence of the magic distributions obtained from $\mathcal{M}_{\gamma}+\mathcal{M}_{Z}$ under Bhabha scattering in the $Z$-resonant limit.}
    \label{fig:BhabhaZR}
\end{figure}

Bhabha scattering in the $Z$-resonance limit, whose magic functions are summarized in Table \ref{tab:BhabhaZRComplete} and graphed in Fig. \ref{fig:BhabhaZR}, exhibits the richest features among all scatterings analyzed. In particular, it counts with the largest set of genuinely new  magic distributions, in contrast to the rest of the processes analyzed which showed slight or mild deviations of already-known distributions functions.

\begin{table}[ht]
    \centering
    \begin{tabularx}{\textwidth}{cXcc}
        \toprule
        Magic distribution & Stabilizer state & $(M_2)_{\text{max}}$ & $\theta_{\text{max}}$ (rad)\\
        \toprule
        $\boldsymbol{\mathcal{F}}_8$ & 1, 2, 39, 40 & 0.632\dots & 0.811\dots \\
        \hline
        $\mathcal{F}_7$ & 3, 4, 43, 44  & $\log(16/9)$ & $\pi/4$, $3\pi/4$ \\
        \hline
       $\boldsymbol{\mathcal{F}}_{Z\text{-}2}$ & 7, 8, 59, 60 & 0.802\dots & 0.587\dots \\
        \hline
        $\mathcal{F}_{Z\text{-}3}$ & 5, 6, 49, 50 & 0.748\dots & 0.586\dots \\
        \hline
        $\mathcal{F}_{Z\text{-}4}$ & 11 & 0.436\dots & 0.605\dots \\
        \hline
        $\mathcal{F}_{Z\text{-}5}$ & 12 & 0.381\dots & 0.786\dots \\
        \hline
        $\mathcal{F}_{Z\text{-}6}$ & 9, 10, 45, 48 & 0.612\dots & 1.50\dots \\
        \hline
        $\boldsymbol{\mathcal{F}}_{Z\text{-}7}$ & 13, 16, 18, 19, 22, 23, 25, 28 & 0.664\dots & 0.624\dots \\
        \hline
        $\boldsymbol{\mathcal{F}}_{Z\text{-}8}$ & 14, 15, 17, 20, 21, 24, 26, 27 & 0.581\dots & 1.04\dots \\
        \hline
        $\boldsymbol{\mathcal{F}}_{Z\text{-}9}$ & 29, 30, 31, 32, 33, 34, 35, 36, 51, 52, 53, 54, 55, 56, 57, 58 & 0.776\dots & 1.14\dots\\
        \hline
        $\mathcal{F}_{Z\text{-}10}$ & 38, 41 & 0.336\dots & 1.54\dots \\
        \hline
        $\boldsymbol{\mathcal{F}}_5$  & 37, 42 & 0.329\dots & 1.17\dots \\
        \hline
        $\boldsymbol{\mathcal{F}}_{Z\text{-}11}$ & 46, 47 & 0.431\dots & 1.09\dots \\
        \bottomrule
    \end{tabularx}
    \caption{Magic distributions and corresponding stabilizers obtained from $\mathcal{M}_{\gamma}+\mathcal{M}_{Z}$ under Bhabha scattering in the $Z$-resonant limit.}
    \label{tab:BhabhaZRComplete}
\end{table}

A key distinction with respect to the high-energy case is that the grouping of stabilizer states is no longer preserved. In the QED and EW high-energy analyses, the same sets of stabilizer states remain grouped together, even though the corresponding magic distributions may differ from one regime to another. At the $Z$-resonance, however, this pattern breaks down and some previously preserved groups split into smaller groups associated with different magic functions. For example, the large class formed by the stabilizer states \#13 to \#28, which remained unchanged in both the QED and EW high-energy regimes, is separated into two different groups.

Despite this acute reorganization of stabilizer classes, a few remain \textit{fixed} across all energy regimes. In particular, the groups corresponding to $\mathcal{F}_5$ and $\mathcal{F}_7$ retain both their angular profiles and maximal values, confirming their insensitivity to both the inclusion of the $Z$-boson and the choice of energy regime. These groups therefore continue to serve as a reference throughout the analysis.

Other groups exhibit more subtle but still noticeable changes. The group containing the stabilizers \#1, \#2, \#39 and \#40, which in the EW case at high energy was promoted to a new function, now maps instead to the already known $\mathcal{F}_8$ rather than generating a completely new structure. This illustrates that, despite the emergence of completely new magic distributions at resonance, not all stabilizer groups change entirely. Similarly, the group corresponding to the stabilizers \#38 and \#41, which yields the trivial magic function $\mathcal{F}_1$ in QED, now develops a profile resembling its high-energy behavior (a nearly flat function with a localized sharp peak around the center).

Overall, the $Z$-resonance regime leads to a departure from the patterns observed at high energy. Unlike the high-energy case, where the inclusion of the $Z$-boson primarily deforms existing magic distributions, the resonant energy induces both the emergence of new angular profiles and a restructuring of the stabilizer groups. This confirms the earlier claim that the full Bhabha scattering amplitude at resonance is the most sensitive to electroweak effects in the context of magic generation.

\subsubsection*{Elastic scattering $e^-\mu^-\rightarrow e^-\mu^-$}

\begin{figure}[ht]
    \centering
    \includegraphics[width=1\linewidth]{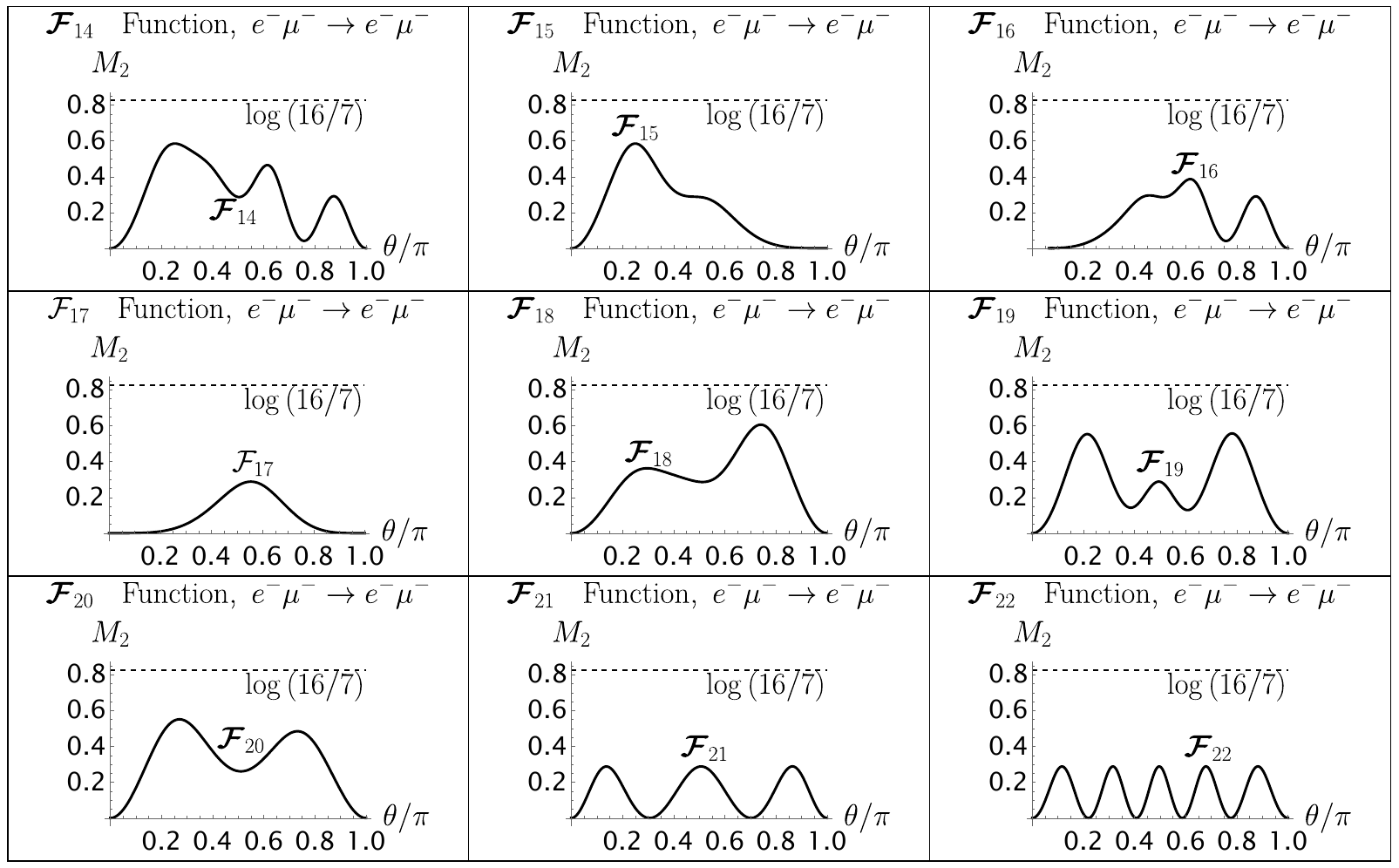}
    \caption{Angular dependence of the magic distributions obtained from $\mathcal{M}_{\gamma}+\mathcal{M}_{Z}$ under elastic scattering $e^-\mu^-\rightarrow e^-\mu^-$ scattering in the $Z$-resonant limit.}
    \label{fig:emuZR}
\end{figure}

\begin{table}[ht]
    \centering
    \begin{tabularx}{\textwidth}{cXcc}
        \toprule
        Magic distribution & Stabilizer state & $(M_2)_{\text{max}}$ & $\theta_{\text{max}}$ (rad)\\
        \toprule
        $\mathcal{F}_{7}$ & 1, 2, 3, 4, 39, 40, 43, 44 & $\log(16/9)$ & $\pi/4$, $3\pi/4$ \\
        \hline
        $\boldsymbol{\mathcal{F}}_{14}$ & 5, 6, 49, 50  & 0.585\dots & 0.793\dots \\
        \hline
        $\boldsymbol{\mathcal{F}}_{15}$ & 7, 8, 59, 60 & 0.585\dots & 0.786\dots \\
        \hline
         $\boldsymbol{\mathcal{F}}_{16}$ & 9, 10 & 0.386\dots & 1.94\dots \\
        \hline
         $\widehat{\mathcal{F}}_{17}$ & 11, 12, 46, 47 & $\log(4/3)$\dots & 1.53\dots \\
        \hline
         $\boldsymbol{\mathcal{F}}_{18}$ & 13, 14, 15, 16, 17, 18, 19, 20, 21, 22, 23, 24, 25, 26, 27, 28 & 0.605\dots & 2.33\dots \\
        \hline
         $\boldsymbol{\mathcal{F}}_{19}$ & 29, 31, 34, 36, 55, 56, 57, 58 & 0.553\dots & $(\pi/2)\pm0.895\dots$ \\
        \hline
         $\boldsymbol{\mathcal{F}}_{20}$ & 30, 32, 33, 35, 51, 52, 53, 54 & 0.514\dots & $(\pi/2)\pm0.715\dots$\dots \\
        \hline
        $\mathcal{F}_1$ & 37, 42 & 0 & Arbitrary \\
        \hline
        $\mathcal{F}_5$ & 38, 41 & $\log(4/3)$ & $\pi/8$, $3\pi/8$, $5\pi/8$, $7\pi/8$ \\
        \hline
         $\boldsymbol{\mathcal{F}}_{21}$ & 45 & 0.288\dots & 0.431\dots, 1.59\dots, 2.72\dots \\
        \hline
        $\boldsymbol{\mathcal{F}}_{22}$ & 48 & 0.288\dots & 0.365\dots, 0.991\dots, 1.56\dots, 2.13\dots, 2.77\dots \\
        \bottomrule
    \end{tabularx}
    \caption{Magic distributions and corresponding stabilizers obtained from $\mathcal{M}_{\gamma}+\mathcal{M}_{Z}$ under elastic scattering $e^-\mu^-\rightarrow e^-\mu^-$ scattering in the $Z$-resonant limit.}
    \label{tab:emuZRComplete}
\end{table}
The magic distributions for $e^-\mu^-\rightarrow e^-\mu^-$ at the $Z$-resonance are listed in Table \ref{tab:emuZRComplete} and graphed in Fig. \ref{fig:emuZR}. These distributions closely follow the behavior observed in the resonant M\o ller scattering discussed before. In particular, the distributions retain the same overall structure as in the corresponding high-energy QED case, illustrated in Fig. \ref{fig:compareemuZR}, with each function preserving the number and approximate location of its extrema.

In contrast with the resonant M\o ller case, however, none of these deformations leads to an enhancement of the maximal magic. While in resonant M\o ller scattering some $M_{2}$ functions exhibited increased maxima due to the electroweak contribution, here all distributions remain bounded by their QED counterparts, showing either comparable or slightly reduced maximal values. We thus concluded that the $Z$-boson exchange does not enhance the magic generation even at resonance.

The only noticeable difference appears in the class corresponding to $\mathcal{F}_{17}$, whose profile remains the same up to a small shift in $\theta$ after electroweak effects. Apart from this effect, the distributions are nearly identical to those of the QED case at high energy.

\section{Fixed stabilizer states}
\label{sec:FixedStabStates}

Through this work, the quantum magic $M_2(\theta)$ was computed for different energy regimes\footnote{The magic distributions were also computed in the low-energy regime. However, in this case the inclusion of electroweak contributions does not generate new structures, and the resulting magic coincides with the QED case.}: high-energy and resonant energy, at zero and first order in their respective expansion parameter. Although the scattering amplitudes are different in each regime, a subset of stabilizer states produces the same magic distribution in either regime. A stabilizer state is called \textit{fixed} if, when taken as an initial state, it induces a $M_2(\theta)$ that looks the same across energy regimes under the same process.

At first glance, the fixed behavior might suggest that the corresponding final states are identical across regimes or that electroweak parameters do not affect their amplitudes. However, a deeper look shows that this interpretation is incorrect. Explicit computation of the normalized (final) states demonstrates that, in general,
\begin{equation*}
    |\psi^{\text{QED}}\rangle\neq |\psi^{\text{EW}}\rangle\neq|\psi^{Z\text{-}\text{Res}}\rangle,
\end{equation*}
and that the electroweak contribution introduces nontrivial dependence on parameters such as $s_W$, $m_Z$, and $\Gamma_Z$. We confirmed this by looking at the sensitivity of the magic distributions under slight variations of these parameters, which cause significant deformations to $M_{2}(\theta)$.

Even when two regimes produce the same $M_{2}(\theta)$, this does not mean that the final states are identical or that the electroweak parameters play no role. Instead, the electroweak corrections modify the final state, but for certain stabilizer inputs and parameter values these modifications do not affect the resulting magic.

The stabilizer states that repeatedly show the fixed behavior mostly belong to the following set 
\begin{equation}
    \{1,2,3,4\}\cup\{37,38,39,40,41,42,43,44\}. \label{eq:FixedSet}
\end{equation}
While not all of the stabilizers above are always fixed across every single energy regime, whenever a magic distribution remains the same across regimes, it always happens for one state in this set. The reason goes as follows: states in Eq. (\ref{eq:FixedSet}) have a simple structure, having either a single nonzero component in the computational basis, or are made of superpositions of at most two basis states such as $|\uparrow\uparrow\rangle\pm|\downarrow\downarrow\rangle$ or $|\uparrow\downarrow\rangle\pm|\downarrow\uparrow\rangle$ up to normalization factors. Due to this, they depend on fewer independent amplitude ratios, which makes the magic less sensitive to changes in the interaction. The extent of this behavior is respectively presented for M\o ller, Bhabha, and elastic scatterings\footnote{Pair annihilation is not included, because for it no fixed stabilizer states are found.} in Tables \ref{tab:Mollerrobust}, \ref{tab:Bhabharobust}, and \ref{tab:emurobust}.

Rather than including all fixed stabilizer states, these tables show those whose magic distribution is sensitive to variations of any of the EW parameters $s_W$, $m_Z$, or $\Gamma_Z$, with a checkmark (cross) indicating that $M_{2}$ is changed (unchanged) under said parameter variation. Meanwhile, a dash indicates the absence of dependence on that parameter, at least at the chosen perturbative order. The purpose of these tables is to show that, for some fixed stabilizer states, the fixed behavior occurs only for specific values of the parameters. In this sense, the fixed structure is tied to the measured, precise values of the electroweak parameters, rather that to the complete absence of parameter dependence.

\begin{table}[ht]
    \centering
    
    \renewcommand{\arraystretch}{1.2}
    
    \begin{tabularx}{\textwidth}{|c|>{\centering\arraybackslash}X>{\centering\arraybackslash}X>{\centering\arraybackslash}X|
                                   >{\centering\arraybackslash}X>{\centering\arraybackslash}X>{\centering\arraybackslash}X|
                                   >{\centering\arraybackslash}X>{\centering\arraybackslash}X>{\centering\arraybackslash}X|}
        \hline
        
        \multirow{2}{*}{\#}
        & \multicolumn{3}{c|}{High energy}
        & \multicolumn{3}{c|}{$Z$-resonance at $O(\lambda_{Z}^{0})$}
        & \multicolumn{3}{c|}{$Z$-resonance at $O(\lambda_{Z}^{-1})$} \\
        
        \cline{2-10}
        
        & $s_W$ & $m_Z$ & $\Gamma_Z$
        & $s_W$ & $m_Z$ & $\Gamma_Z$
        & $s_W$ & $m_Z$ & $\Gamma_Z$ \\
        
        \hline
        
        43
        & $\checkmark$ & --- & ---
        & $\checkmark$ & --- & ---
        & $\checkmark$ & $\checkmark$ & $\times$ \\
        \hline
        
        44
        & $\checkmark$ & --- & ---
        & $\checkmark$ & --- & ---
        & $\checkmark$ & $\checkmark$ & $\times$ \\
        \hline
        
        42
        & $\checkmark$ & --- & ---
        & $\checkmark$ & --- & ---
        & $\checkmark$ & $\times$ & $\times$ \\
        \hline
        
        41
        & $\checkmark$ & --- & ---
        & $\checkmark$ & --- & ---
        & $\checkmark$ & $\times$ & $\times$ \\
        \hline
        
    \end{tabularx}
    
    \caption{Parameter sensitivity of stabilizer states displaying fixed behavior across energy regimes under M\o ller scattering ($\checkmark$ indicates sensitivity, $\times$ no sensitivity, and --- parameter absence).}
    \label{tab:Mollerrobust}
\end{table}

\begin{table}[ht]
    \centering
    
    \renewcommand{\arraystretch}{1.2}
    
    \begin{tabularx}{\textwidth}{|c|>{\centering\arraybackslash}X>{\centering\arraybackslash}X>{\centering\arraybackslash}X|
                                   >{\centering\arraybackslash}X>{\centering\arraybackslash}X>{\centering\arraybackslash}X|
                                   >{\centering\arraybackslash}X>{\centering\arraybackslash}X>{\centering\arraybackslash}X|}
        \hline
        
        \multirow{2}{*}{\#}
        & \multicolumn{3}{c|}{High energy}
        & \multicolumn{3}{c|}{$Z$-resonance at $O(\lambda_{Z}^{0})$}
        & \multicolumn{3}{c|}{$Z$-resonance at $O(\lambda_{Z}^{-1})$} \\
        
        \cline{2-10}
        
        & $s_W$ & $m_Z$ & $\Gamma_Z$
        & $s_W$ & $m_Z$ & $\Gamma_Z$
        & $s_W$ & $m_Z$ & $\Gamma_Z$ \\
        
        \hline
        
        3
        & $\times$ & --- & ---
        & $\times$ & $\times$ & $\times$
        & $\checkmark$ & $\checkmark$ & $\checkmark$ \\
        \hline
        
        4
        & $\times$ & --- & ---
        & $\times$ & $\times$ & $\times$
        & $\checkmark$ & $\checkmark$ & $\checkmark$ \\
        \hline
        
        43
        & $\times$ & --- & ---
        & $\times$ & $\times$ & $\times$
        & $\checkmark$ & $\checkmark$ & $\checkmark$ \\
        \hline
        
        44
        & $\times$ & --- & ---
        & $\times$ & $\times$ & $\times$
        & $\checkmark$ & $\checkmark$ & $\checkmark$ \\
        \hline

        37
        & $\checkmark$ & --- & ---
        & $\checkmark$ & $\times$ & $\times$
        & $\checkmark$ & $\times$ & $\times$ \\
        \hline

        42
        & $\times$ & --- & ---
        & $\times$ & $\times$ & $\times$
        & $\checkmark$ & $\times$ & $\checkmark$ \\
        \hline
        
    \end{tabularx}
    
    \caption{Analogue of Table \ref{tab:Mollerrobust} for Bhabha scattering.}
    \label{tab:Bhabharobust}
\end{table}

\begin{table}[ht]
    \centering
    
    \renewcommand{\arraystretch}{1.2}
    
    \begin{tabularx}{\textwidth}{|c|>{\centering\arraybackslash}X>{\centering\arraybackslash}X>{\centering\arraybackslash}X|
                                   >{\centering\arraybackslash}X>{\centering\arraybackslash}X>{\centering\arraybackslash}X|
                                   >{\centering\arraybackslash}X>{\centering\arraybackslash}X>{\centering\arraybackslash}X|}
        \hline
        
        \multirow{2}{*}{\#}
        & \multicolumn{3}{c|}{High energy}
        & \multicolumn{3}{c|}{$Z$-resonance at $O(\lambda_{Z}^{0})$}
        & \multicolumn{3}{c|}{$Z$-resonance at $O(\lambda_{Z}^{-1})$} \\
        
        \cline{2-10}
        
        & $s_W$ & $m_Z$ & $\Gamma_Z$
        & $s_W$ & $m_Z$ & $\Gamma_Z$
        & $s_W$ & $m_Z$ & $\Gamma_Z$ \\
        
        \hline
        
        43
        & $\checkmark$ & --- & ---
        & $\checkmark$ & --- & ---
        & $\checkmark$ & $\times$ & $\times$ \\
        \hline
        
        44
        & $\checkmark$ & --- & ---
        & $\checkmark$ & --- & ---
        & $\checkmark$ & $\times$ & $\times$ \\
        \hline
        
        42
        & $\checkmark$ & --- & ---
        & $\checkmark$ & --- & ---
        & $\checkmark$ & $\times$ & $\times$ \\
        \hline
        
        41
        & $\checkmark$ & --- & ---
        & $\checkmark$ & --- & ---
        & $\checkmark$ & $\times$ & $\times$ \\
        \hline
        
    \end{tabularx}
    
    \caption{Analogue of Table \ref{tab:Mollerrobust} for electron-muon scattering.}
    \label{tab:emurobust}
\end{table}

\begin{figure}[ht]
    \centering
    \includegraphics[width=1\linewidth,trim={0 250 0 0}]{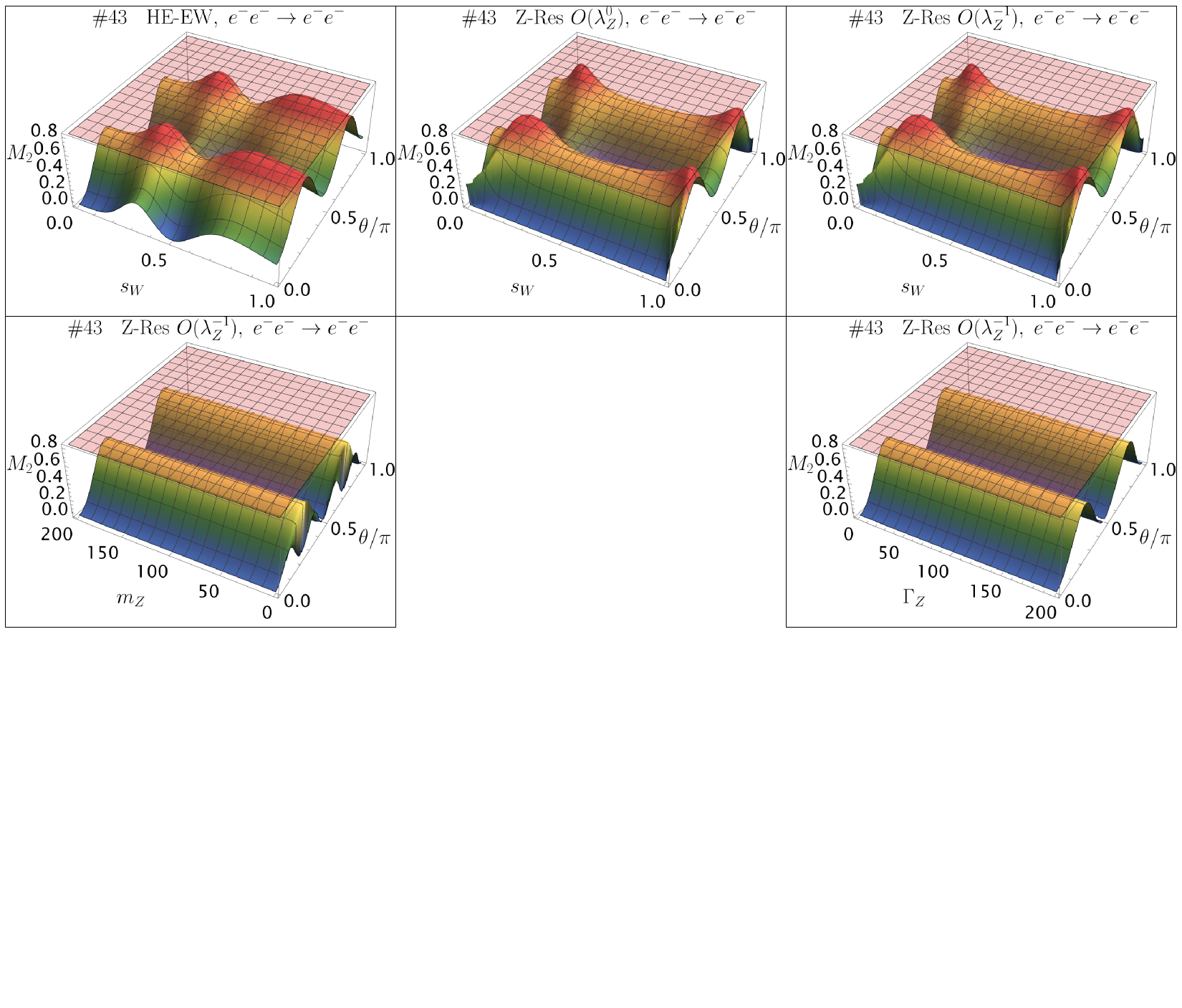}
    \caption{Magic distributions for stabilizer state $\#43$ in M\o ller scattering across the high energy and $Z$-resonance regimes, showing the dependence (or lack of) on $s_W$, $m_Z$, and $\Gamma_Z$. The red plane shows the maximal magic allowed, and dimensionful quantities are taken in GeV.}
    \label{fig:Moller43}
\end{figure}

To further illustrate the behavior summarized in the tables above,  Fig. \ref{fig:Moller43} shows the M\o ller scattering magic distributions for the fixed stabilizer state $\#43$, across energy regimes. As can be seen, the distributions obtained in the different regimes are in general different when varying the parameters. However, for specific values of $s_W$ and for most values of $m_Z$, the plots coincide and reproduce the same angular dependence. This clearly shows that the agreement between regimes is not generic, but occurs only for particular parameter values. It is also interesting to note that the maximal value of the magic is reached only for certain values of $s_W$, indicating that the electroweak parameters can enhance or reduce the amount of quantum magic even within the same stabilizer sector.

\section{Dark $U(1)$ extension}
\label{sec:DarkU1}

Dark photons have been an attractive candidate for Beyond-the-Standard-Model (BSM) physics over the past few decades due to their rich phenomenology prospects and role in several theories, namely supersymmetry \cite{photino}, string theory \cite{photoverse}, and grand unified theories \cite{Gherghetta_2019}, posing as mediators and/or dark matter candidates themselves \cite{DMdarkphoton}. Recently, dark photons playing these roles have been explored in the context of cosmological histories, lending viability to complex models such as Strongly Interacting Massive Particles \cite{SIMPs}, Elastically Decoupling Relics \cite{ELDERs}, or Inelastic Dark Matter \cite{IDM}. Accordingly, due to their multiple appearances in literature, one could ask what possible implications they have in the generation of magic states.

As opposed to the $Z$ boson, a yet-undiscovered dark photon has its gauge coupling, mass and width as free parameters, at least up to a variety of experimental constraints, and when strictly acting as a mediator. Given this freedom, scattering mediated by a dark photon may result in deviations from the QED $M_{2}$ functions discussed in Ref. \cite{magic_in_qed}, or even those of full EW exchange in the previous sections. To address this possibility, we have considered a rather minimal dark scenario consisting on a $U(1)$ symmetry with an unspecified mass-generating mechanism for its gauge boson $A_{\mu}'$, and a dark Dirac fermion $\chi$ charged under it.

In the general treatment one starts with two gauge bosons $A_{a}^{\mu}$ and $A_{b}^{\mu}$, kinetically mixed, and having respective mass parameters $M_{a}$ and $M_{b}$. With primed quantities referring to the dark sector, after the transformation that diagonalizes the gauge kinetic terms the gauge bosons couple to matter through the currents below \cite{Fabbrichesi_2021}
\begin{equation}
 \mathcal{L} \supset \frac{1}{\sqrt{1-2\delta\epsilon + \delta^2}}\left[ \frac{e'(1-\delta\epsilon)}{\sqrt{1-\epsilon^2}}J_\mu' + \frac{e(\delta - \epsilon)}{\sqrt{1-\epsilon^2}}J_\mu\right]A'^\mu + \frac{1}{\sqrt{1-2\delta\epsilon + \delta^2}} \left[ eJ_\mu - \delta e' J_\mu' \right]A^\mu \ , 
 \label{eq:General_Lagrangian}
\end{equation}
where the new basis states $\{ A^{\mu},A^{'\mu} \}$ have canonical field strengths, $\delta = M_b/M_a$ is the mass parameter ratio, $\epsilon$ is the kinetic mixing parameter, and $e,~e'$ are the gauge couplings of each Abelian symmetry. 

As a first approach in the lookout for new behavior, one commits to the scenario where kinetic mixing is manifest between the QED $U(1)$ and dark $U(1)$ symmetries\footnote{A complete top-down treatment, out of the scope of the present work, is that where kinetic mixing is present before electroweak spontaneous symmetry breaking.} so that one of the two gauge bosons is taken massless ($\delta = 0$) and identified with the photon \cite{Fabbrichesi_2021},
\begin{equation}
    \mathcal{L} \supset \frac{1}{\sqrt{1-\epsilon^2}}(e' J'_\mu -e\epsilon J_\mu)A'^\mu + (eJ_\mu)A^\mu.
    \label{eq:dark-lagrangean}
\end{equation}

The interaction above enables the dark photon $A_{\mu}'$ to interact with SM ($f$) and dark Dirac fermions ($\chi$) via respective currents $J_{\mu}=\bar{f}\gamma^\mu f$ and $J_{\mu}'=\bar{\chi}\gamma^\mu\chi$. In principle, this sets off the production of magic states through $2 \rightarrow 2$ scatterings like those analyzed in earlier sections. Unlike QED, $A'$ exchange occurs via a massive propagator 
\begin{equation}
    \frac{-i(g_{\mu \nu} - q_\nu q_\mu/m_{A'}^2)}{q^2 -m_{A'}^2 + i m_{A'} \Gamma_{A'}}.
    \label{eq:dark_Propagator_fake}
\end{equation}
where $m_{A'}$ and $\Gamma_{A'}$ are the mass and total decay width of the dark photon. For simplicity, $m_{A'}\gg \Gamma_{A'}$ will be assumed\footnote{In this work terms of order $O(\Gamma_{A'}^2/m_{A'}^2)$ are neglected in the propagator. After all, this term only impacts in M\o ller-like and Bhabha-like scatterings, generating functions respectively depending on $\Gamma_{A'}, m_\chi$, and $\Gamma_{A'}, m_\chi, \theta$, none of which generates a maximal magic configuration.}. Nevertheless, for dark fermions with strictly vectorial coupling to $A'$, the $q_\mu q_\nu$ piece can be shown to not contribute to the scattering amplitude which leaves the propagator as
\begin{equation}
        \frac{-ig_{\mu \nu} }{q^2 -m_{A'}^2 + i m_{A'} \Gamma_{A'}}.
    \label{eq:dark_Propagator_true}
\end{equation}

\section{Dark magic production}
\label{sec:DarkMagic}

Although the $2\to2$ scatterings mediated by $A'$ resemble those of QED and EW cases, some differences may be expected in their $M_{2}$ functions due to the massive propagator in Eq.~\eqref{eq:dark_Propagator_true} and the vectorial coupling of the dark fermions.

As indicated in Sec.~\ref{subsubsec:PairAnniZ}, global factors do not influence the production of magic states, implying that in processes where more than one scattering channel is present (i.e. M\o ller-like and Bhabha-like scatterings), the propagator will not be factored out and $\Gamma_{A'}$ plays a role. On the other hand, in processes with only one available channel (i.e. annihilation, inverse annihilation, elastic scattering) the propagator is indeed factored out and so is its $\Gamma_{A'}$ dependence. In either case, we checked that the QED results are recovered when $m_{A'} \rightarrow 0$.

The study of magic generation in the dark $U(1)$ scenario proceeds as in the previous analyses, where one looks at the production of $M_{2}$ via the exchange of a dark photon between dark and SM fermions. A dimensionless expansion parameter $\mu_{\text{D}} = |\vec{p}|/m_\chi$ is defined, where $m_\chi$ is the mass of the dark fermion and $\vec{p}$ is the momentum of either initial state particle in the center-of-mass frame. Likewise, mass ratios $\lambda_\chi = m_f/m_\chi$ and $\lambda_a = m_{A'}/m_\chi$ are defined to help organize results, where we distinguish between dark Photons Lighter-than-dark Fermions (PLF) ($\lambda_a \ll 1$) and the (SM-like) case of dark Photons-Heavier-than-dark Fermions (PHF) ($\lambda_a \gg 1$).

In the following subsection, when going over the different scatterings between $f$ and $\chi$, we will restrict ourselves to the low-energy regime,  $E\rightarrow m_{\text{initial}}$. This is so because in the high-energy limit, $E\rightarrow |\vec{p}_{\text{initial}}|$, the four-momentum of the propagator in 
Eq.~\eqref{eq:dark_Propagator_true} dominates over the $A'$ mass, reducing it to
\begin{equation}
    \frac{-ig_{\mu \nu} }{q^2}.
    \label{eq:QED_propagator}
\end{equation}
Since the propagator above resembles its QED counterpart, the high-energy regime of the dark $U(1)$ scatterings will reproduce the same angular magic distributions as in QED, up to the feeble effects of $\lambda_{\chi}$.

\subsection{Low-energy limit}
\label{sec:LEL_DARK}

As previously stated, the production of magic states in the low-energy limit defined by $E\rightarrow m_{\text{initial}}$ is described by the vertex and propagator in Eqs.~\eqref{eq:dark-lagrangean} and~\eqref{eq:dark_Propagator_true}. As will be shown, this will ultimately yield a magic distribution that departs from the QED ones.

To organize the various scenarios, some prior considerations are pertinent. Now that the masses of $A'$ and $\chi$ are free parameters, processes involving kinematic thresholds (say pair production) will be looked at separately for the $m_f > m_\chi$ and $m_\chi > m_f$ fermion mass orderings. We note that processes admitting a power expansion in the ratio $\mu_{\text{D}}$, such as M\o ller-like, Bhabha-like, and elastic scattering, are expected to scale up two orders of magnitude due to the dark photon mass moving the pole in the propagator term from 0 to $\pm \sqrt{m^2_{A'}-i\Gamma_{A'}m_{A'}}$. Then after expanding in $\mu_{\text{D}} \to 0$, the leading-order contribution in the propagator is proportional to $1/(-m_{A'}^2 + i\Gamma_{A'}m_{A'})$ instead of $1/\mu_{\text{D}}^2$. 

\subsubsection*{Pair annihilation $f \bar{f} \rightarrow \chi \bar{\chi}$ with $m_{f}<m_{\chi}$}

\begin{figure}[ht]
    \centering
    \includegraphics[width=1\linewidth]{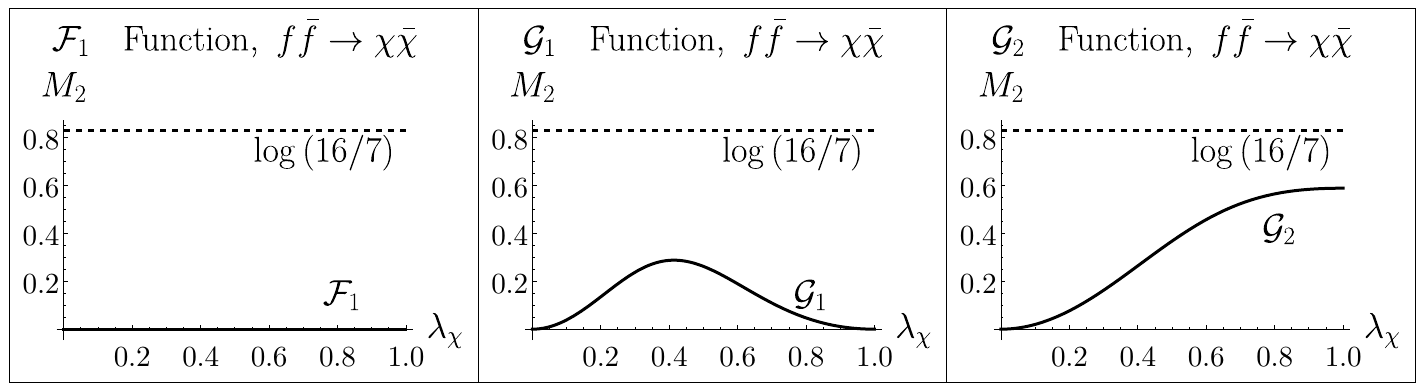}
    \caption{Parameter dependence of the magic distribution functions featured in Table~\ref{tab:D_Anni_SML}, obtained for pair annihilation $f  \bar{f} \rightarrow \chi \bar{\chi}$ in the low-energy regime and $\lambda_{\chi}=m_{f}/m_{\chi}\leq 1$. The dotted line shows the maximal magic allowed.}
    \label{fig:D_Anni_SML}
\end{figure}

Consider dark fermions heavier than SM fermions, $m_{f}<m_{\chi}$. Similar to Sec.~\ref{subsubsec:PairAnniZ}, the kinematic threshold of the process is given by $\sqrt{s} \geq 2m_\chi$, and the three-momentum of the initial states is expressed as $|\vec{p}| \rightarrow m_\chi \sqrt{1 - \lambda_\chi^2}$ with final states produced at rest (the scattering angle is undefined). The corresponding amplitude is shown in Eq.~\eqref{eq:Scat_DAnni1}, and is consistent with QED up to a global factor. Although this process goes through a single channel, QED results are expected because the $s$ Mandelstam becomes the global factor $4m_\chi^2$. This is seen in Table~\ref{tab:D_Anni_SML} and Fig.~\ref{fig:D_Anni_SML}, where the free parameter $\lambda_\chi$ satisfies $0 \leq \lambda_\chi \leq 1$.

\begin{table}[ht]
    \centering
    \begin{tabularx}{\textwidth}{cXcc}
        \toprule
        Magic distribution & Stabilizer state & $(M_2)_{\text{max}}$ & $\theta_{\text{max}}$ or $(\lambda_\chi)_{\text{max}}$\\
        \toprule
        $\mathcal{F}_1$ & 1, 2, 3, 4, 5, 6, 9, 10, 37, 38, 39, 40, 42, 43, 44, 45, 48, 49, 50 & 0 & ---\\
        \hline
        $\mathcal{G}_1$ & 7, 8, 11, 12, 46, 47, 59, 60  & $\log(4/3)$ & $-1 + \sqrt{2}$\\
        \hline
        $\mathcal{G}_2$ & 13, 14, 15, 16, 17, 18, 19, 20, 21, 22, 23, 24, 25, 26, 27, 28, 29, 30, 31, 32, 33, 34, 35, 36, 51, 52, 53, 54, 55, 56, 57, 58 & $\log(9/5)$ & 1 \\
        \hline
        Vanishing & 41 & --- & ---\\
        \bottomrule
    \end{tabularx} 
    \caption{
    $M_{2}$ distributions and corresponding stabilizers obtained in pair annihilation of SM fermions ($f$) into dark fermions ($\chi$) through dark photon exchange, in the low-energy regime and with mass ordering $\lambda_{\chi}=m_{f}/m_{\chi}\leq 1$. Corresponding plots are shown in Fig.~\ref{fig:D_Anni_SML}.}.
    \label{tab:D_Anni_SML}
\end{table}

\subsubsection*{Pair annihilation $\chi \bar{\chi} \rightarrow f \bar{f}$ with $m_{\chi}< m_{f}$}

In the reverse scattering, if SM fermions are heavier than dark fermions, then the kinematic threshold becomes $\sqrt{s}\geq 2m_f$. In this case the momentum of the initial states is expressed as $|\vec{p}| \rightarrow m_\chi\sqrt{\lambda_\chi^2 -1}$, and the mass ratio $\lambda_\chi$ satisfies $\lambda_\chi \geq 1$.

\begin{figure}[ht]
    \centering
    \includegraphics[width=1\linewidth]{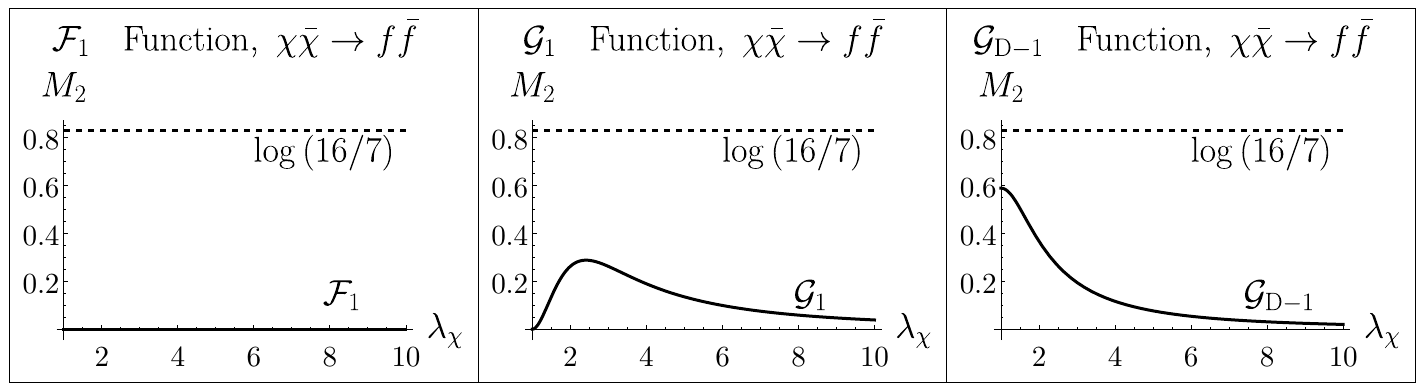}
    \caption{Parameter dependence of the magic distribution functions featuring in Table \ref{tab:D_Anni_SMH}, obtained for pair annihilation $\chi \bar{\chi}\rightarrow f\bar{f}$ in the low-energy regime and with $\lambda_{\chi}=m_{f}/m_{\chi}\geq 1$. The dotted line shows the maximal magic allowed.}
    \label{fig:D_Anni_SMH}
\end{figure}

The role reversal of the fermions, whose amplitude matrix is portrayed in Eq.~\eqref{eq:Scat_DAnni1_inv}, leads to similar magic distribution functions than in $f\bar{f}\to \bar{\chi}\chi$ with $\chi$ heavier, see Table~\ref{tab:D_Anni_SMH} and Fig.~\ref{fig:D_Anni_SMH}. The primary distinction with respect to QED is that the stabilizers whose $M_{2}$ distribution was characterized by the $\mathcal{G}_{2}$ function back in Table~\ref{tab:D_Anni_SML} now produce a new function named $\mathcal{G}_{\text{D}-1}$. Interestingly enough, $\mathcal{G}_{\text{D}-1}$ presents the same maximum magic as $\mathcal{G}_{2}$ at the same $\lambda_\chi$ value.

\begin{table}[ht]
    \centering
    \begin{tabularx}{\textwidth}{cXcc}
        \toprule
        Magic distribution & Stabilizer state & $(M_2)_{\text{max}}$ & $\theta_{\text{max}}$ or $(\lambda_\chi)_{\text{max}}$\\
        \toprule
        $\mathcal{F}_1$ & 1, 2, 3, 4, 5, 6, 9, 10, 37, 38, 39, 40, 42, 43, 44, 45, 48, 49, 50 & 0 & ---\\
        \hline
        $\mathcal{G}_1$ & 7, 8, 11, 12, 46, 47, 59, 60   & $\log(4/3)$ & $1 + \sqrt{2}$\\
        \hline
        $\mathcal{G}_{\text{D}-1}$ & 13, 14, 15, 16, 17, 18, 19, 20, 21, 22, 23, 24, 25, 26, 27, 28, 29, 30, 31, 32, 33, 34, 35, 36, 51, 52, 53, 54, 55, 56, 57, 58 & $\log(9/5)$ & 1 \\
        \hline
        Vanishing & 41 & --- & ---\\
        \bottomrule
    \end{tabularx}
    \caption{$M_{2}$ distributions and corresponding stabilizers obtained in pair annihilation of dark fermions ($\chi$) into SM fermions ($f$) through dark photon exchange, in the low-energy regime and with mass ordering $\lambda_{\chi}=m_{f}/m_{\chi}\geq 1$. The dotted line shows the maximal magic allowed. Corresponding plots are shown in Fig. \ref{fig:D_Anni_SMH}.}
    \label{tab:D_Anni_SMH}
\end{table}

\subsubsection*{M\o ller-like scattering $\chi\chi\to\chi\chi$} 

For M\o ller-like scattering ($\chi \chi \to \chi \chi$) in the low-energy regime ($\mu_{\text{D}} \to 0$), one expands the scattering amplitude at leading order in $\mu_{\text{D}}$, see Eqs.~\eqref{eq:Scat_DMoller_PLF} and~\eqref{eq:Scat_DMoller_PHF}. As discussed in Sec.~\ref{sec:LEL_DARK}, this results in a scaling $O(\mu_{\text{D}}^0)$ in contrast to the $O(\mu_{\text{D}}^{-2})$ QED scaling (due to some amplitude matrix entries vanishing at $O(\mu_{\text{D}}^0)$ for certain stabilizers, terms up to order $O(\mu_{\text{D}}^2)$ are kept). To understand the resulting $M_{2}$ better, the scattering amplitude is also expanded first in the gauge-to-fermion mass ratio $\lambda_a$ for the limiting cases of dark Photons-Lighter-than-dark Fermions (PLF) where $\lambda_a\rightarrow 0$, and later in $\lambda_{a}^{-1}$ for dark Photons-Heavier-than-dark Fermion (PHF), where instead $\lambda_a \rightarrow \infty$.

\begin{figure}[ht]
    \centering
    \includegraphics[width=1\linewidth]{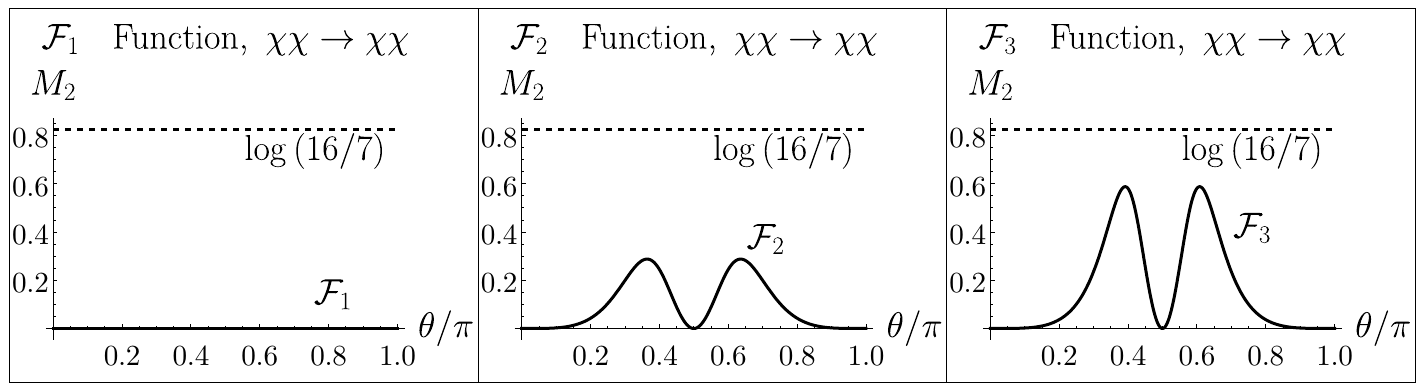}
    \caption{Parameter dependence of the magic distribution functions featuring in Table \ref{tab:D_Moller_PLF}, obtained for M\o ller-like scattering ($\chi \chi \to \chi \chi$) in the low-energy regime, and with dark Photons-Lighter-than-dark Fermions (PLF). The dotted line shows the maximal magic allowed.}
    \label{fig:D_Moller_PLF}
\end{figure}

\begin{figure}[ht]
    \centering
    \includegraphics[width=1\linewidth]{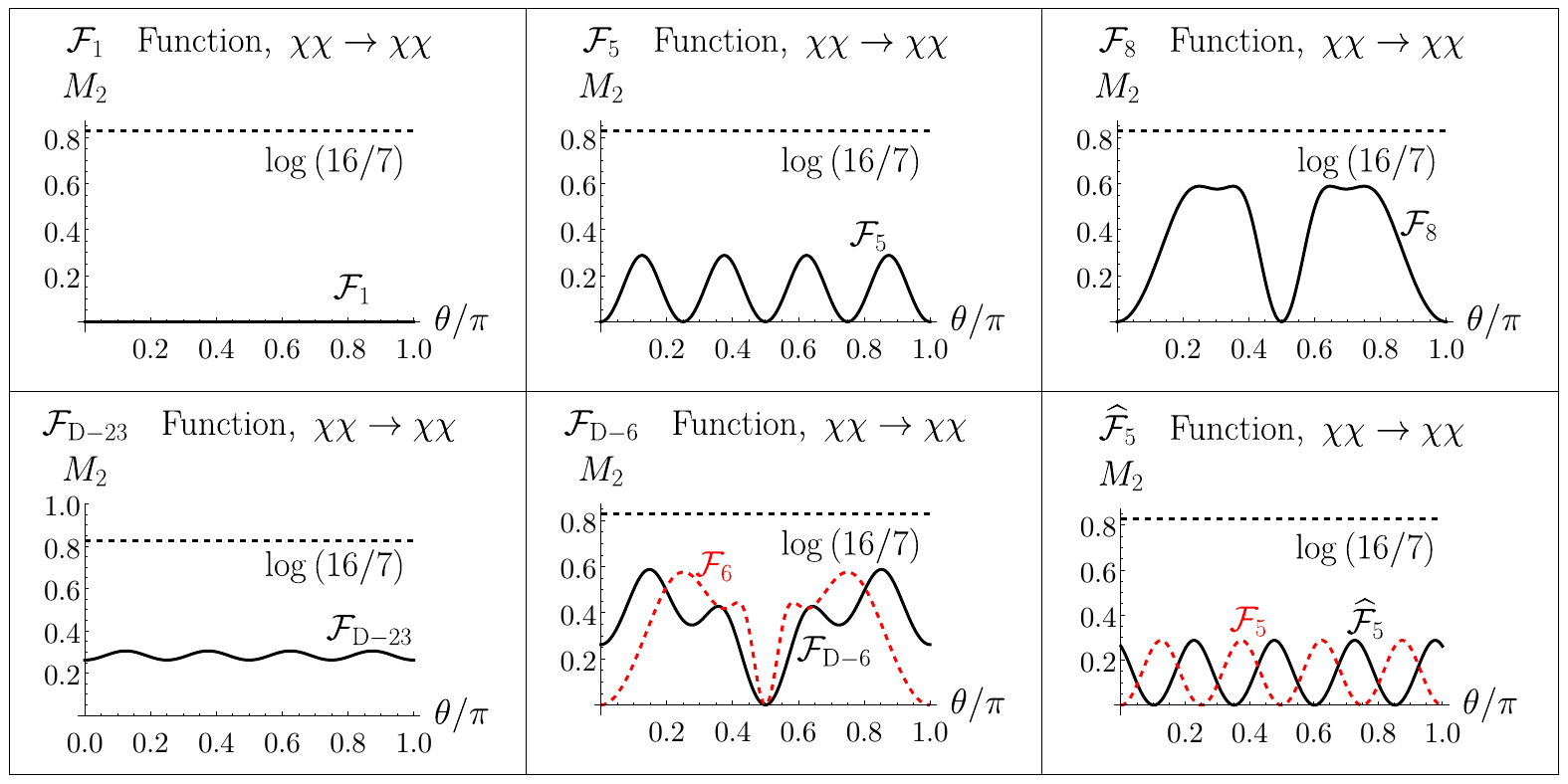}
    \caption{Parameter dependence of the magic distribution functions featuring in Table \ref{tab:D_Moller_PHF}, obtained for M\o ller-like scattering in the low-energy regime, and with dark Photons-Higher-than-dark Fermions (PHF). The dotted line shows the maximal magic allowed. Similar-looking QED magic distributions are shown as red dotted lines.}
    \label{fig:D_Moller_PHF}
\end{figure}

When expanding the scattering amplitude in $\lambda_a$ at leading order, a scaling of order $O(\lambda_a^{-2})$ is found for both PLF and PHF cases. The results~\footnote{Be aware that the family of $\mathcal{F}_1$ functions is composed of 1 and sign$(\cos(\theta))^8$ which are identical except for $\theta= \pi/2$.} for the PLF limit are shown in Table \ref{tab:D_Moller_PLF} and Fig.~\ref{fig:D_Moller_PLF}, whose resemblance to the QED case is expected because the limit $m_{A'}\rightarrow 0$ is SM-like. As for the PHF case, new angular magic distribution functions appear, see Table~\ref{tab:D_Moller_PHF} and Fig.~\ref{fig:D_Moller_PHF}. Notice that even though three new magic distribution functions originate, $\widehat{\mathcal{F}}_5$ is nothing but a $\theta$ shift of the $\mathcal{F}_5$ function previously obtained in QED. Meanwhile, $\mathcal{F}_{\text{D}-6}$ follows a similar shape to $\mathcal{F}_6$, reaching the same maximal value. The $\mathcal{F}_{\text{D}-23}$ function, however, is truly new.

\begin{table}[ht]
    \centering
    \begin{tabularx}{\textwidth}{cXcc}
        \toprule
        Magic distribution & Stabilizer state & $(M_2)_{\text{max}}$ & $\theta_{\text{max}}$ or $(\lambda_\chi)_{\text{max}}$\\
        \toprule
        $\mathcal{F}_1$ & 1, 2, 5, 6, 9, 10, 37, 38, 39, 40, 41, 42, 45, 48, 49, 50 & 0 & ---\\
        \hline
        $\mathcal{F}_2$ & 3, 4, 7, 8, 11, 12, 43, 44, 46, 47, 59, 60 & $\log(4/3)$ & $\begin{matrix} 2\arctan(\sqrt{\sqrt{2}-1}), \\ \pi -2\arctan(\sqrt{\sqrt{2}-1}) \end{matrix} $\\
        \hline
        $\mathcal{F}_3$ & 13, 14, 15, 16, 17, 18, 19, 20, 21, 22, 23, 24, 25, 26, 27, 28, 29, 30, 31, 32, 33, 34, 35, 36,  51, 52, 53, 54, 55, 56, 57, 58 & $\log(9/5)$ & $\begin{matrix} \arctan(2\sqrt{2}), \\ \pi-\arctan(2\sqrt{2})\end{matrix}  $\\
        \bottomrule
    \end{tabularx}
    \caption{$M_{2}$ distributions and corresponding stabilizers obtained in M\o ller-like scattering ($\chi \chi \to \chi \chi$) through dark photon exchange, in the low-energy regime and with dark Photons-Lighter-than-dark Fermions (PLF).}
    \label{tab:D_Moller_PLF}
\end{table}

\begin{table}[ht]
    \centering
    \begin{tabularx}{\textwidth}{cXcc}
        \toprule
        Magic distribution & Stabilizer state & $(M_2)_{max}$ & $\theta_{\text{max}}$ or $(\lambda_\chi)_{max}$\\
        \toprule
        $\mathcal{F}_1$ & 3, 4, 7, 8, 11, 12, 13, 14, 15, 16, 17, 18, 19, 20, 21, 22, 23, 24, 25, 26, 27, 28, 29, 30, 31, 32, 33, 34, 35, 36, 37, 42, 43, 44, 46, 47, 51, 52, 53, 54, 55, 56, 57, 58, 59, 60 & 0 & ---\\
        \hline
        $\mathcal{F}_5$ & 38, 41 & $\log(4/3)$ & $\begin{matrix}
            n\pi/8, \\
            n={1,3,5,7}
        \end{matrix} $\\
        \hline
        $\widehat{\mathcal{F}}_{5,1}$ & 45 & $\log(4/3)$ & $\begin{matrix}
            
        1/4 (n\pi - \arctan[7/24]), \\ n = {1,2,3,4}\end{matrix}$\\
        \hline
        $\widehat{\mathcal{F}}_{5,2}$ & 48 & $\log(4/3)$ & $\begin{matrix}
            
        1/4 (n\pi + \arctan[7/24]), \\ n = {0,1,2,3}\end{matrix}$\\
        \hline
        $\mathcal{F}_{\text{D}-6}$ & 5, 6, 49, 50 & $\log(9/5)$ & $\begin{matrix}
            \pi - \arctan(1/2) \\
            \arctan(1/2)
        \end{matrix}$\\
        \hline
        $\mathcal{F}_8$ & 1, 2, 39, 40 & $\log(9/5)$ & $\begin{matrix}
            \pi/4, \quad 3\pi/4, \\
            2\arctan(\frac{1}{2}[-1+\sqrt{5}]), \\
            2\arctan(\frac{1}{2}[1+\sqrt{5}])
        \end{matrix}$ \\
        \hline
        $\mathcal{F}_{\text{D}-23}$ & 9, 10 & $\log(2500/1843)$ & $\begin{matrix}
            n\pi/8, \\
            n={1,3,5,7}
        \end{matrix} $ \\
        \bottomrule
    \end{tabularx}
    \caption{$M_{2}$ distributions and corresponding stabilizers obtained in in M\o ller-like scattering ($\chi \chi \to \chi \chi$) through dark photon exchange, in the low-energy regime and with dark Photons-Heavier-than-dark Fermions (PHF). Corresponding plots are shown in Fig. \ref{fig:D_Moller_PHF}. Hatted functions are shifted in $\theta$ with respect to those without a hat.}
    \label{tab:D_Moller_PHF}
\end{table}

\subsubsection*{Bhabha-like scattering $\chi\bar{\chi}\to \chi\bar{\chi}$}

\begin{figure}[ht]
    \centering
    \includegraphics[width=1\linewidth]{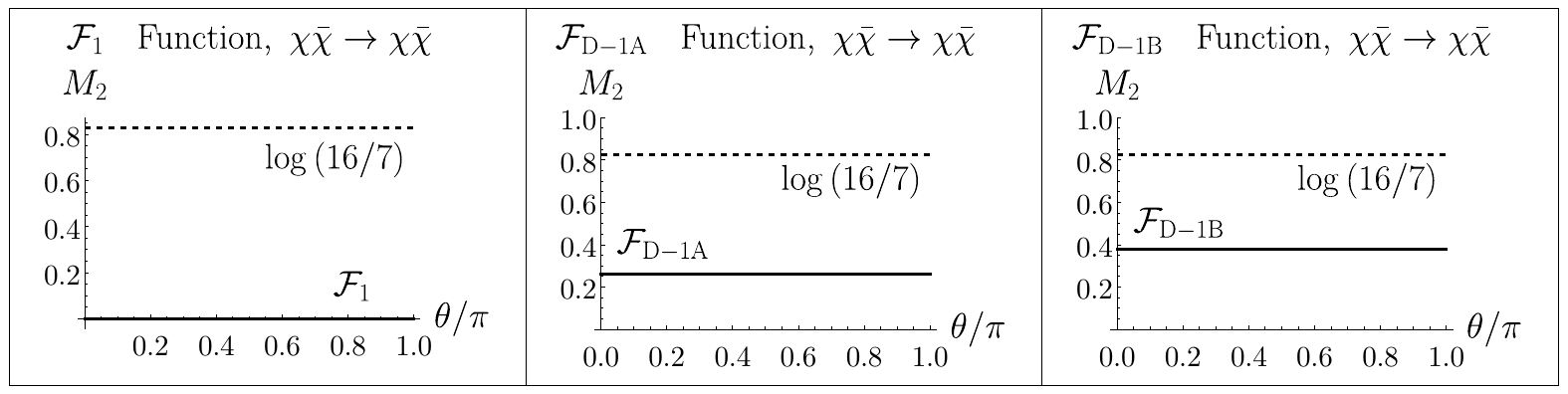}
    \caption{Constant magic distribution functions featured in Table \ref{tab:D_Bhabha_PHF}, obtained for Bhabha-like scattering ($\chi\bar{\chi}\to\chi\bar{\chi}$) in the low-energy regime, and with dark Photons-Higher-than-dark Fermions (PHF). The dotted line shows the maximal magic allowed.}
    \label{fig:D_Bhabha_PHF}
\end{figure}

Similar to the M\o ller-like case, expanding the scattering amplitude of Bhabha-like scattering ($\chi\bar{\chi}\to\chi\bar{\chi}$) at leading order in $\mu_{\text{D}} \rightarrow 0$ results in a $O(\mu_{\text{D}}^0)$ scaling in contrast with the $O(\mu_{\text{D}}^{-2})$ QED behavior. When expanding in $\lambda_a$ at leading order, the PLF case yields a diagonal scattering amplitude matrix with a scaling $O(\lambda_a^{-1})$, see Eq.~\eqref{eq:Scat_bhabha_PLF}. This mimics the QED results and generates no magic states, as it is expected because the PLF matches the SM-like case $m_{A'}\to0$. Meanwhile, the PHF case generates a non-diagonal scattering amplitude matrix whose entries carry a global factor going as $O(\lambda_a^{-2})$, shown in Eq.~\eqref{eq:Scat_bhabha_PHF}. These new, constant magic distributions are collected in Fig.~\ref{fig:D_Bhabha_PHF} and Table~\ref{tab:D_Bhabha_PHF}.

\begin{table}[ht]
    \centering
    \begin{tabularx}{\textwidth}{cXcc}
        \toprule
        Magic distribution & Stabilizer state & $(M_2)_{\text{max}}$ & $\theta_{\text{max}}$ or $(\lambda_\chi)_{\text{max}}$\\
        \toprule
        $\mathcal{F}_1$ & 1, 2, 7, 8, 11, 12, 37, 38, 39, 40, 41, 42, 46, 47, 59, 60 & 0 & ---\\
        \hline
        $\mathcal{F}_{\text{D}-1A}$ & 3, 4, 5, 6, 9, 10, 43, 44, 45, 48, 49, 50 & $\log(625/481)$ & ---\\
        \hline
        $\mathcal{F}_{\text{D}-1B}$ & 13, 14, 15, 16, 17, 18, 19, 20, 21, 22, 23, 24, 25, 26, 27, 28, 29, 30, 31, 32, 33, 34, 35, 36, 51, 52, 53, 54, 55, 56, 57, 58 & $\log(343/235)$ & ---\\ 
        \bottomrule
    \end{tabularx}
    \caption{$M_{2}$ distributions and corresponding stabilizers obtained in Bhabha scattering ($\chi\bar{\chi}\to \chi\bar{\chi}$), in the low-energy regime and with dark Photons-Heavier-than-dark Fermions (PHF). Corresponding plots are shown in Fig. \ref{fig:D_Bhabha_PHF}.}
    \label{tab:D_Bhabha_PHF}
\end{table}

\subsubsection*{Elastic scattering $\chi f \to \chi f$}

In dark-visible elastic scattering, $\chi f \to \chi f$, the leading-order expansion yields a scaling $O(\mu_{\text{D}}^0)$, as opposed to the QED $O(\mu_{\text{D}}^{-2})$ dependence. Nonetheless, the scattering amplitude matrix is diagonal, as shown in Eq.~\eqref{eq:Dark_fermion_scattering}. Thus, just as in QED, no magic is generated regardless of the dark sector masses.

\subsubsection*{Inverse pair annihilation $\chi\bar{\chi} \rightarrow f \bar{f}$ with $m_{\chi}>m_{f}$ }

\begin{figure}[ht]
    \centering
    \includegraphics[width=1\linewidth]{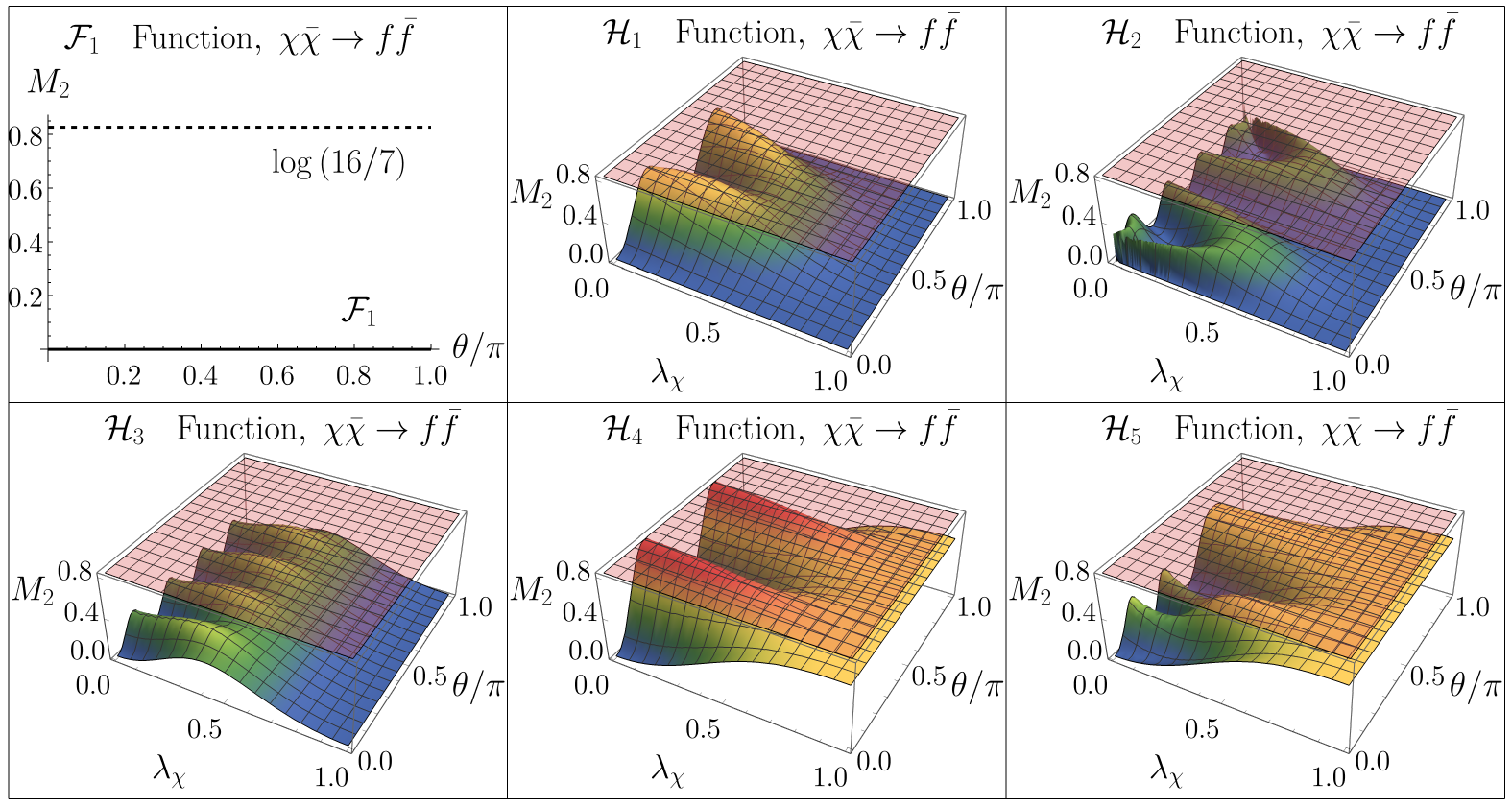}
    \caption{Parameter dependence of the magic distribution functions featuring in Table \ref{tab:D_Inv_Anni_SML}, obtained for inverse pair annihilation in the low-energy regime, for $\chi$ heavier than $f$. The black dotted line and red plane show the maximal magic allowed.}
    \label{fig:D_Inv_Anni_SML}
\end{figure}

As mentioned in Sec. \ref{sec:LEL_DARK}, the dark fermion mass being a free parameter brings up two scenarios, dark fermions heavier or lighter than SM fermions. When $m_{\chi}>m_{f}$, the expansion at leading order is $O(\mu_{\text{D}}^0)$, and displays simultaneous angular ($\theta$) and mass ratio ($\lambda_\chi$) dependence, see Eqs. (\ref{eq:entries_Inv_Pair_Anni_SML}). New magic distributions functions $\mathcal{H}(\lambda_\chi,\theta)$ appear, all presented in Table \ref{tab:D_Inv_Anni_SML} and Fig. \ref{fig:D_Inv_Anni_SML}, where $ 0\leq\lambda_\chi \leq 1$ is satisfied.

\begin{table}[ht]
    \centering
    \begin{tabularx}{\textwidth}{cXcc}
        \toprule
        Magic distribution & Stabilizer state & $(M_2)_{\text{max}}$ & $\theta_{\text{max}}$ or $(\lambda_\chi)_{\text{max}}$\\
        \toprule
        $\mathcal{F}_1$ & 9, 10, 38, 45, 48 & 0 & ---\\
        \hline
        $\mathcal{H}_1$ & 1, 2, 39, 40& 0.587... & 0.785...(0.000410954...)\\
        \hline
        $\widehat{\mathcal{H}}_1$ &7, 8, 59, 60 & 0.587... & 0.785...(0.000413034...)\\
        \hline
        $\mathcal{H}_2$ & 3, 4, 42, 43, 44 & 0.287... &0.878...(0.372743...)\\
        \hline
        $\widehat{\mathcal{H}}_{2,1}$ & 5,6,37,49,50  & 0.287682 & 0.563...(0.31951...)\\
        \hline
        $\widehat{\mathcal{H}}_{2,2}$ & 46 & 0.287... & 1.190...(0.409699...)\\
        \hline
        $\widehat{\mathcal{H}}_{2,3}$ & 47 & 0.287... & 0.256...(0.23501...)\\
        \hline
        $\mathcal{H}_3$ & 11, 12 & 0.405... & 0.392...(0.317837...)\\
        \hline
        $\mathcal{H}_4$ & 13, 14, 15, 16, 17, 18, 19, 20, 21, 22, 23, 24, 25, 26, 27, 28 & 0.826678... & 0.785...(0.000649697...)\\
        \hline
        $\mathcal{H}_5$ & 29, 30, 31, 32, 52, 54, 57, 58 & 0.587... & 0.819...(0.999108...)\\
        \hline
        $\widehat{\mathcal{H}}_5$ & 33, 34, 35, 36, 51, 53, 55, 56 & 0.587... & 0.785...(0.575904...)\\
        \hline
        Vanishes & 41 & 0 & ---\\
        \bottomrule
        \end{tabularx}
    \caption{$M_{2}$ distributions and corresponding stabilizers obtained in inverse pair annihilation $\chi\bar{\chi} \rightarrow f \bar{f}$, in the low-energy regime and with $\chi$ heavier than $f$. Corresponding plots are shown in Fig. \ref{fig:D_Inv_Anni_SML}. Once again, hatted functions are $\theta$-shifts of their counterparts without a hat. These shifts in $\theta$ are listed in Table \ref{tab:Phases}.}
    \label{tab:D_Inv_Anni_SML}
\end{table}

\subsubsection*{Inverse pair annihilation $f \bar{f} \rightarrow \chi\bar{\chi}$ with $m_{f}>m_{\chi}$}

\begin{figure}[ht]
    \centering
    \includegraphics[width=1\linewidth]{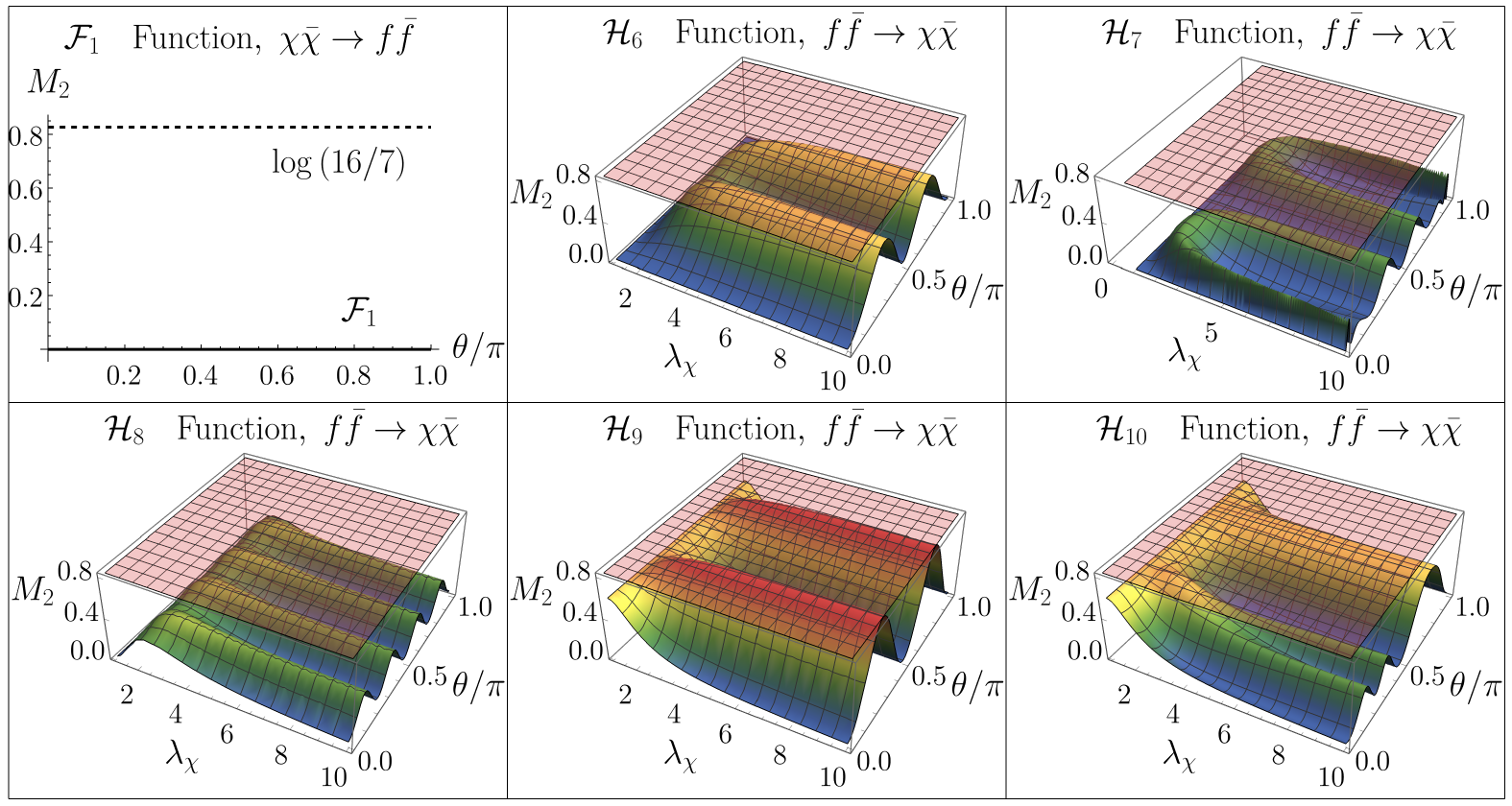}
    \caption{
    Parameter dependence of the magic distribution functions featuring in Table \ref{tab:D_Inv_Anni_SMH}, obtained for inverse pair annihilation in the low-energy regime, for $f$ heavier than $\chi$. The black dotted line and red plane shows the maximal magic allowed.}
    \label{fig:D_Inv_Anni_SMH}
\end{figure}

When SM fermions are heavier than dark fermions, the leading order terms in $f\bar{f}\to \chi\bar{\chi}$ are $O(\mu_{\text{D}}^0)$, depending only on the scattering angle $\theta$ and mass ratio $\lambda_{\chi}$ (constrained by $\lambda_\chi \geq 1$). This amplitude, with entries listed in Eq. (\ref{eq:entries_Inv_Pair_Anni_SMH}), yields new $\mathcal{H}(\lambda_\chi,\theta)$ functions, summarized in Table \ref{tab:D_Inv_Anni_SMH} and Fig. \ref{fig:D_Inv_Anni_SMH}.

The $\mathcal{H}(\lambda_\chi,\theta)$ functions have been plotted up to $\lambda_\chi = 10$ in order to show their monotonic behavior at $\lambda_\chi \gg1$. It is seen that their most interesting behavior happens at small mass ratio values, $\lambda_\chi=O(1)$.

\begin{table}[ht]
    \centering
    \begin{tabularx}{\textwidth}{cXcc}
        \toprule
        Magic distribution & Stabilizer state & $(M_2)_{\text{max}}$ & $\theta_{\text{max}}$ or $(\lambda_\chi)_{\text{max}}$\\
        \toprule
        $\mathcal{F}_1$ & 9, 10, 38, 45, 48 & 0 & ---\\
        \hline
        $\mathcal{H}_6$ & 1, 2, 39, 40& 0.587...& ---\\
        \hline
         $\widehat{\mathcal{H}}_6$ & 7, 8, 59, 60& 0.587.. & ---\\
        \hline
        $\mathcal{H}_7$ &3, 4, 42, 43, 44 & 0.287...& ---\\
        \hline
         $\widehat{\mathcal{H}}_{7,1}$ &5, 6, 37, 49, 50 & 0.287... & ---\\
        \hline
        $\widehat{\mathcal{H}}_{7,2}$ & 46 & 0.287... & ---\\
        \hline
        $\widehat{\mathcal{H}}_{7,3}$ & 47 & 0.287... & ---\\
        \hline
        $\mathcal{H}_8$ & 11, 12& 0.405... & ---\\
        \hline
        $\mathcal{H}_9$ & 13, 14, 15, 16, 17, 18, 19, 20, 21, 22, 23, 24, 25, 26, 27, 28 & 0.826... & ---\\
        \hline
        $\mathcal{H}_{10}$ & 29, 30, 31, 32, 52, 54, 57, 58 & 0.587... & ---\\
        \hline
        $\widehat{\mathcal{H}}_{10}$ & 33, 34, 35, 36, 51, 53, 55, 56& 0.587... & ---\\
        \hline
        Vanishes & 41 & --- & ---\\ 
        \bottomrule
    \end{tabularx}
    \caption{
    $M_{2}$ distributions and corresponding stabilizers obtained in inverse pair annihilation $f \bar{f}\rightarrow \chi\bar{\chi}$, in the low-energy regime, with $f$ heavier than $\chi$. Corresponding plots are shown in Fig. \ref{fig:D_Inv_Anni_SMH}. Once again, hatted functions are $\theta$-shifts of their counterparts without a hat. These shifts in $\theta$ are listed in Table \ref{tab:Phases}.
    }
    \label{tab:D_Inv_Anni_SMH}
\end{table}

\section{Maximal dark magic generation}
\label{sec:MaxMagic}

The former analysis allows to identify two new magic distribution functions that saturate the maximum magic in the inverse annihilation process in the low energy regime, whether for dark fermions are heavier or lighter than SM fermions. For dark fermions heavier than SM fermions (i.e $0\leq \lambda_\chi \leq 1$), it is the function $\mathcal{H}_4(\lambda_\chi,\theta)$, while for dark fermions lighter than SM fermions (i.e $\lambda_\chi \geq 1$), such function is $\mathcal{H}_9(\lambda_\chi,\theta)$. Both functions, with analytical forms
\begin{equation}
    \begin{split}
    \mathcal{H}_4(\lambda_\chi,\theta) = \frac{1}{8(2+\lambda_\chi^2)^4}&\Big( 89 + 8\lambda_\chi + 64\lambda_\chi ^2 + 8\lambda_\chi^3 +180\lambda_\chi^4 + 4\lambda_\chi^6 + 7\lambda_\chi^8 + 36(-1+\lambda_\chi)^2\cos(4\theta)\\
    & + (-1+\lambda_\chi)^4(3+4\lambda_\chi+6\lambda_\chi^2+4\lambda_\chi^3+\lambda_\chi^4)\cos(8\theta)\Big),
    \end{split}
    \label{eq:H4}
\end{equation}

\begin{equation}
    \begin{split}
         \mathcal{H}_9(\lambda_\chi,\theta) &=\frac{1}{8(1+2\lambda_\chi^2)^4}\Big(7 + 4\lambda_\chi^2 + 180\lambda_\chi^4 +8\lambda_\chi^5+64\lambda_\chi^6+8\lambda_\chi^7+89\lambda_\chi^8\\
         & + 36\lambda_\chi^4(-1+\lambda_\chi^2)^2\cos(4\theta) + (-1 + \lambda_\chi^2)^4(1 + 4\lambda_\chi + 6\lambda_\chi^2 + 4\lambda_\chi^3 + 3\lambda_\chi^4)\cos(8\theta)\Big),
    \end{split}    
    \label{eq:H9}
\end{equation}
reach their maxima at the same angular value(s), $\theta \to \pi/4, \, 3\pi/4$, but at different $\lambda_\chi$, being $\lambda_\chi \to 0$ for $\mathcal{H}_4$ and $\lambda_\chi \to \infty$ for $\mathcal{H}_9$. This imposes a rather loose restriction on the maximization of the production of dark magic states: for $\mathcal{H}_4$, we found that $\lambda_\chi^{-1} \sim \mathcal{O}(1)-\mathcal{O}(2)$ is enough, while for $\mathcal{H}_9$ one must take $\lambda_\chi \sim \mathcal{O}(1)- \mathcal{O}(2)$, depending on the specific SM fermion involved in the annihilation.

\section{Discussion}
\label{sec:discussion}

This work studied the production of magic states via $2 \rightarrow 2$ fermion tree-level scattering processes mediated by neutral electroweak bosons ($\gamma, Z$) and a massive gauge boson ($A'$) under a $U(1)$ dark symmetry, taking the fermion spins as qubits. In the EW sector's low-energy regime, the absence of new magic distributions with respect to QED can be understood directly from the behavior of the propagators involved. When $q^2\ll m_Z^2$, the $Z$-boson propagator is effectively reduced to a term proportional to $1/m_Z^2$, whereas the photon propagator still scales as $1/s$. Since in this regime $s\ll m_Z^2$, the neutral-current contribution is strongly suppressed with respect to the photon exchange. As a result, the full scattering amplitude is dominated by its QED component, and the spin structure relevant for magic production remains unchanged.

In the $Z$-resonance regime, a relevant observation arises: it is found that, with the only exception of pair annihilation, the resulting magic distributions remain unchanged when calculating them up to order $O(\lambda_Z^{-1})$. The reason is that, for pair annihilation, several amplitudes vanish at leading order and the first nontrivial contribution appears only until that order. For M\o ller and elastic scattering, the leading-order amplitudes simplify considerably because the explicit dependence on the $Z$-boson mass and width drops out of the scattering matrix, leaving only the angular dependence and sensitivity to the specific values of the weak angle. In Bhabha scattering, however, even at order $O(\lambda_Z^{0})$, the amplitudes still retain explicit dependence on the $Z$-boson mass and width. Nevertheless, despite this distinction, the corresponding magic distributions are unchanged when the analysis is extended to $O(\lambda_Z^{-1})$ showing that the leading-order structure already captures the relevant features of magic generation in the resonant regime.

Regarding the dark $U(1)$ sector, we studied the effect of a massive mediator exchanged between SM fermions and a Dirac dark fermion. First of all, the QED results are recovered in the high-energy regime since in this limit the propagator is dominated by the squared (inverse) center-of-mass energy $1/s^{2}$. Meanwhile, in the low-energy limit some effects are expected to alter the generation of magic states in comparison with QED. Specifically, the change in the propagator scales down the sizes of the M\o ller-like, Bhabha-like, elastic scattering, and inverse pair annihilation amplitudes by two powers of the momentum-to-fermion mass ratio $\mu_{\text{D}}$, but they may generate magic (through a non-diagonal scattering matrix) depending on the relative mass ratios between the dark photon and dark fermion. Processes where two or more channels were present, for instance M\o ller-like and Bhabha-like scatterings, are affected by the decay width dependence too.

QED results have been mostly recovered for pair annihilation process since neither the $\mu_{\text{D}}$ expansion nor the dark photon decay width $\Gamma_{A'}$ play a major role, and the only extra effect is sourced by the extra degree of freedom in $\lambda_\chi$ (bound within $0 \leq \lambda_\chi \leq 1$ or $\lambda_\chi \leq 1$ depending on the case).

Concerning M\o ller-like and Bhabha-like scatterings, new magic distribution functions appear when dark photons are heavier than dark fermions, whereas the results reduce to those of QED when the dark photons are instead lighter than the dark fermions, unsurprising since this matches the $m_{A'}\to0$, SM-like case. The QED case is recovered too for elastic fermion scattering, for which the amplitude is still diagonal.

Finally, we obtained genuinely new magic distribution functions for $\chi\bar{\chi}\to f\bar{f}$ and $f\bar{f}\to\chi\bar{\chi}$ scatterings, which display stabilizer classes that reach the maximal magic value depending on the SM-to-dark fermion mass ratio, but only at specific scattering angles. In $\chi\bar{\chi}\to f\bar{f}$, when dark fermions are heavier than SM fermions, a subset of stabilizers lead to the maximal magic value for scattering angles of $\theta\to \pi/4,~3\pi/4$ provided that the SM-to-dark mass fermion ratio decreases. On the other hand, in the reverse scattering $f\bar{f}\to\chi\bar{\chi}$, that same subset of stabilizers maximizes its magic as long as the the SM-to-dark mass ratio increases (at the same scattering angle values).

\section*{Acknowledgments}
\label{sec:ack}
A.A. and Y.L. acknowledge support from SNII-SECIHTI (M\'exico). C.B. acknowledges support from  ANID-Chile FONDECYT/Regular 1241855 and ANID-Chile FONDECYT/Regular 1261103. We thank Antonio Delgado for valuable comments about qubits in the context of high-energy scattering.


\appendix

\section{EW magic scattering amplitudes}
\label{appendix:EW_amplitudes}

Due to their length, only three representative cases are included here. These correspond to the pair annihilation process $e^-e^+\rightarrow \mu^-\mu^+$ in the three energy regimes studied previously. In order to simplify and keep a shorthand notation, the next definitions should be introduced for all three scattering matrices,
\begin{equation}
a=1-4s_W^2, \quad b= 1-2s_W^2, \quad c=1+8s_W^2, \quad d=ab, \quad x=\cos{\theta}, \quad y=\sin{\theta}.
\end{equation}
Below are presented the individual entries of the three scattering amplitude matrices.

\subsection*{Low-energy limit}

\begin{align}
&\mathcal{M}_{\uparrow\uparrow\rightarrow\uparrow\uparrow}
= -2-\frac{\rho}{2s_W^2}a\left( a-\beta\right), \notag \\
&\mathcal{M}_{\downarrow\downarrow\rightarrow\downarrow\downarrow}
= -2-\frac{\rho}{2s_W^2}a\left(a+\beta \right), \notag\\
&\mathcal{M}_{\uparrow\downarrow\rightarrow\uparrow\downarrow}
=\mathcal{M}_{\downarrow\uparrow\rightarrow\downarrow\uparrow}
=\lambda\left( -1+2\rho b \right), \notag\\
&\mathcal{M}_{\uparrow\downarrow\rightarrow\downarrow\uparrow}
=\mathcal{M}_{\downarrow\uparrow\rightarrow\uparrow\downarrow}
= \lambda\left[ 1+\frac{\rho}{2s_W^2}\left( b^2+4s_W^4 \right) \right],
\end{align}
where, 
\begin{equation}
\lambda=\dfrac{m_{e}}{m_{\mu}}, \quad\rho=\frac{m_\mu^2}{c_W^2 m_Z^2}, \quad \beta= \sqrt{1-\lambda^2}.
\end{equation}
Any matrix entry not listed above vanishes.

\subsection*{High-energy limit}

Under the same definitions given before, the high-energy amplitude matrix looks like
\begin{align}
&\mathcal{M}_{\uparrow\uparrow\rightarrow\uparrow\uparrow}
= -\frac{3\left( 1+\mu \right) +2\mu x+3\left(\mu-1 \right)x^2}{4c_W^2\mu}, \notag\\
&\mathcal{M}_{\uparrow\uparrow\rightarrow\uparrow\downarrow}
= -\mathcal{M}_{\uparrow\uparrow\rightarrow\downarrow\uparrow} =\frac{\left[ \mu+3\left(\mu-1 \right)x \right]y}{4c_W^2 \mu}, \notag\\
&\mathcal{M}_{\uparrow\uparrow\rightarrow\downarrow\downarrow}
=\frac{3\left( \mu-1 \right)y^2}{4c_W^2 \mu}, \notag \\
&\mathcal{M}_{\uparrow\downarrow\rightarrow\uparrow\uparrow}
=-\mathcal{M}_{\downarrow\uparrow\rightarrow\uparrow\uparrow}
= \frac{\lambda\left( cx-a \right)y}{16c_W^2 s_W^2 \mu}, \notag\\
&\mathcal{M}_{\uparrow\downarrow\rightarrow\uparrow\downarrow}
=\mathcal{M}_{\downarrow\uparrow\rightarrow\downarrow\uparrow}
=-\mathcal{M}_{\uparrow\downarrow\rightarrow\downarrow\uparrow}=-\mathcal{M}_{\downarrow\uparrow\rightarrow\uparrow\downarrow}=- \frac{c\lambda y^2}{16c_W^2 s_W^2 \mu}, \notag\\
&\mathcal{M}_{\uparrow\downarrow\rightarrow\downarrow\downarrow}
=-\mathcal{M}_{\downarrow\uparrow\rightarrow\downarrow\downarrow}
= \frac{\lambda\left(a+cx \right)y}{16c_W^2 s_W^2 \mu}, \notag\\
&\mathcal{M}_{\downarrow\downarrow\rightarrow\uparrow\uparrow}
=\frac{\left(1+2s_W^2 \right) \left( \mu-1\right)y^2}{8c_W^2 s_W^2 \mu}, \notag\\
&\mathcal{M}_{\downarrow\downarrow\rightarrow\uparrow\downarrow}
=-\mathcal{M}_{\downarrow\downarrow\rightarrow\downarrow\uparrow}
= \frac{\left[ \mu b+\left( 1+2s_W^2 \right) \left( \mu-1\right)x \right]y}{8c_W^2 s_W^2 \mu}, \notag \\
&\mathcal{M}_{\downarrow\downarrow\rightarrow\downarrow\downarrow}
=-\frac{\left(1+2s_W^2 \right) \left[\left(1+\mu\right) +
\left( \mu-1 \right)x^2\right]+2\mu bx}{8c_W^2 s_W^2 \mu}.
\end{align}

\subsection*{$Z$-resonance limit}

As for the resonant regime of the $Z$ boson,
\begin{align}
&\mathcal{M}_{\uparrow\uparrow\rightarrow\uparrow\uparrow}
= -\frac{\Delta}{2}+\kappa\left(4x-a\Delta \right),\notag\\
&\mathcal{M}_{\uparrow\uparrow\rightarrow\uparrow\downarrow}
= -\mathcal{M}_{\uparrow\uparrow\rightarrow\downarrow\uparrow} =y\left[ \eta x-2\kappa\left( 1-a\delta x \right) \right],\notag\\
&\mathcal{M}_{\uparrow\uparrow\rightarrow\downarrow\downarrow}
=\eta\left( 1+2a\kappa \right)y^2, \notag\\
&\mathcal{M}_{\uparrow\downarrow\rightarrow\uparrow\uparrow}
=-\mathcal{M}_{\downarrow\uparrow\rightarrow\uparrow\uparrow}
= \frac{\lambda y}{\lambda_Z}\left[ 2x+a\tilde{\kappa}\left( 1-ax \right) \right],\notag\\
&\mathcal{M}_{\uparrow\downarrow\rightarrow\uparrow\downarrow}
=\mathcal{M}_{\downarrow\uparrow\rightarrow\downarrow\uparrow}
=-\mathcal{M}_{\uparrow\downarrow\rightarrow\downarrow\uparrow}=-\mathcal{M}_{\downarrow\uparrow\rightarrow\uparrow\downarrow}=\frac{\lambda y^2}{\lambda_Z}\left( -2+a^2\tilde{\kappa}\right),\notag\\
&\mathcal{M}_{\uparrow\downarrow\rightarrow\downarrow\downarrow}
=-\mathcal{M}_{\downarrow\uparrow\rightarrow\downarrow\downarrow}
=\frac{\lambda y}{\lambda_Z}\left[ 2x-a\tilde{\kappa}\left(1+ax \right) \right],\notag\\
&\mathcal{M}_{\downarrow\downarrow\rightarrow\uparrow\uparrow}
=\eta \left( 1-d\tilde{\kappa} \right)y^2, \notag\\
&\mathcal{M}_{\downarrow\downarrow\rightarrow\uparrow\downarrow}
=-\mathcal{M}_{\downarrow\downarrow\rightarrow\downarrow\uparrow}
= y\left[ \eta x-b\tilde{\kappa}\left( 1+a\delta x \right) \right],\notag\\
&\mathcal{M}_{\downarrow\downarrow\rightarrow\downarrow\downarrow}
=-\frac{\Delta}{2}+\frac{b\tilde{\kappa}}{2}\left(4x+a\Delta\right),
\end{align}
where
\begin{equation}
    \eta=1-2\Lambda, \quad \Delta=3+2\Lambda+\eta\cos{2\theta}, \quad \kappa=\frac{im_Z}{8c_W^2\Gamma_Z}, \quad \tilde{\kappa}=\frac{im_Z}{8c_W^2s_W^2\Gamma_Z}.
\end{equation}

\section{Dark magic scattering amplitudes}
\label{appendix:Dark_amplitudes}

The matrices for scattering amplitudes, in the basis of the different initial- and final-state spin combinations, are listed below. The different $a_i$, which are global factors, are shown in Table \ref{tab:Global_factors}.

\subsection*{Pair annihilation $f \bar{f} \rightarrow\chi\bar{\chi}$ with $m_{f} < m_{\chi}$}
\begin{equation}
    \mathcal{M} = a_1\begin{pmatrix}
        2 & 0 & 0 & 0 \\ 0 & \lambda_\chi &  -\lambda_\chi & 0 \\ 0 & -\lambda_\chi &  \lambda_\chi & 0\\ 0 & 0 & 0 &2
    \end{pmatrix}
    \label{eq:Scat_DAnni1}
\end{equation}

\subsection*{Pair annihilation $\chi \bar{\chi} \rightarrow f \bar{f}$ with $m_{\chi} < m_{f}$}

\begin{equation}
    \mathcal{M} = a_2\begin{pmatrix}
        2\lambda_\chi & 0 & 0 & 0 \\ 0 & 1 & -1 & 0 \\ 0 & -1 & 1  & 0\\ 0 & 0 & 0 &2\lambda_\chi
    \end{pmatrix}
    \label{eq:Scat_DAnni1_inv}
\end{equation}

\subsection*{M\o ller-like scattering $\chi \chi \rightarrow\chi\chi$}

\begin{equation}
    \mathcal{M}_{\text{PLF}} = a_3 \begin{pmatrix}
        -2\cos(\theta) & 0 & 0 & 0 \\ 0 & 2\sin^2(\theta/2) & -1-\cos(\theta) & 0 \\ 0 & -1-\cos(\theta) &  2\sin^2(\theta/2) & 0 \\ 0 & 0 & 0 & -2\cos(\theta)
        \label{eq:Scat_DMoller_PLF}
    \end{pmatrix}
\end{equation}

\begin{equation}
    \mathcal{M}_{\text{PHF}} = a_4 \begin{pmatrix}
        4\mu_{\text{D}}^2\cos(\theta) & -4\mu_{\text{D}}^2\sin(\theta) & -4\mu_{\text{D}}^2\sin(\theta) & 0 \\ 2\mu_{\text{D}}^2\sin(\theta) & 2+3\mu_{\text{D}}^2+4\mu_{\text{D}}^2\cos(\theta) & -2-3\mu_{\text{D}}^2+4\mu_{\text{D}}^2\cos(\theta) & -2\mu_{\text{D}}^2\sin(\theta) \\ 2\mu_{\text{D}}^2\sin(\theta) & -2-3\mu_{\text{D}}^2+4\mu_{\text{D}}^2\cos(\theta) &  2+3\mu_{\text{D}}^2+4\mu_{\text{D}}^2\cos(\theta) & -2\mu_{\text{D}}^2\sin(\theta) \\ 0 & 4\mu_{\text{D}}^2\sin(\theta) & 4\mu_{\text{D}}^2\sin(\theta) & 4\mu_{\text{D}}^2\cos(\theta)
        \label{eq:Scat_DMoller_PHF}
    \end{pmatrix}
\end{equation}

\subsection*{Bhabha-like scattering $\chi\bar{\chi} \rightarrow \chi\bar{\chi}$}

\begin{equation}
    \mathcal{M}_{\text{PLF}} = a_5 I_{4\times4}
    \label{eq:Scat_bhabha_PLF}
\end{equation}

\begin{equation}
    \mathcal{M}_{\text{PHF}} = a_6 \begin{pmatrix}
        -3 & 0 & 0 & 0 \\ 0 & -2 & 1 & 0 \\ 0 & 1 & -2 & 0 \\ 0 & 0 & 0 & -3
    \end{pmatrix}
    \label{eq:Scat_bhabha_PHF}
\end{equation}

\subsection*{Elastic scattering $\chi f \to \chi f$}

\begin{equation}
    \mathcal{M} =a_7I_{4\times4}
    \label{eq:Dark_fermion_scattering}
\end{equation}

\subsection*{Inverse pair annihilation $\chi\bar{\chi} \rightarrow f \bar{f}$ with $m_\chi > m_f$}
With an overall factor of $a_{8}$, the amplitude matrix entries for this process are
\begin{align}
    \mathcal{M}_{\uparrow\uparrow \rightarrow \uparrow \uparrow} &= \mathcal{M}_{\downarrow\downarrow \rightarrow \downarrow \downarrow} = -3 - \lambda_\chi + (-1 + \lambda_\chi) \cos(2 \theta) \notag \\
    \mathcal{M}_{\uparrow\uparrow \rightarrow \downarrow \downarrow} &= \mathcal{M}_{\downarrow\downarrow \rightarrow \uparrow \uparrow} = -2 (-1 + \lambda_\chi) \sin(\theta)^2 \notag \\
    \mathcal{M}_{\uparrow\uparrow \rightarrow \uparrow \downarrow} &= - \mathcal{M}_{\uparrow\uparrow \rightarrow \downarrow \uparrow} = \mathcal{M}_{\downarrow\downarrow \rightarrow \uparrow \downarrow} = -\mathcal{M}_{\downarrow\downarrow \rightarrow \downarrow \uparrow} = 
    \mathcal{M}_{\uparrow\downarrow \rightarrow \uparrow \uparrow} \notag \\ &= -\mathcal{M}_{\downarrow\uparrow \rightarrow \uparrow \uparrow} = \mathcal{M}_{\uparrow\downarrow \rightarrow \downarrow \downarrow} = -\mathcal{M}_{\downarrow\uparrow \rightarrow \downarrow \downarrow} = -2 (-1 + \lambda_\chi) \cos(\theta) \sin(\theta) \notag \\
    \mathcal{M}_{\uparrow\downarrow \rightarrow \uparrow \downarrow} &= -\mathcal{M}_{\uparrow\downarrow \rightarrow \downarrow \uparrow} =\mathcal{M}_{\downarrow\uparrow \rightarrow \downarrow \uparrow}= -\mathcal{M}_{\downarrow\uparrow \rightarrow \uparrow \downarrow}   = -1 - \lambda_\chi - (-1 + \lambda_\chi) \cos(2 \theta)
    \label{eq:entries_Inv_Pair_Anni_SML}
\end{align}

\subsection*{Inverse pair annihilation $f \bar{f}\rightarrow\chi\bar{\chi}$ with $m_{f} > m_{\chi}$}
With $a_{9}$ as a prefactor,
\begin{align}
    \mathcal{M}_{\uparrow\uparrow \rightarrow \uparrow \uparrow} &= \mathcal{M}_{\downarrow\downarrow \rightarrow \downarrow \downarrow} = 2\lambda_\chi(1 + \lambda_\chi +(-1+\lambda_\chi)\cos^2(\theta)) \notag \\
    \mathcal{M}_{\uparrow\uparrow \rightarrow \downarrow \downarrow} &= \mathcal{M}_{\downarrow\downarrow \rightarrow \uparrow \uparrow} = -2 (-1 + \lambda_\chi) \lambda_\chi \sin^2(\theta) \notag \\
    \mathcal{M}_{\uparrow\uparrow \rightarrow \uparrow \downarrow} &= -\mathcal{M}_{\uparrow\uparrow \rightarrow \downarrow \uparrow} =\mathcal{M}_{\downarrow\downarrow \rightarrow \uparrow \downarrow} = -\mathcal{M}_{\downarrow\downarrow \rightarrow \downarrow \uparrow} = \mathcal{M}_{\uparrow\downarrow \rightarrow \uparrow \uparrow} \notag \\
    &=  -\mathcal{M}_{\downarrow\uparrow \rightarrow \uparrow \uparrow}= \mathcal{M}_{\uparrow\downarrow \rightarrow \downarrow \downarrow} = -\mathcal{M}_{\downarrow\uparrow \rightarrow \downarrow \downarrow} = -(-1 + \lambda_\chi) \lambda_\chi \sin(2 \theta) \notag \\
    \mathcal{M}_{\uparrow\downarrow \rightarrow \uparrow \downarrow} & = -\mathcal{M}_{\uparrow\downarrow \rightarrow \downarrow \uparrow} = \mathcal{M}_{\downarrow\uparrow \rightarrow \downarrow \uparrow} = -\mathcal{M}_{\downarrow\uparrow \rightarrow \uparrow \downarrow} = \lambda_\chi(1 + \lambda_\chi - (-1+\lambda_\chi)\cos(2\theta))
    \label{eq:entries_Inv_Pair_Anni_SMH}
\end{align}

\subsection*{Global factors}
The $a_{i}$ in front of the dark sector amplitudes, which drop in the computation of the $M_{2}$ distributions, are listed in Table \ref{tab:Global_factors}.
\begin{table}[h!]
    \centering
    \begin{tabular}{c|c}
    \toprule          
              & \\[-1em]
        $a_1$ & $\dfrac{4e'e\epsilon \,m_\chi^2}{(-1+\epsilon^2)(4m_\chi^2-m_{A'}^2+im_{A'}\Gamma_{A'})}$ \\
              & \\[-0.8em]
        \hline
              & \\[-1em]
        $a_2$ & $\dfrac{4e'e\epsilon\ m_{\chi}^{2} \lambda_\chi}{(-1+\epsilon^2)(4m_\chi^2\lambda_\chi^2-m_{A'}^2+im_{A'}\Gamma_{A'})}$ \\
              & \\[-0.8em]
        \hline
              & \\[-1em]
        $a_3$ & $\dfrac{8e'm_\chi^2 \mu_{\text{D}}^2}{\Gamma_{A'}^2(1-\epsilon^2)\lambda_a^2}$ \\
              & \\[-0.8em]
        \hline
              & \\[-1em]
        $a_4$ & $\dfrac{2{e'}^2 }{(1-\epsilon^2)\lambda_a^2}$ \\
              & \\[-0.8em]
        \hline
              & \\[-1em]
        $a_5$ & $-\dfrac{4i {e'}^2 m_\chi}{(1-\epsilon^2)\lambda_a\Gamma_{A'}}$ \\
              & \\[-0.8em]
        \hline
              & \\[-1em]
        $a_6$ & $\dfrac{4 {e'}^2 }{(1-\epsilon^2)\lambda_a}$ \\
              & \\[-0.8em]
        \hline
              & \\[-1em]
        $a_7$ & $\dfrac{4 e' e \,m_\chi^2 \epsilon^2 \lambda_\chi}{(-m_{A'}^2 +im_{A'} \Gamma_{A'})(1-\epsilon^2)}$ \\
              & \\[-0.8em]
        \hline
              & \\[-1em]
        $a_8$ & $\dfrac{2 e' e \epsilon\, m_\chi^2}{(1-\epsilon^2)(4m_\chi^2 - m_{A'}^2+i \Gamma_{A'} m_{A'} )}$ \\
              & \\[-0.8em]
        \hline
              & \\[-1em]
        $a_9$ & $\dfrac{-2 e' e \epsilon m_\chi^2}{(m_{A'}^2 - 4 m_\chi^2 \lambda_\chi^2 - i m_{A'} \Gamma_{A'}) (-1 + \epsilon^2)}$ \\
              & \\[-0.8em]
        \bottomrule
    \end{tabular}
    \caption{Global factors in the dark $U(1)$ scattering amplitudes.}
    \label{tab:Global_factors}
\end{table}

\section{Angular shifts in magic distributions}
\label{appendix:Angular_shifts}

Table \ref{tab:Phases} shows the angular shifts $\alpha_{i}$ of the hat functions defined through $\mathcal{F}_{n}({\theta}) = \widehat{\mathcal{F}}_{n,i}(\theta-\alpha_i)$ and $\mathcal{H}_{n,i}({\lambda_\chi,\theta}) = \widehat{\mathcal{H}}_{n,i}(\lambda_\chi,\theta-\alpha_i)$

\begin{table}[ht]
    \centering
    \begin{tabular}{cc}
        \toprule
       Magic distribution function  & $\alpha_i$ \\
       \toprule
        $\widehat{\mathcal{F}}_{5,1}$ & $\frac{1}{4} (\pi - \arctan(24/7))$\\
        \hline
        $\widehat{\mathcal{F}}_{5,2}$ & $\frac{1}{4} (\pi + \arctan(24/7))$\\
        \hline
        $\widehat{\boldsymbol{\mathcal{F}}}_{12}$ & $\frac{\pi}{2}$ \\
        \hline
        $\widehat{\boldsymbol{\mathcal{F}}}_{13}$ & $\frac{\pi}{2}$ \\
        \hline
        $\widehat{\mathcal{F}}_{17}$ & $\frac{\pi}{2}$ \\
        \hline
        $\widehat{\mathcal{F}}_{\text{HE -}8}$ & $\frac{\pi}{2}$ \\
        \hline
        $\widehat{\mathcal{F}}_{\text{HE -}9}$ & $\frac{\pi}{2}$ \\
        \hline
        $\widehat{\mathcal{H}}_{1}$ & $\frac{\pi}{2}$\\
        \hline
        $\widehat{\mathcal{H}}_{2,1}$ & $\frac{\pi}{2}$\\
        
        \hline
        $\widehat{\mathcal{H}}_{2,2}$ & $\frac{\pi}{4}$\\
        \hline
        $\widehat{\mathcal{H}}_{2,3}$ & $\frac{3\pi}{4}$\\
        \hline
        $\widehat{\mathcal{H}}_{5}$ & $\frac{\pi}{2}$\\
        \hline
        $\widehat{\mathcal{H}}_{6}$ & $\frac{\pi}{2}$\\
        \hline
        $\widehat{\mathcal{H}}_{7,1}$ & $\frac{\pi}{2}$\\
        \hline
        $\widehat{\mathcal{H}}_{7,2}$ & $\frac{\pi}{4}$\\
        \hline
        $\widehat{\mathcal{H}}_{7,3}$ & $\frac{3\pi}{4}$\\
        \hline
        $\widehat{\mathcal{H}}_{10}$ & $\frac{\pi}{2}$\\
        \bottomrule
    \end{tabular}
    \caption{Angular shifts for the hat functions obtained in both the EW and $U(1)$ dark sectors, where $\mathcal{F}_{n}({\theta}) = \widehat{\mathcal{F}}_{n,i}(\theta-\alpha_i)$ and $\mathcal{H}_{n,i}({\lambda_\chi,\theta}) = \widehat{\mathcal{H}}_{n,i}(\lambda_\chi,\theta-\alpha_i)$.}
    \label{tab:Phases}
\end{table}

\section{Variations in magic distribution functions}
\label{appendix:Variations_magic}
\begin{figure}[ht]
    \centering
    \includegraphics[width=0.7\linewidth]{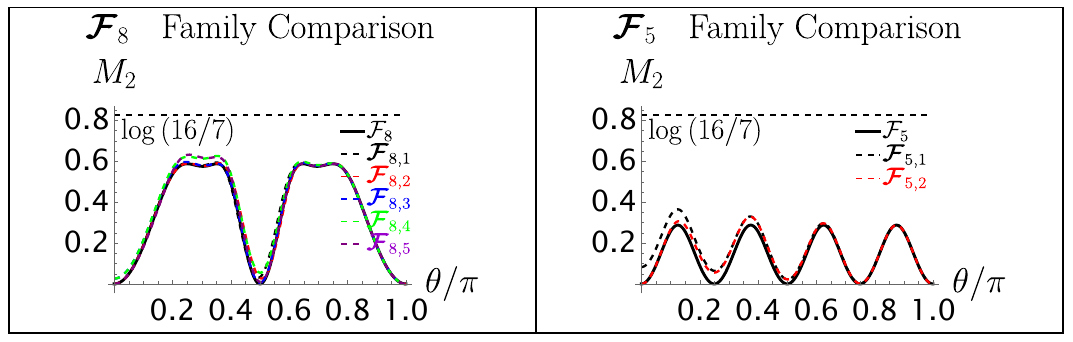}
    \caption{Family comparison plots between the different magic functions in pair annihilation scattering $e^-e^+\rightarrow \mu^-\mu^+$ at the $Z$-resonance.}
    \label{fig:compareAnniZR}
\end{figure}

\begin{figure}[h!]
    \centering
    \includegraphics[width=1\linewidth,trim={0 205 0 0}]{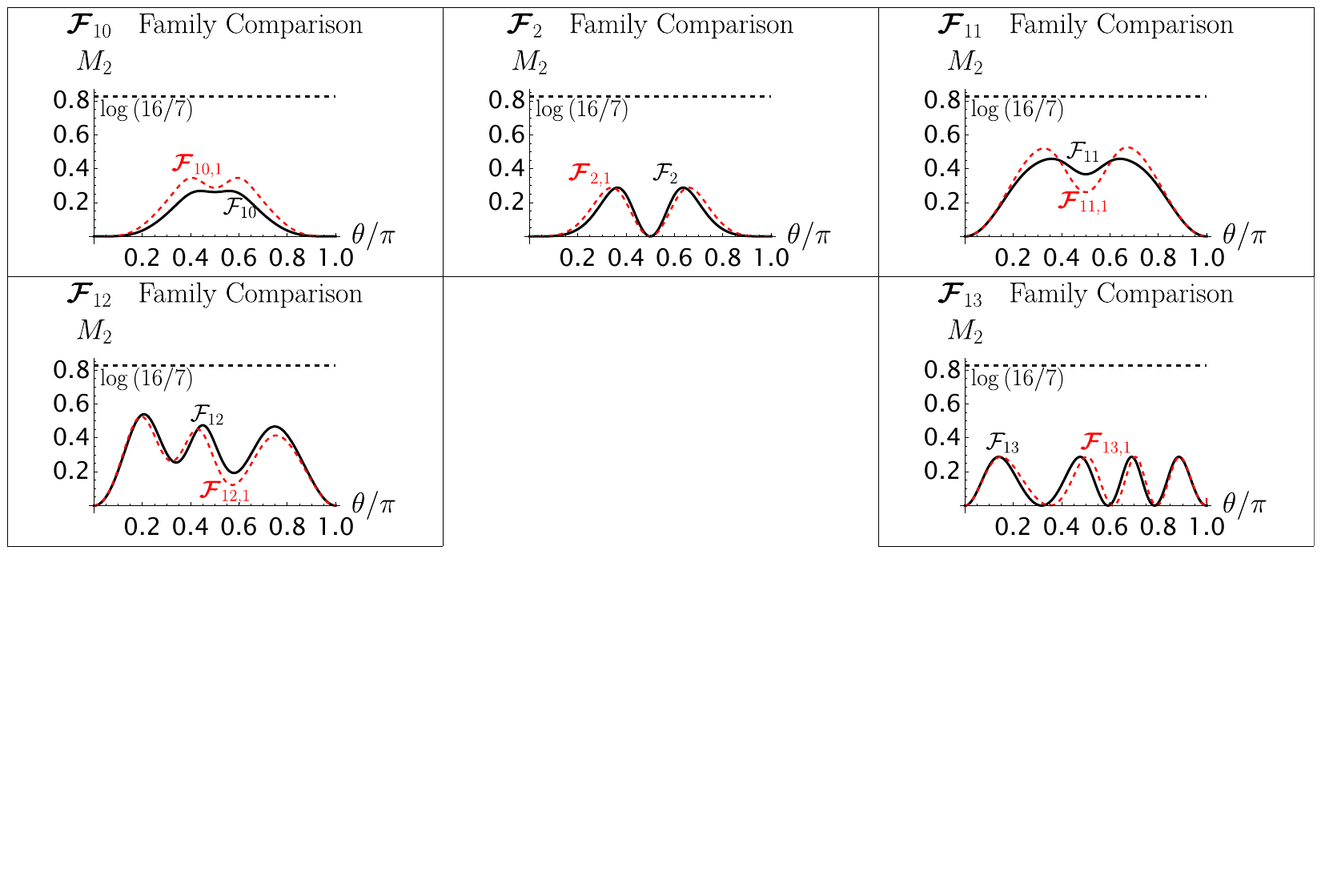}
    \caption{Family comparison plots between the different magic functions in M\o ller scattering $e^-e^-\rightarrow e^-e^-$ at the $Z$-resonance.}
    \label{fig:compareMollerZR}
\end{figure}

\begin{figure}[h!]
    \centering
    \includegraphics[width=1\linewidth,trim={0 225 0 0}]{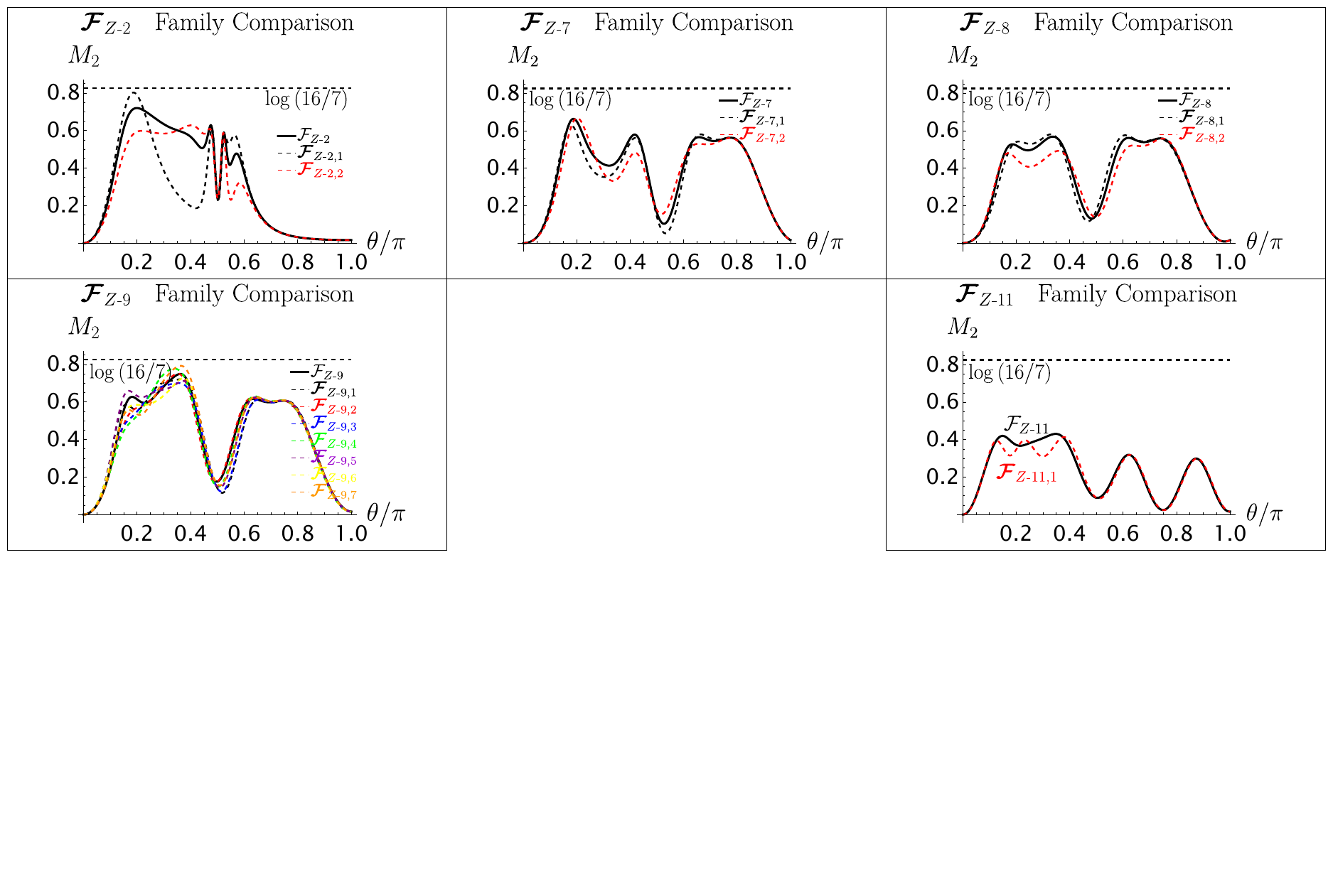}
    \caption{Family comparison plots between the different magic functions in Bhabha scattering $e^-e^+\rightarrow e^-e^+$ at the $Z$-resonance.}
    \label{fig:compareBhabhaZR}
\end{figure}

\begin{figure}[h!]
    \centering
    \includegraphics[width=1\linewidth]{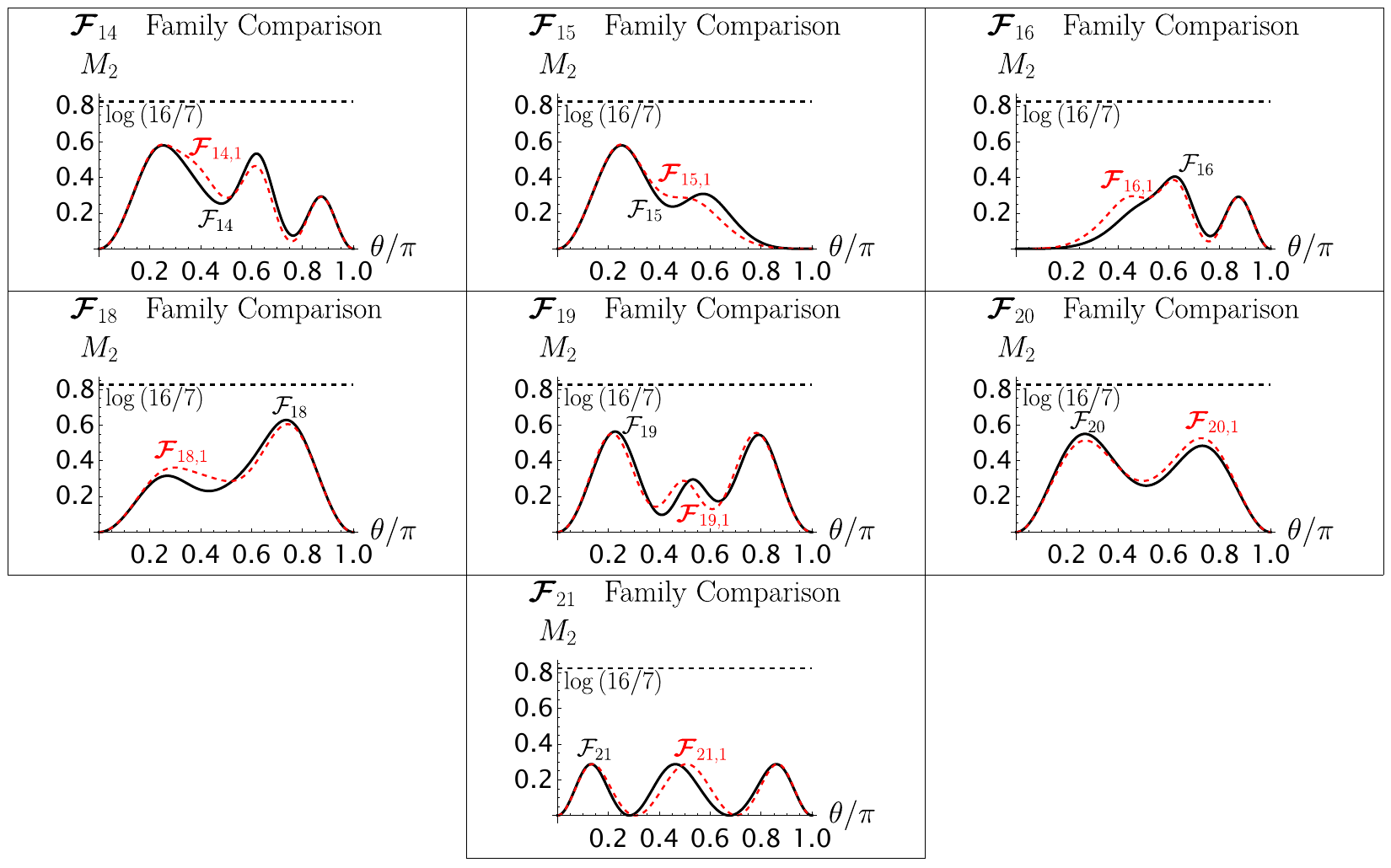}
    \caption{Family comparison plots between the different magic functions in elastic scattering $e^-\mu^-\rightarrow e^-\mu^-$ at the $Z$-resonance.}
    \label{fig:compareemuZR}
\end{figure}

As discussed in the $Z$-resonance section, the corresponding magic distribution functions often acquire complicated analytic expressions while still preserving very similar qualitative angular behavior. For this reason, they are organized into families, denoted by $\boldsymbol{\mathcal{F}}_{n,i}$, where the index $n$ labels the general family and the additional index $i$ distinguishes the different variations within that family. In this way, functions carrying the same first index share the same characteristic structure, while the second index identifies small deformations of said functions. In this Appendix, tables \ref{tab:varAnniZR} to \ref{tab:varemuZR} and their corresponding plots in Figs. \ref{fig:compareAnniZR} to \ref{fig:compareemuZR} accordingly organize those functions. In each case, the first column in the tables lists the representative variation $\boldsymbol{\mathcal{F}}_{n,i}$, while the second column indicates the stabilizer states whose corresponding magic distributions generate that particular variation. Thus, for each process the tables group all the functions that vary according to their parent family. The plots then allow the different members of a given family to be compared directly, making visible both their shared structure and their subtle differences.

\begin{table}[ht]
    \centering
    \begin{tabular}{cc}
        \toprule
        Magic distribution & Stabilizer state\\
        \toprule
        $\boldsymbol{\mathcal{F}}_{8,1}$ & 1, 21, 24, 25, 28, 29, 32, 33, 36  \\
        \hline
         $\boldsymbol{\mathcal{F}}_{8,2}$ & 2, 22, 23, 26, 27, 30, 31, 34, 35  \\
        \hline
         $\boldsymbol{\mathcal{F}}_{8,3}$ & 13, 15, 18, 20, 39, 53, 54, 55, 57 \\
        \hline
         $\boldsymbol{\mathcal{F}}_{8,4}$ & 14, 16, 17, 19, 40, 51, 52, 56, 58 \\
        \hline
         $\boldsymbol{\mathcal{F}}_{5,1}$ & 5, 6, 11, 12, 37, 46, 47, 49, 50 \\
        \bottomrule
    \end{tabular}
    \caption{Angular magic distributions variations obtained for $e^{-}e^{+}\to \mu^{-}\mu^{+}$ in the $Z$ resonance regime.}
    \label{tab:varAnniZR}
\end{table}

\begin{table}[h!]
    \centering
    \begin{tabular}{cc}
        \toprule
        Magic distribution & Stabilizer state\\
        \toprule
        $\boldsymbol{\mathcal{F}}_{10}$ & 9, 10  \\
        \hline
         $\boldsymbol{\mathcal{F}}_{2}$ & 11, 12, 46, 47  \\
        \hline
        $\boldsymbol{\mathcal{F}}_{11}$ & 13, 14, 15, 16, 17, 18, 19, 20, 21, 22, 23, 24, 25, 26, 27, 28 \\
        \hline
        $\boldsymbol{\mathcal{F}}_{12}$ & 29, 31, 34, 36, 55, 56, 57, 58 \\
        \hline
         $\boldsymbol{\widehat{\mathcal{F}}}_{12}$ & 30, 32, 33, 35, 51, 52, 53, 54 \\
        \hline
        $\boldsymbol{\mathcal{F}}_{13}$ & 45 \\
        \hline
         $\boldsymbol{\widehat{\mathcal{F}}}_{13}$ & 48 \\
        \bottomrule
    \end{tabular}
    \caption{Angular magic distributions variations obtained for M\o ller scattering in the $Z$ resonance regime.}
    \label{tab:varMollerZR}
\end{table}

\begin{table}[h!]
    \centering
    \begin{tabular}{cc}
        \toprule
        Magic distribution & Stabilizer state\\
        \toprule
        $\boldsymbol{{\mathcal{F}}}_{8,5}$ & 39  \\
        \hline
          $\boldsymbol{\mathcal{F}}_{Z\text{-}2,1}$ & 59 \\
        \hline
        $\boldsymbol{\mathcal{F}}_{Z\text{-}2,2}$ & 60 \\
        \hline
        $\boldsymbol{\mathcal{F}}_{Z\text{-}7,1}$ & 22, 23 \\
        \hline
         $\boldsymbol{\mathcal{F}}_{Z\text{-}7,2}$ & 25, 28 \\
        \hline
        $\boldsymbol{\mathcal{F}}_{Z\text{-}8,1}$ & 21, 24 \\
        \hline
         $\boldsymbol{\mathcal{F}}_{Z\text{-}8,2}$ & 26, 27 \\
        \hline
        $\boldsymbol{\mathcal{F}}_{Z\text{-}9,1}$ & 30, 31 \\
        \hline
        $\boldsymbol{\mathcal{F}}_{Z\text{-}9,2}$ & 33, 36 \\
        \hline
        $\boldsymbol{\mathcal{F}}_{Z\text{-}9,3}$ & 34, 35 \\
        \hline
        $\boldsymbol{\mathcal{F}}_{Z\text{-}9,4}$ & 51, 56 \\
        \hline
        $\boldsymbol{\mathcal{F}}_{Z\text{-}9,5}$ & 52, 58 \\
        \hline
        $\boldsymbol{\mathcal{F}}_{Z\text{-}9,6}$ & 53, 55 \\
        \hline
        $\boldsymbol{\mathcal{F}}_{Z\text{-}9,7}$ & 54, 57 \\
        \hline

        $\boldsymbol{{\mathcal{F}}}_{5,2}$ & 37  \\
        \hline
         $\boldsymbol{\mathcal{F}}_{Z\text{-}11}$ & 47 \\
        \bottomrule
    \end{tabular}
    \caption{Angular magic distributions variations obtained for Bhabha scattering in the $Z$ resonance regime.}
    \label{tab:varBhabhaZR}
\end{table}

\begin{table}[h!]
    \centering
    \begin{tabular}{|c|c|}
        \hline
        Magic distribution & Stabilizer state\\
        \hline
        $\boldsymbol{\mathcal{F}}_{14}$ & 5, 6, 49, 50  \\
        \hline
          $\boldsymbol{\mathcal{F}}_{15}$ & 7, 8, 59, 60 \\
        \hline
        $\boldsymbol{\mathcal{F}}_{16}$ & 9, 10 \\
        \hline
        $\boldsymbol{\mathcal{F}}_{18}$ & 13, 14, 15, 16, 17, 18, 19, 20, 21, 22, 23, 24, 25, 26, 27, 28 \\
        \hline
        $\boldsymbol{\mathcal{F}}_{19}$ & 29, 31, 34, 36, 55, 56, 57, 58 \\
        \hline
         $\boldsymbol{\mathcal{F}}_{20}$ & 30, 32, 33, 35, 51, 52, 53, 54 \\
        \hline
        $\boldsymbol{\mathcal{F}}_{21}$ & 45 \\
        \hline
         $\boldsymbol{\mathcal{F}}_{22}$ & 48 \\
        \hline
    \end{tabular}
    \caption{Angular magic distributions variations obtained for elastic scattering $e^-\mu^-\rightarrow e^-\mu^-$ in the $Z$ resonance regime.}
    \label{tab:varemuZR}
\end{table}

\clearpage

\bibliography{References}

\end{document}